\documentclass[twocolumn,numberedappendix,iop]{openjournal}
\usepackage{lineno}
\usepackage{graphicx,amsmath,amssymb,amstext}
\usepackage{amsbsy,amsfonts,amsthm,color}
\usepackage[colorlinks,linkcolor=blue,citecolor=blue,urlcolor=blue]{hyperref}
\usepackage[utf8]{inputenc}

\usepackage{xcolor}
\usepackage{ulem}
\usepackage[T1]{fontenc}
\usepackage[title]{appendix}
\usepackage{savesym}
\savesymbol{tablenum}
\usepackage{subcaption}  
\usepackage{graphicx}
\usepackage{subcaption}
\usepackage{siunitx} 
\restoresymbol{SIX}{tablenum}
\DeclareSIUnit\h{\text{$h$}}

\newcommand{\rossi}[1]{\textcolor{black}{  #1}}

\expandafter\let\csname longtable*\endcsname\relax
\expandafter\let\csname endlongtable*\endcsname\relax

\begin{document}
\title{Clustering redshift distribution calibration of weak lensing surveys \\ using the DESI-DR1 spectroscopic dataset}
\shorttitle{DESI-DR1 clustering redshift}

\author{
R.~Ruggeri,$^{1,*}$
C.~Blake,$^{2}$
J.~Aguilar,$^{3}$
S.~Ahlen,$^{4}$
D.~Bianchi,$^{5,6}$
D.~Brooks,$^{7}$
F.~J.~Castander,$^{8,9}$
T.~Claybaugh,$^{3}$
A.~Cuceu,$^{3}$
K.~S.~Dawson,$^{10}$
A.~de la Macorra,$^{11}$
B.~Dey,$^{12,13}$
P.~Doel,$^{7}$
A.~Elliott,$^{14,15}$
N.~Emas,$^{2}$
S.~Ferraro,$^{3,16}$
A.~Font-Ribera,$^{17}$
J.~E.~Forero-Romero,$^{18,19}$
C.~Garcia-Quintero,$^{20}$
E.~Gaztañaga,$^{8,21,9}$
S.~Gontcho A Gontcho,$^{3,22}$
G.~Gutierrez,$^{23}$
J.~Guy,$^{3}$
B.~Hadzhiyska,$^{24,16}$
H.~K.~Herrera-Alcantar,$^{25,26}$
S.~Heydenreich,$^{27}$
K.~Honscheid,$^{28,14,15}$
C.~Howlett,$^{29}$
D.~Huterer,$^{30,31}$
M.~Ishak,$^{32}$
S.~Joudaki,$^{33}$
R.~Joyce,$^{34}$
D.~Kirkby,$^{35}$
A.~Krolewski,$^{36,37,38}$
O.~Lahav,$^{7}$
C.~Lamman,$^{15}$
M.~Landriau,$^{3}$
J.~U.~Lange,$^{39}$
A.~Leauthaud,$^{27,40}$
M.~E.~Levi,$^{3}$
M.~Manera,$^{41,17}$
A.~Meisner,$^{34}$
R.~Miquel,$^{42,17}$
J.~Moustakas,$^{43}$
S.~Nadathur,$^{21}$
J.~ A.~Newman,$^{13}$
W.~J.~Percival,$^{36,37,38}$
C.~Poppett,$^{3,44,16}$
A.~Porredon,$^{33,45,46,15}$
F.~Prada,$^{47}$
I.~P\'erez-R\`afols,$^{48}$
A.~Robertson,$^{34}$
G.~Rossi,$^{49}$
E.~Sanchez,$^{33}$
C.~Saulder,$^{50}$
D.~Schlegel,$^{3}$
M.~Schubnell,$^{30,31}$
A.~Semenaite,$^{2}$
H.~Seo,$^{51}$
J.~Silber,$^{3}$
D.~Sprayberry,$^{34}$
G.~Tarl\'{e},$^{31}$
B.~A.~Weaver,$^{34}$
P.~Zarrouk,$^{52}$
R.~Zhou,$^{3}$
and H.~Zou$^{53}$
\\
{\it (Affiliations can be found after the references)}
}
\thanks{$^*$E-mail: rruggeri@qut.edu.au}
\shortauthors{R.~Ruggeri et al.}

\begin{abstract}
We estimate the source redshift distribution of current weak lensing surveys by applying the clustering-based redshift calibration technique, using the galaxy redshift sample provided by the Dark Energy Spectroscopic Instrument Data Release 1 (DESI-DR1).  We cross-correlate the Bright Galaxy Survey (BGS), Luminous Red Galaxies (LRGs) and Emission Line Galaxies (ELGs) from DESI, within the redshift range $0.1 < z < 1.6$, with overlapping tomographic source samples from the Dark Energy Survey (DES), Kilo-Degree Survey (KiDS), and Hyper Suprime-Cam (HSC) survey. 
\rossi{Using realistic mock catalogues, we test the stability 
of the clustering redshift signal to fitting scale, reference-sample choice, and the evolution of source galaxy bias. We calibrate a source galaxy bias correction from simulations with conservative (factor-of-two inflated) priors to account for modelling uncertainties, and we explicitly model and marginalise over magnification contributions, which become non-negligible at $z \gtrsim 1$ due to the depth of the DESI ELG sample. }
We then compare the resulting bias-weighted redshift distributions to \rossi{previously calibrated redshift distributions from the literature, based on self-organising map (SOM) techniques}, finding agreement within uncertainties for all surveys and tomographic bins. Our results demonstrate that clustering redshifts enabled by DESI's unprecedented spectroscopic sample provides a robust, complementary, and independent constraint capable of reducing one of the dominant systematic uncertainties in weak lensing cosmology.
  \\[1em]
\end{abstract}

\maketitle

\section{Introduction}

A key requirement for precision cosmology with weak gravitational lensing is the accurate estimation of redshift distributions for source galaxies. Errors in these estimates can propagate into significant biases in inferred cosmological parameters, especially those probing the amplitude of structure growth and the nature of dark energy
\citep[e.g.,][]{2006MNRAS.366..101H, 2015APh....63...81N, 2019MNRAS.486.2730A, 2023MNRAS.518..709Z, 2024PhRvD.109h3528L, 2025JCAP...03..064A}. Despite progress in photometric redshift (photo-$z$) methodologies \citep[e.g.,][]{2015ApJ...813...53M, 2018MNRAS.478..592H, 2019NatAs...3..212S, 2022ARA&A..60..363N, 2024MNRAS.531.2582M}, current approaches are still limited by degeneracies in galaxy colours, the impact of dust, and the non-representativity of spectroscopic samples used for training. These limitations can potentially lead to systematic uncertainties in the inferred redshift distributions.  Indeed, redshift-distribution calibration remains a key issue for interpreting evidence for ``tensions'' in cosmological parameter determinations between analyses of weak lensing surveys and the cosmic microwave background \citep[for recent discussions, see,][]{2025A&A...703A.144W, 2026PhRvD.113j3530Z}.

As a complementary approach for constraining source redshift distributions, ``clustering-based redshift'' (clustering-$z$) methods infer the redshift distribution of a photometric sample through angular cross-correlations with a spectroscopic reference sample of known redshifts \citep[e.g.,][]{Newman2008, 2010ApJ...721..456M, Menard2013, 2013MNRAS.431.3307S, 2013MNRAS.433.2857M, Rahman2015}.  Crucially, the reference sample does not need to match the photometric dataset in galaxy type or colour -- only to overlap in redshift and area, hence may be constructed from the brightest or most observationally-accessible galaxies at a given redshift.  The potential of this technique has been widely \rossi{demonstrated} in simulations and real surveys \citep[e.g.][]{Johnson2017, 2017MNRAS.467.3576M, 2018MNRAS.474.3921S, 2020JCAP...05..047K, 2020A&A...642A.200V, 2022MNRAS.510.1223G, 2022MNRAS.513.5517C, 2025A&A...702A.155D}, establishing clustering-$z$ as a credible supplement to standard photo-$z$ techniques, such as direct calibration \citep[e.g.,][]{2008MNRAS.390..118L, 2020A&A...633A..69H} via neural networks or Self-Organizing Map (SOM)-based methods \citep[e.g.,][]{2019MNRAS.489..820B, 2020A&A...637A.100W}.  Direct photo-$z$ approaches and clustering-based redshift estimation may be combined in hybrid pipelines \citep[e.g.,][]{2021A&A...647A.124H, 2021MNRAS.505.4249M, 2023MNRAS.524.5109R, 2025A&A...703A.144W, 2025OJAp....846159A, 2026PhRvD.113l3520L}, which are widely recognized as a promising strategy for robust source redshift calibration.

The Dark Energy Spectroscopic Instrument (DESI) is a next-generation spectroscopic survey designed to map the large-scale structure of the Universe with unprecedented precision \citep{desi_collaboration_desi_2016, desi_collaboration_desi_2016b, 2022AJ....164..207D, 2024AJ....167...62D}.  During its eight-year survey, DESI is obtaining high-precision redshifts for more than 60 million galaxies and quasars across a footprint of more than $17{,}000$ deg$^2$, spanning the redshift range $0 < z < 3.5$.  The survey components include the Bright Galaxy Survey \citep[BGS,][]{2023AJ....165..253H} at low redshift, Luminous Red Galaxies \citep[LRGs,][]{2023AJ....165...58Z} at intermediate redshifts, and Emission Line Galaxies \citep[ELGs,][]{2023AJ....165..126R} at higher redshifts, providing dense and well-characterised spectroscopic samples for cosmological inference.

Although DESI's primary science goal is to measure baryon acoustic oscillations and redshift space distortions to constrain the expansion history and growth of structure \citep{2025JCAP...07..017A, desi_fs}, the breadth and depth of its spectroscopic dataset make it an ideal reference for cross-survey analyses.  In particular, DESI's significant overlap with weak lensing surveys such as the Dark Energy Survey (DES; \citealt{Abbott2018, 2021ApJS..254...24S}), the Kilo-Degree Survey (KiDS; \citealt{Kuijken2015, 2019A&A...625A...2K}), and the Hyper Suprime-Cam (HSC) lensing survey \citep{Aihara2018, 2022PASJ...74..247A} unlocks synergies between imaging and spectroscopy that are essential for redshift calibration, systematics mitigation, and next-generation $3 \times 2$-pt cosmology.  As part of this program,  DESI provides an optimal reference for clustering redshift analyses due to its spectroscopic precision, statistical power, and wide redshift coverage.

The goal of this work is to use the DESI Data Release 1 (DR1) dataset to test and apply clustering-$z$ techniques to calibrate the redshift distributions of tomographic weak lensing samples from DES, KiDS, and HSC.  We aim to benchmark and compare clustering-$z$ calibrations against those based on SOM techniques, using DESI spectroscopy as the reference sample \citep{2026ApJ...998..263L, 2025arXiv251215964B, 2026JCAP...05..073C}.  This study is hence part of a broader program to exploit DESI as a spectroscopic backbone for redshift calibration across multiple imaging surveys.  We begin by validating the clustering-$z$ methodology using simulated catalogues, assessing fitting strategies, magnification contributions, and the impact of source galaxy bias. We then apply the clustering-$z$ method to cross-correlate DESI BGS, LRG, and ELG samples with weak lensing datasets, to estimate the source redshift distributions along with associated uncertainties.   The methodology and results presented here are relevant not only for current weak lensing studies, but also for future Stage-IV surveys such as the Vera C. Rubin Observatory Legacy Survey of Space and Time (LSST; \citealt{2019ApJ...873..111I}) and the \textit{Euclid} mission (\citealt{2011arXiv1110.3193L, 2025A&A...697A...1E}), where redshift calibration will play a pivotal role in controlling systematics and achieving the required cosmological precision.

Our paper is structured as follows. Sec.~\ref{sec:methodology} introduces the clustering-$z$ methodology, formalism, and modelling choices, and our mitigation strategies for systematic effects \rossi{associated} with magnification, source bias, and cosmology dependence.  Sec.~\ref{sec:datasets} outlines the spectroscopic and photometric datasets employed, including DESI lenses and public imaging survey data. Sec.~\ref{sec:measurements} presents the angular clustering measurements, while Sec.~\ref{sec:fits} details the fitting procedure used to extract redshift distributions and the tests performed on the mocks. Sec.~\ref{sec:results} then summarises the recovered redshift distributions, corrections, and results for DES, KiDS, and HSC. Finally, Sec.~\ref{sec:conclu} concludes and discusses the implications of these findings.
 
\section{Methodology}
\label{sec:methodology}

In this section we present the methodology used for the clustering-based redshift estimation performed in this work.  The technique of clustering-$z$ or ``cross-correlation redshifts'', was originally proposed by \citet{Newman2008} and further developed by studies such as \citet{2010ApJ...721..456M, Menard2013, 2013MNRAS.431.3307S, 2013MNRAS.433.2857M, Rahman2015}.  It has since been applied in a number of survey contexts, including KiDS \citep[e.g.,][]{Johnson2017, 2020A&A...642A.200V, 2025A&A...703A.144W}, DES \citep[e.g.,][]{2022MNRAS.510.1223G, 2022MNRAS.513.5517C, 2026PhRvD.113l3549D}, and HSC \citep[e.g.,][]{2023MNRAS.524.5109R}.  In our work, we particularly follow the approach presented by \cite{2020A&A...642A.200V} and \cite{2022MNRAS.510.1223G}.

Our pipeline differs from those adopted in previous clustering-$z$ analyses, where the effects of lensing magnification were often neglected or assumed to be relevant only for the very high-redshift tail of the source distribution. In practice, magnification alters the observed cross-correlation signal through flux and size selection effects, and its impact grows with the depth of the reference sample \citep{Menard2010}. Since our reference galaxies extend to significantly higher redshifts (up to $z \sim 1.6$ with the DESI ELG sample), the contribution from magnification is potentially no longer negligible, and must be incorporated directly in the modelling.  We therefore build a likelihood that explicitly includes magnification terms in the theoretical templates. To account for the associated uncertainty, we introduce additional free parameters connected to the galaxy–magnification angular power spectra, which we marginalise over in the inference.  This approach allows us to propagate magnification-induced systematics into the redshift calibration errors in a self-consistent way.

\subsection{Formalism}
\label{sec:formalism}

The clustering-$z$ approach is based on the cross-correlation signal between two overlapping samples: a sample with \textit{unknown} redshift probability distribution, $p_\mathrm{u}(z)$, such as a photometric imaging sample, and a \textit{reference} sample for which the redshift distribution, $p_\mathrm{r} (z)$, is accurately measured through spectroscopy.   The objective is to infer the redshift probability distribution of the unknown sample.   The method employs the property that, neglecting magnification effects, the cross-correlation signal between the two samples is non-zero only if the two are physically overlapping in redshift.  Further, provided we know the underlying cosmological model, we can model the signal of this cross-correlation and use it to infer the unknown redshift distribution.

To illustrate the process, we start by modelling the angular cross-power spectrum between the unknown and reference samples assuming the Limber approximation
\rossi{ and linear, scale-independent galaxy bias over the scales considered, }
\begin{equation}
  C_{\ell,ur} = \int \mathrm{d}z \, \frac{p_\mathrm{u}(z) p_\mathrm{r} (z)}{\chi^2(z) \, \frac{\mathrm{d} \chi}{\mathrm{d}z}} b_\mathrm{u}(z) b_\mathrm{r}(z) \, P_m \left( \frac{\ell}{\chi(z)},z \right) ,
\label{eq:crosspow}
\end{equation}
where $b_\mathrm{u}(z)$ and $b_\mathrm{r}(z)$  denote the linear galaxy bias of the unknown and reference samples as a function of redshift $z$, $\chi(z)$ is the comoving distance, and $P_m(k,z)$ is the matter power spectrum.  If we assume a ``top-hat'' distribution for the reference sample $p_\mathrm{r} (z) = 1/\Delta z$ in a narrow range $z_\mathrm{r} \pm \Delta z/2$, we can approximate Eq.~\ref{eq:crosspow} as,
\begin{equation}
  C_{\ell,ur} \approx \frac{p_\mathrm{u}(z_\mathrm{r}) \, b_\mathrm{u}(z_\mathrm{r}) \, b_\mathrm{r}(z_\mathrm{r}) \, P_m \left( \frac{\ell}{\chi(z_\mathrm{r})}, z_\mathrm{r} \right)}{\chi^2(z_\mathrm{r}) \, \frac{\mathrm{d} \chi}{\mathrm{d}z}(z_\mathrm{r})} ,
\end{equation}
noting this is a valid approximation when the spectroscopic redshift range of the reference sample is narrow compared to the redshift range over which the galaxy bias and matter power spectrum vary significantly.

The corresponding angular galaxy cross-correlation function $w_{ur}(\theta)$ is related to the above angular power spectrum as,
\begin{equation}
  w_{ur}(\theta) = \int \frac{\mathrm{d} \ell \, \ell}{2\pi} \, C_{\ell,ur} \, J_0(\ell \theta) ,
\label{wmm}
\end{equation}
and hence, under the above assumptions, can be written,
\begin{equation}
\begin{split}
  w_{ur}(\theta) &= \frac{p_\mathrm{u}(z_\mathrm{r}) \, b_\mathrm{u}(z_\mathrm{r}) \, b_\mathrm{r}(z_\mathrm{r})}{\chi^2(z_\mathrm{r}) \, \frac{\mathrm{d} \chi}{\mathrm{d}z}(z_\mathrm{r})} \\ &\times \int \frac{\mathrm{d} \ell \, \ell}{2\pi} \, J_0(\ell \theta) \, P_m\left( \frac{\ell}{\chi(z_\mathrm{r})}, z_\mathrm{r} \right) .
\label{eq:crosscorr}
\end{split}
\end{equation}
To produce these models, we integrated the angular power spectrum across multipoles $1 \le \ell \le 10^{5}$.

The angular auto-power spectrum of the reference sample is given by,
\begin{equation}
  C_{\ell,rr} = \int \mathrm{d}z \, \frac{p_\mathrm{r} ^2(z)}{\chi^2(z) \, \mathrm{d} \chi/\mathrm{d}z} \,
  b_\mathrm{r}^2(z) \, P_m \left( \frac{\ell}{\chi(z)}, z \right) .
\end{equation}
Assuming the same top-hat reference-sample distribution as before, we find,
\begin{equation}
  C_{\ell,rr} = \frac{b_\mathrm{r}^2(z_\mathrm{r}) \, P_m \left( \frac{\ell}{\chi(z_\mathrm{r})}, z_\mathrm{r}    \right)}{\Delta z \, \chi^2(z_\mathrm{r}) \, \frac{\mathrm{d} \chi}{\mathrm{d}z}(z_\mathrm{r})} .
\end{equation}
Hence, the angular auto-correlation function of the reference sample can be written as
\begin{equation}
\label{eq:wll}
\begin{split}
 w_{rr}(\theta) &= \frac{b_\mathrm{r}^2(z_\mathrm{r})}{ \Delta z \, \chi^2(z_\mathrm{r}) \, \frac{\mathrm{d} \chi}{\mathrm{d}z}(z_\mathrm{r})} \\ &\times \int \frac{\mathrm{d} \ell \, \ell}{2\pi} \, J_0(\ell\,\theta)\, P_m \left( \frac{\ell}{\chi(z_\mathrm{r})}, z_\mathrm{r} \right) .
\end{split}
\end{equation}
If we define the matter auto-correlation function $w_m(\theta)$ in the same redshift range as the reference sample by setting $b_\mathrm{r} = 1$ in Eq.~\ref{eq:wll}, we can express the \rossi{reference-unknown sample cross-correlation and reference sample auto-correlation functions as},
\begin{equation}
  w_{ur}(\theta) = p_\mathrm{u}(z_\mathrm{r}) \, b_\mathrm{u}(z_\mathrm{r}) \, b_\mathrm{r}(z_\mathrm{r}) \, \Delta z \, w_m(\theta) , 
\end{equation}
and,
\begin{equation}
  w_{rr}(\theta) = b_\mathrm{r}^2(z_\mathrm{r}) \, w_m(\theta).
  \label{eq:wurr}
\end{equation}
Hence, if we fit the correlation functions with amplitudes $w_{ur}(\theta) = A_{ur} \, w_m(\theta)$ and $w_{rr}(\theta) = A_{rr} \, w_m(\theta)$, we can use these amplitudes $A_{ur} = p_\mathrm{u} b_\mathrm{u} b_\mathrm{r} \Delta z$ and $A_{rr} = b_\mathrm{r}^2$ to determine the product of the unknown redshift distribution and galaxy bias as a function of redshift,
\begin{equation}
  b_\mathrm{u}(z_\mathrm{r}) p_\mathrm{u}(z_\mathrm{r}) = \frac{A_{ur}(z_\mathrm{r})}{\sqrt{A_{rr}(z_\mathrm{r})} \, \Delta z} .
\label{eq:bupu}
\end{equation}

\subsection{Magnification effects}
\label{sec:magnification} 

Magnification bias arises because weak gravitational lensing by large-scale structure alters both the observed fluxes and the apparent sizes of galaxies. As a result, galaxies that would otherwise fall below a survey’s flux limit can be magnified into the sample, while others may be demagnified out of it.  The net effect depends on the slope of the galaxy number counts as a function of magnitude: if the slope is steep, magnification increases the observed number density of galaxies, whereas for shallower slopes, it may decrease it. 
\rossi{In addition to this selection effect, magnification changes the observed solid angle and hence dilutes the apparent number density of sources. The effective magnification response therefore contains both the response of the sample selection to magnification and this geometrical area-dilution contribution. For a simple flux-limited sample, the latter appears as an additional constant term in the usual number-count expression. More generally, when the sample selection depends on multiple observed properties, the effective magnification response need not be given by the derivative of the number counts with respect to a single magnitude threshold \citep[e.g.][]{2026PhRvD.113l3520L}.}

This change in the observed density introduces correlations between physically unassociated populations at different redshifts, contaminating the clustering signal used to calibrate redshift distributions.  Magnification hence introduces an additional contribution to the observed clustering signal, which becomes increasingly important at higher redshifts; most previous studies neglected this effect because the redshift range considered was relatively low \citep[for an exception, see][]{2020JCAP...05..047K}. However, since we include higher redshift samples, these effects potentially become non-negligible, and we therefore apply a model to incorporate these effects.

The contribution of magnification to the angular cross-correlation function between a reference sample $r$ in a narrow redshift bin, and an unknown sample $u$ in a broad tomographic bin, is given by,
 \begin{equation}
w^{\rm mag,tot}_{ur}(\theta)
= \int \frac{{\rm d}\ell\,\ell}{2\pi}
C_{\rm mag}(\ell)J_0(\ell\theta).
\end{equation}
where \citep[following][]{2022MNRAS.510.1223G},
\begin{equation}
C_\mathrm{mag}(\ell) = b_\mathrm{r} \alpha_\mathrm{u} \, C_{g\kappa}^{ru}(\ell) + b_\mathrm{u} \alpha_\mathrm{r}  \, C_{g\kappa}^{ur}(\ell). 
\end{equation}
In this equation, $C_{g\kappa}^{ru}(\ell)$ is the galaxy–convergence angular power spectrum with the reference sample acting as the lenses and the unknown sample as the sources (evaluated with $b_\mathrm{r} = 1$), whilst $C_{g\kappa}^{ur}(\ell)$ is the corresponding galaxy–convergence angular power spectrum with the unknown sample as the lenses and the reference sample as the sources (evaluated with $b_\mathrm{u} = 1$); $\alpha_{r}$ and $\alpha_{u}$ denote the number-count slope parameters of the reference and unknown samples, respectively.  In both cases, the galaxy-convergence cross-power spectrum takes the form,
\begin{equation}
\begin{split}
    &C^{ru}_{g\kappa}(\ell) = \frac{3\,\Omega_m\,H_0^2}{2\,c^2} \\ &\times \int_{0}^{\infty} \frac{(1+z)}{\chi(z)}\,P_{gm}\left(\frac{\ell}{\chi(z)},z\right)\, p_\mathrm{r}(z)\, W_s(z)\,\mathrm{d}z,
\end{split}
\label{eq:clcross}
\end{equation}
in which the lensing efficiency $W_s(z)$ is given by,
\begin{equation}
\label{eq:lenseff}
    W_s(z) = \int_{z}^{\infty}p_\mathrm{u}(z')\, \left[ \frac{\chi(z')-\chi(z)}{\chi(z')} \right] \,\mathrm{d}z'.
\end{equation}
In Eq.~\ref{eq:clcross}, $P_{gm}(k,z)$ denotes the galaxy-matter cross-power spectrum, and the functions $p_\mathrm{u}(z)$ and $p_\mathrm{r}(z)$ represent the normalised redshift distributions of the source and lens samples, respectively.  The cross-correlation function due to magnification can then be expressed in terms of these two contributions as,
\begin{equation}
w^{\rm mag,tot}_{ur}(\theta) = b_\mathrm{r} \alpha_\mathrm{u} \, w_{ru}^{\mathrm{mag}}(\theta) + b_\mathrm{u} \alpha_\mathrm{r}  \, w_{ur}^{\mathrm{mag}}(\theta).
\end{equation}

Our model for the total cross-correlation function, including the contribution from clustering, is hence,
\begin{equation}
w_{ur}(\theta) = A_{ur} w_m(\theta) + p \, w_{ru}^{\mathrm{mag}}(\theta) + q \, w_{ur}^{\mathrm{mag}}(\theta),
\label{eq:wcross}
\end{equation}
where
\rossi{$p$ and $q$ are effective nuisance parameters representing the corresponding combinations of galaxy bias and magnification response. We do not fix these amplitudes using the raw number-count slopes, but instead marginalise over them in the fits and test that the inferred clustering amplitudes are stable when substantially broadening the priors adopted for $p$ and $q$. We note that this treatment does not constitute a full forward model, but rather an effective phenomenological correction.} 

To correct for magnification, for each narrow redshift slice we:
\begin{enumerate}
    \item Use the narrow redshift bounds of the reference sample, and the fiducial cosmological model defined below, to determine $w_m(\theta)$;
    \item Use the fiducial redshift distribution of the unknown sample (determined for example by direct calibration), and the narrow redshift bounds of the reference sample, to determine $w_{ru}^{\mathrm{mag}}(\theta)$ and $w_{ur}^{\mathrm{mag}}(\theta)$;
    \item Fit for $A_{ur}$, $p$, and $q$ as free parameters, using priors on $p$ and $q$;
    \item Deduce $b_\mathrm{u} p_\mathrm{u} = A_{ur}/[\sqrt{A_{rr}} \Delta z]$ as before.
\end{enumerate}
\subsection{Fiducial cosmology}

We model the angular correlation functions using a fiducial cosmological model consistent with the \textsc{Buzzard} simulation mock catalogues \citep{2019arXiv190102401D} that are further described below.  Specifically, we adopt a flat $\Lambda$CDM cosmology with matter density $\Omega_m = 0.286$, baryon density $\Omega_b = 0.048$, Hubble parameter $h = 0.7$, spectral index $n_s = 0.96$, scalar amplitude $A_s = 2.145 \times 10^{-9}$.  This fiducial cosmology is consistently applied across all theoretical predictions for both clustering and magnification contributions.
We generate the linear model power spectrum $P(k)$ using \texttt{CAMB} \citep{lewis2000}, with non-linear corrections based on \texttt{halofit} \citep{takahashi2012}.

While the correlation function templates depend on the assumed cosmological model (through the matter power spectrum, growth function, and comoving distances), the clustering-$z$ technique itself is largely cosmology-independent. As shown in previous studies such as \citet{2022MNRAS.510.1223G}, the dependence on cosmology cancels out in the ratio that defines the clustering-$z$ signal, particularly on large angular scales where the signal is linear and the bias evolution is smooth (the dependence on $\sigma_8$ does not affect the determination of the mean source redshift and cancels out when the redshift distribution is normalised).  We tested the robustness of this assumption in our context by re-computing the theoretical templates with alternate cosmological parameter values. These tests showed negligible variation in the derived amplitudes and the resulting $b_\mathrm{u}(z) \, p_\mathrm{u}(z)$ signal. 
\rossi{This insensitivity is expected because changes in the overall clustering amplitude due to the assumed cosmology are largely absorbed by the fitted normalisation. The recovered redshift distribution mainly depends on the evolution of the clustering signal with redshift, rather than on its absolute amplitude.}

We note that our analysis is performed on relatively small angular scales, where non-linear clustering, baryonic feedback, and scale-dependent galaxy bias can, in principle, affect the shape and amplitude of the angular correlation functions. Nevertheless, the impact of the fiducial cosmology on our final results remains limited. This is because the clustering-$z$ signal depends primarily on the relative amplitude of cross- and auto-correlations in matched redshift bins, and not on their absolute values.

\section{Datasets}
\label{sec:datasets}

This study aims to calibrate the photometric redshift distributions of weak lensing samples from KiDS-1000, DES-Y3, HSC-Y1 and HSC-Y3, through cross-correlation with spectroscopic galaxies from the DESI-DR1 dataset.  We use galaxies from the Bright Galaxy Survey (BGS), Luminous Red Galaxy (LRG), and Emission Line Galaxy (ELG) samples as the reference population.  In this section, we briefly summarise these datasets.

\subsection{DESI-DR1}

The Dark Energy Spectroscopic Instrument (DESI) is a multi-fibre spectrograph located at the 4-metre Mayall Telescope at Kitt Peak National Observatory.  The instrument design is summarised by \cite{2022AJ....164..207D}, including the optical corrector \citep{2024AJ....168...95M}, a robotic positioner \citep{2023AJ....165....9S} that can assign up to 5000 fibres to targets in the focal plane \citep{2024AJ....168..245P}, and the DESI spectroscopic pipeline \citep{2023AJ....165..144G} which processes the resulting data.  Over its planned main survey program \citep{2023AJ....166..259S}, which commenced in May 2021, DESI will increase the size of existing large-scale structure samples by more than an order-of-magnitude, collecting over 60 million spectra of galaxies and quasars across $17{,}000$\,deg$^2$ \citep{desi_collaboration_desi_2016, desi_collaboration_desi_2016b, 2024AJ....167...62D}.  In our study, we use the DESI Data Release 1 \citep[DR1,][]{2026AJ....171..285D}, which consists of all data acquired during the first 13 months of the main survey up to June 2022, including high-confidence redshifts for 13.1M galaxies.  The DESI survey obtains spectra for four principal target classes, photometrically-selected from the DESI Legacy optical imaging surveys \citep{2019AJ....157..168D}.  We use spectroscopic redshifts from the DESI BGS, LRG, and ELG samples to construct the reference population for clustering-$z$ measurements.

The DESI BGS targets a magnitude-limited sample of galaxies across redshifts $z < 0.5$, and serves as the primary reference for cross-correlations with low-redshift photometric bins.  The selection criteria of the BGS are described by \cite{2023AJ....165..253H}, resulting in a sample with density 854 deg$^{-2}$ containing redshift measurements for over 6.3 million galaxies in DR1 \citep{2026AJ....171..285D}. The BGS consists of a magnitude-limited Bright sample with $r < 19.5$, and a colour-selected Faint component with $19.5 < r < 20.175$. For this study, we will only consider the BGS Bright sample, as the Faint sample suffers from complications regarding incompleteness and systematics \citep{2023AJ....165..253H}.  We also cut the sample to the low-redshift range $0.1 < z < 0.4$, following the DR1 Baryon Acoustic Oscillation and full-shape redshift bins \citep{2025JCAP...07..017A, desi_fs}.

The DESI LRG sample spans the redshift range $0.4 < z < 1.1$, selected through colour and magnitude cuts in $g$, $r$, and $z$ bands \citep{2023AJ....165...58Z}, resulting in a sample with density 605 deg$^{-2}$ containing redshift measurements for over 2.8 million galaxies in DR1 \citep{2026AJ....171..285D}.  LRGs have strong clustering amplitudes and high spectroscopic success rates, making them ideal for calibrating intermediate-redshift photometric distributions.

The DESI ELG sample provides spectroscopic coverage in the redshift range $0.8 < z < 1.6$ suitable for calibrating higher-redshift photometric distributions, targetting galaxies with strong [OII] emission using $g$-band selections and morphology cuts \citep{2023AJ....165..126R}.  We note that [OII] emission is shifted out of the DESI spectral range for $z > 1.62$.  The main subset of the ELG selection has a density of 1940 deg$^{-2}$, resulting in 3.9 million spectroscopic redshift in DR1 \citep{2026AJ....171..285D}.  We note that we are using the sample internally referred to as ``ELG\_LOPnotqso'' which favours targets in the redshift range $1.1 < z < 1.6$ and excludes quasars \citep{2023AJ....165..126R}.  Whilst the redshift range $0.8 < z < 1.1$ allows both LRG and ELG selections, we find that LRGs have superior signal-to-noise performance for correlation analysis, and hence we only utilise ELGs in redshift range $1.1 < z < 1.6$ for our main analysis, although we check that both samples yield consistent results.

All three spectroscopic samples are corrected for observational systematics using imaging weights, which account for variations in seeing, depth, stellar density, and Galactic extinction, as described by \cite{2025JCAP...01..125R} and \cite{2025JCAP...07..017A}, and we assume the samples are effectively homogeneous after these corrections.  On small angular scales, where fibre collisions lead to missing pairs, we also apply the Pairwise Inverse Probability (PIP) weights \citep{2025JCAP...04..074B} for auto-correlation measurements. This technique upweights pairs based on their observational probability and is essential for recovering unbiased clustering measurements below the fibre collision scale.  For auto-correlation measurements of the reference samples, we use the full DESI footprint.  For cross-correlations between the reference and unknown samples, we trim spectroscopic sample to match the footprint of each imaging survey. This ensures consistency in mask application and proper normalization of the random catalogues used in the estimator.  For cross-correlation measurements we also apply individual inverse probability (IIP) weights, as discussed further in Sec.~\ref{sec:measurements}.

\subsection{DES-Y3}

We calibrate the photometric imaging datasets of three weak lensing surveys in our analysis.  First, we apply our clustering-$z$ method to the Dark Energy Survey Year 3 (DES-Y3) shear catalogue \citep{2021MNRAS.504.4312G}.  The Dark Energy Survey has used the Dark Energy Camera at the Blanco 4-metre Telescope to create the largest existing lensing survey.  The DES-Y3 catalogue is based on imaging data from the first three years of DES operation between 2013 and 2016, and contains more than $10^8$ objects over $4{,}143$ deg$^2$.  The shear catalogue has a weighted source number density of 5.6 arcmin$^{-2}$, and was created from this imaging dataset with the \textsc{metacalibration} pipeline \citep{2017ApJ...841...24S}.  We analysed DES-Y3 sources in the four tomographic bins defined by the DES collaboration \citep{2021MNRAS.505.4249M}, 
\rossi{constructed by grouping SOM cells with photo-$z$ bin edges $[0.00, 0.36, 0.63, 0.87, 2.00]$.}
 The area of overlap of DES-Y3 and DESI-DR1 is 851.3 deg$^2$.

\subsection{KiDS-1000}

We also include in our analysis the Kilo-Degree Survey (KiDS-1000) weak lensing catalogue \citep{2021A&A...645A.105G}.  KiDS has utilised the OmegaCAM instrument on the Very Large Telescope (VLT) Survey Telescope at the Paranal Observatory to image the sky in optical filters.  Overlapping near-infrared imaging has been provided by the VISTA-VIKING survey, resulting in a nine-band dataset enabling improved photometric redshift calibration \citep{2021A&A...647A.124H}.  KiDS-1000 contains 21 million galaxies across $1{,}006$ deg$^2$ with an effective number density 6.2 arcmin$^{-2}$, with source shapes measured using the {\it lens}fit model fitting technique \citep{2013MNRAS.429.2858M, 2017MNRAS.467.1627F}.  Our analysis uses the five tomographic source samples created by the KiDS collaboration, which are divided by photometric redshifts $[0.1, 0.3, 0.5, 0.7, 0.9, 1.2]$, where the redshift distribution of the sources is calibrated as described by \cite{2021A&A...647A.124H}.  The KiDS-1000 catalogue has an overlap of 446.8 deg$^2$ with the DESI-DR1 footprint.  Whilst our work was in progress, cosmological analysis of the extended KiDS-Legacy survey was presented by \cite{2025A&A...703A.158W}.

\subsection{HSC-Y1 and HSC-Y3}

The final weak lensing datasets we include in our study have been released by the Hyper Suprime-Cam survey, where we include both the Year 1 sample \citep[HSC-Y1,][]{2018PASJ...70S..25M} and the Year 3 catalogue \citep[HSC-Y3,][]{2022PASJ...74..421L}.  The HSC survey is using the 8.2-metre Subaru Telescope to image several regions of sky and hence create the deepest current weak lensing dataset.  The HSC-Y1 dataset is based on data acquired between 2014 and 2016, covering 137 deg$^2$ of sky with weighted source density 21.8 arcmin$^{-2}$.  The HSC-Y3 catalogue is compiled from data taken up to 2019, covering 434 deg$^2$ of sky with source density 19.9 arcmin$^{-2}$.  The catalogues are based on source shape measurements performed using a re-Gaussianization PSF correction method \citep{2018PASJ...70S..25M}, further calibrated for HSC-Y3 by realistic image simulations \citep{2022PASJ...74..421L}.  Our work uses the four tomographic source samples defined by the HSC collaboration, with photometric redshift divisions $[0.3, 0.6, 0.9, 1.2, 1.5]$, and redshift distributions inferred by \cite{2019PASJ...71...43H} for HSC-Y1, and \cite{2023MNRAS.524.5109R} for HSC-Y3.  The HSC datasets have almost complete overlap with DESI-DR1.

\subsection{Mock samples}
\label{sec:mocks}

We test and validate our analysis pipelines using mock catalogues sampled from the \textsc{Buzzard} N-body simulations \citep{2019arXiv190102401D}.  These datasets have already been used in the context of DESI-Lensing to quantify astrophysical systematics in galaxy-galaxy lensing analyses by \cite{2024OJAp....7E..57L}, and to perform a large-scale mock challenge analysis of covariance and cosmological-parameter recovery by \cite{2025OJAp....8E..24B}.  As described by these papers, the mock catalogues were tuned to match the redshift distribution, photometric-redshift scatter, source weights and tomographic binning of the KiDS-1000, DES-Y3 and HSC-Y1 datasets (the HSC-Y3 dataset was unavailable at the time).  Sources were identified and weights were assigned by sampling from the closest neighbours to the real lensing data catalogues in apparent magnitude and colour space, based on a {\sc KDTree} match.  Furthermore, mock DESI BGS, LRG and ELG samples were populated by halo occupation distribution recipes to match the clustering and number density of the DESI Early Data Release dataset \citep{2024AJ....168...58D}.
\rossi{The DESI mocks have uniform angular selection and therefore do not capture the spatial completeness variations present in the DESI DR1 data, which we note as a limitation of the mock validation.}

For the purpose of our validation tests, we used a single \textsc{Buzzard} simulation which we divided into closely-packed regions approximating the separate overlap areas of DESI-Y1 and KiDS-1000, DES-Y3 and HSC-Y1 \citep[as described by][]{2025OJAp....8E..24B}, where these regions contain area $483$, $806$ and $161$ deg$^2$, respectively.  Given that the spectroscopic redshifts of the mock source populations are known, we may use these simulations to test the recovery of source redshift distributions by the clustering-$z$ method, and to infer the source galaxy bias of the mock populations.

\section{Clustering measurements}
\label{sec:measurements}

In the clustering-$z$ method, we measure angular auto- and cross-correlations between the reference dataset in narrow redshift bins of width $\Delta z$, and the samples with an unknown redshift distribution.  The formalism assumes that the galaxy bias of the samples can be described by an effective linear scaling of the underlying matter fluctuations, as described in Sec.~\ref{sec:formalism}.  This assumption may not be reliable across a range of scales, especially on the small scales which are commonly employed by the clustering-$z$ method in order to maximise the signal-to-noise of the amplitude determination.  For this reason, we employ scale cuts to define a range of separations for which an ``effective'' linear bias may be considered an acceptable approximation.  We choose fiducial scale cuts corresponding to transverse comoving distances between $R_{\mathrm{min}} = 1.5 \, h^{-1}\mathrm{Mpc}$ and $R_{\mathrm{max}} = 5 \,h^{-1}\mathrm{Mpc}$.
\rossi{This choice is similar to that adopted in previous clustering-$z$ analyses \citep[e.g.,][]{2022MNRAS.510.1223G}, facilitating comparison with earlier work, and provides a compromise between high signal-to-noise and reduced sensitivity to small-scale nonlinearities and observational systematics.}
For each narrow redshift bin with edges $z_i$ and $z_{i+1}$, we convert this co-moving separation to an angular range using the comoving distance at the bin midpoint, $\bar z=(z_i+z_{i+1})/2$,
\begin{equation}
\theta_{\min}(\bar z)=\frac{R_{\min}}{\chi(\bar z)}, \qquad \theta_{\max}(\bar z)=\frac{R_{\max}}{\chi(\bar z)} .
\end{equation}
We perform angular correlation function measurements in $30$ logarithmic bins spanning $0.003 < \theta < 3^\circ$.  Within each redshift slice, we retain only the $\theta$–bins whose centres satisfy $\theta_{\min}(\bar z)\le\theta\le\theta_{\max}(\bar z)$, resulting in between 6 and 8 angular bins.

In addition to this fiducial configuration we tested two alternative scale cuts: (1) a large-scale configuration using only separations above $5\,h^{-1}\mathrm{Mpc}$ (limited by the maximum angular separation $3^\circ$ of the correlation function measurements), and (2) a mid-scale configuration defined by $R_{\mathrm{min}} = 3\,h^{-1}\mathrm{Mpc}$ and $R_{\mathrm{max}} = 20\,h^{-1}\mathrm{Mpc}$. The recovered redshift distributions in these alternative cases are consistent with those from the fiducial configuration. However, jackknife errors on these larger scales are less reliable due to the limited number of independent regions available at such angular separations. For this reason, and to maximize signal-to-noise in the cross-correlation signal, we focus on the small-scale configuration described above.

\subsection{Auto-correlation measurements}

We measured the angular auto-correlation function, $w_{rr}(\theta)$, for the DESI reference samples in narrow redshift bins of width $\Delta z = 0.05$ in the range $0.1 < z < 1.6$, using the full dataset across the survey footprint.  We performed these measurements using the DESI collaboration code based on the {\sc pycorr} library \citep{2025JCAP...07..017A}, which supports pairwise weighting and provides a fast computation of two-point statistics. We used the Landy–Szalay estimator,
\begin{equation}
w_{rr}(\theta) = \frac{DD - 2DR + RR}{RR},
\end{equation}
where $DD$, $DR$, and $RR$ are the normalized data–data, data–random, and random–random pair counts, respectively.

To correct for fibre collision systematics, we applied the Pairwise Inverse Probability (PIP) weights developed by \citet{2025JCAP...04..074B}. \rossi{The performance of these corrections, including their scale dependence, is validated in detail in \citet{2025JCAP...04..074B}}.
These weights are incorporated directly into the angular pair counts, and are essential for recovering the true correlation function at small scales, where fibre assignment incompleteness can strongly bias the measurements (with \citet{2025JCAP...04..074B} demonstrating factor-of-2 effects for angular separations $\theta < 0.05$ deg).  In addition to PIP weights, we also applied systematics weights to the DESI sample. These weights account for observational variations across the footprint such as seeing, depth, and Galactic extinction \citep{2025JCAP...07..017A}.

We estimated the covariance matrix using jackknife resampling. \rossi{We neglected the covariance between different reference redshift bins, which is expected to be small because the bins are narrow and non-overlapping.}
We tested different jackknife partitions, and found that using 60 regions provides a stable and converged estimate of the covariance matrix.  We also ensured that the region diameter was significantly larger than the maximum angular scale of interest, to avoid underestimating large-scale variance.  In all cases, we applied the mode–projection correction of \cite{2022MNRAS.514.1289M}, as implemented in the \textsc{xirunpc} software, to remove survey-geometry–induced large-scale modes from the two-point measurements prior to fitting.  To assess the robustness of the resulting covariance structure, we repeated the full analysis using only the diagonal elements of the jackknife covariance matrices.  As expected, suppressing the off-diagonal terms leads to a modest change in the fitted auto-correlation amplitudes $A_{rr}$, but we find that the impact on the inferred $b_\mathrm{u}(z_\mathrm{r}) p_\mathrm{u}(z_\mathrm{r})$ and on the final source redshift distributions is negligible.  For both the \textsc{Buzzard} mocks and the real data, we find that the differences between using the full covariance and the diagonal-only covariance remain well below the percent level.  The similarity between the diagonal and full-covariance results indicates that the non-diagonal structure of the jackknife covariance does not drive the fits.

\subsection{Cross-correlation measurements}

We also measured the angular cross-correlation between the DESI reference samples in narrow redshift bins of width $\Delta z = 0.05$ in the range $0.1 < z < 1.6$, and each of the tomographic sub-samples of every weak lensing survey.  For the cross-correlation analyses, we cut the catalogues to the joint overlap area.  We performed the measurement using the {\sc treecorr} correlation code\footnote{\url{https://github.com/rmjarvis/TreeCorr}} \citep{2004MNRAS.352..338J}.  Since random catalogues are available for the DESI reference sample, but not the photometric imaging surveys, we use the \rossi{Davis--Peebles estimator \citep{1983ApJ...267..465D},}
\begin{equation}
w(\theta) = \frac{D_\mathrm{u}D_\mathrm{r} (\theta)}{D_\mathrm{u}R_\mathrm{r} (\theta)} - 1,
\end{equation}
where $D_\mathrm{u}$ represents the unknown (photometric) sample, $D_\mathrm{r}$ the reference (DESI) sample, and $R_\mathrm{r}$ the random catalogues corresponding to the reference sample. \rossi{We use the DESI DR1 LSS random catalogues associated with each spectroscopic tracer, where the size of the random sample is around a factor of 30 larger than the data sample. The same redshift selections are applied to the random catalogues as to the corresponding reference samples.}

In our estimators, each DESI galaxy is assigned a total weight,
\begin{equation}
w_i^r = w_i^{\mathrm{sys}} \cdot w_i^{\mathrm{IIP}},
\end{equation}
where $w_i^{\mathrm{sys}}$ corrects for imaging systematics and redshift failures, and $w_i^{\mathrm{IIP}}$ is the individual inverse probability weight for fibre assignment incompleteness \citep{2025JCAP...04..074B}.  This correction is appropriate for cross-correlation measurements, where only one side of the galaxy pair is subject to fibre collisions, and avoids the need for pairwise corrections.  Photometric sources were assigned the same weights as used in lensing shear measurements, which are provided with each lensing catalogue.  Hence, we are deriving the weighted source redshift distributions, which are the appropriate quantities for interpreting lensing correlations.

We again determined the covariance of the cross-correlation function measurements using jackknife re-sampling over angular regions, confirming that our choice of 60 regions ensured stable covariance estimation.  We note that since the auto-correlation is computed over the full DESI footprint and the cross-correlation is limited to the overlap with each imaging survey, it is not possible to match jackknife regions across the two measurements. Nonetheless, this issue does not significantly affect our results because the error budget is dominated by the cross-correlation, and the auto-correlation is measured over a substantially different region of sky.


\section{Redshift distribution fits}
\label{sec:fits}

In this section, we summarize the methodology we apply to fit the amplitudes of the measured auto- and cross-correlation functions, and hence determine the redshift probability distribution of the unknown sample.  In summary, we perform independent fits in each narrow reference redshift bin for the auto-correlation ($A_{rr}$) and cross-correlation ($A_{ur}$) amplitudes defined in Sec.~\ref{sec:formalism}, assuming no cross-correlation between different redshift bins and between the auto- and cross-correlation measurements.  \rossi{Although the auto- and cross-correlation measurements are not strictly independent, we expect their covariance to be subdominant because $w_{ur}$ is measured over a significantly smaller overlapping area, while $w_{rr}$ is measured over the full DESI footprint.}

\subsection{Auto-correlation amplitude \texorpdfstring{$A_{rr}$}{Arr}}

\begin{figure*}[t]
\centering

\begin{subfigure}[t]{0.32\textwidth}
  \centering
  \includegraphics[width=\linewidth]{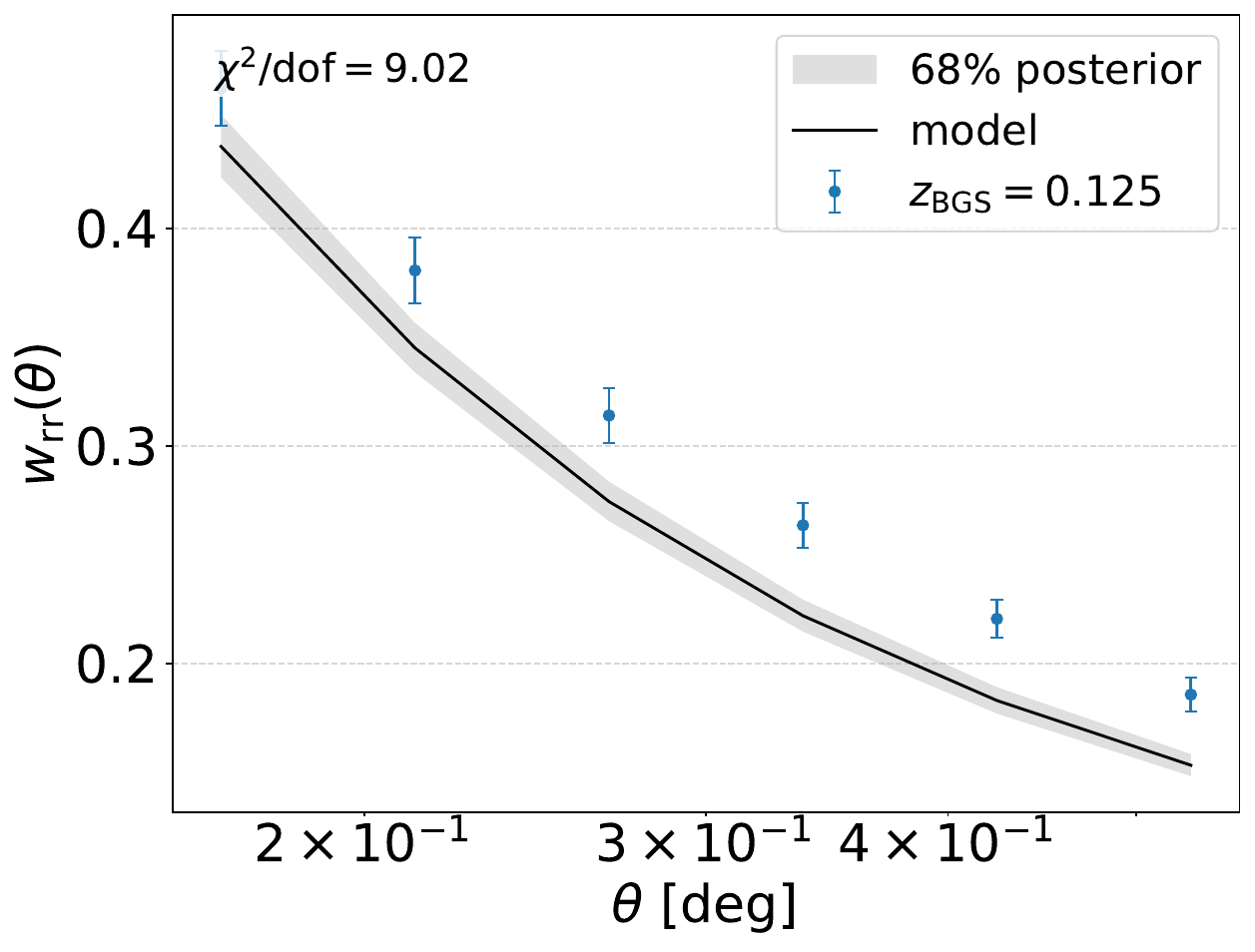}
\end{subfigure}\hfill
\begin{subfigure}[t]{0.32\textwidth}
  \centering
  \includegraphics[width=\linewidth]{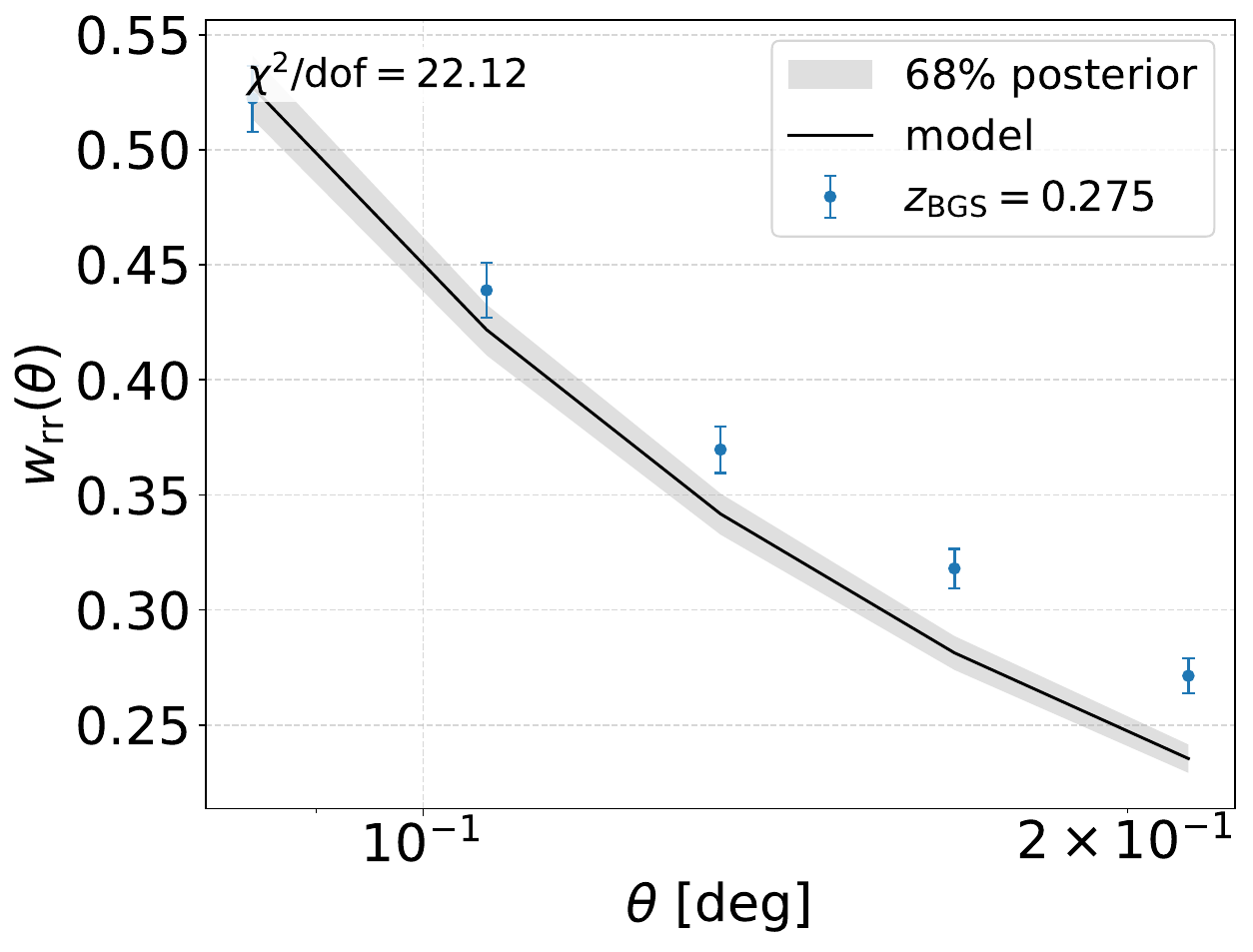}
\end{subfigure}\hfill
\begin{subfigure}[t]{0.32\textwidth}
  \centering
  \includegraphics[width=\linewidth]{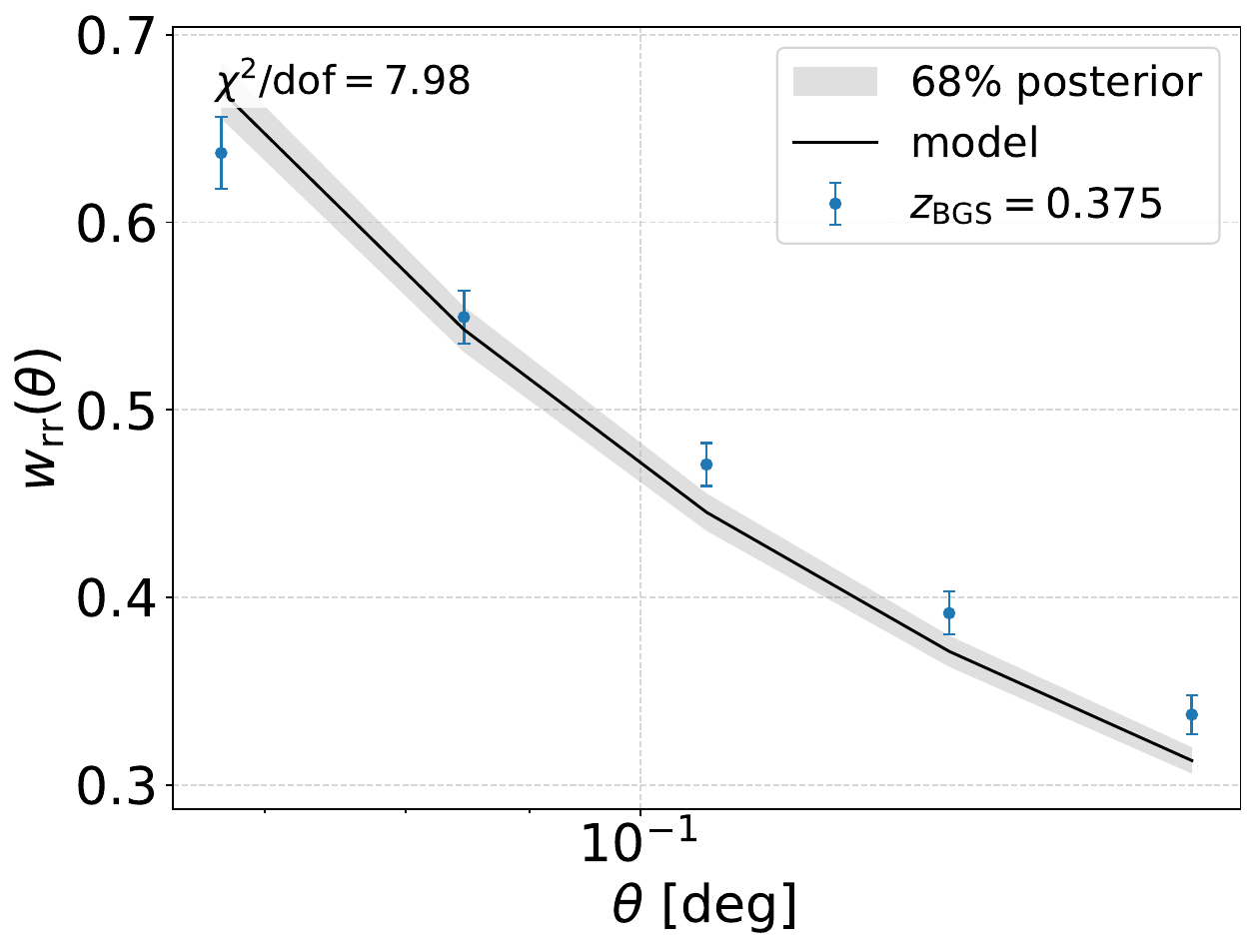}
\end{subfigure}

\begin{subfigure}[t]{0.32\textwidth}
  \centering
  \includegraphics[width=\linewidth]{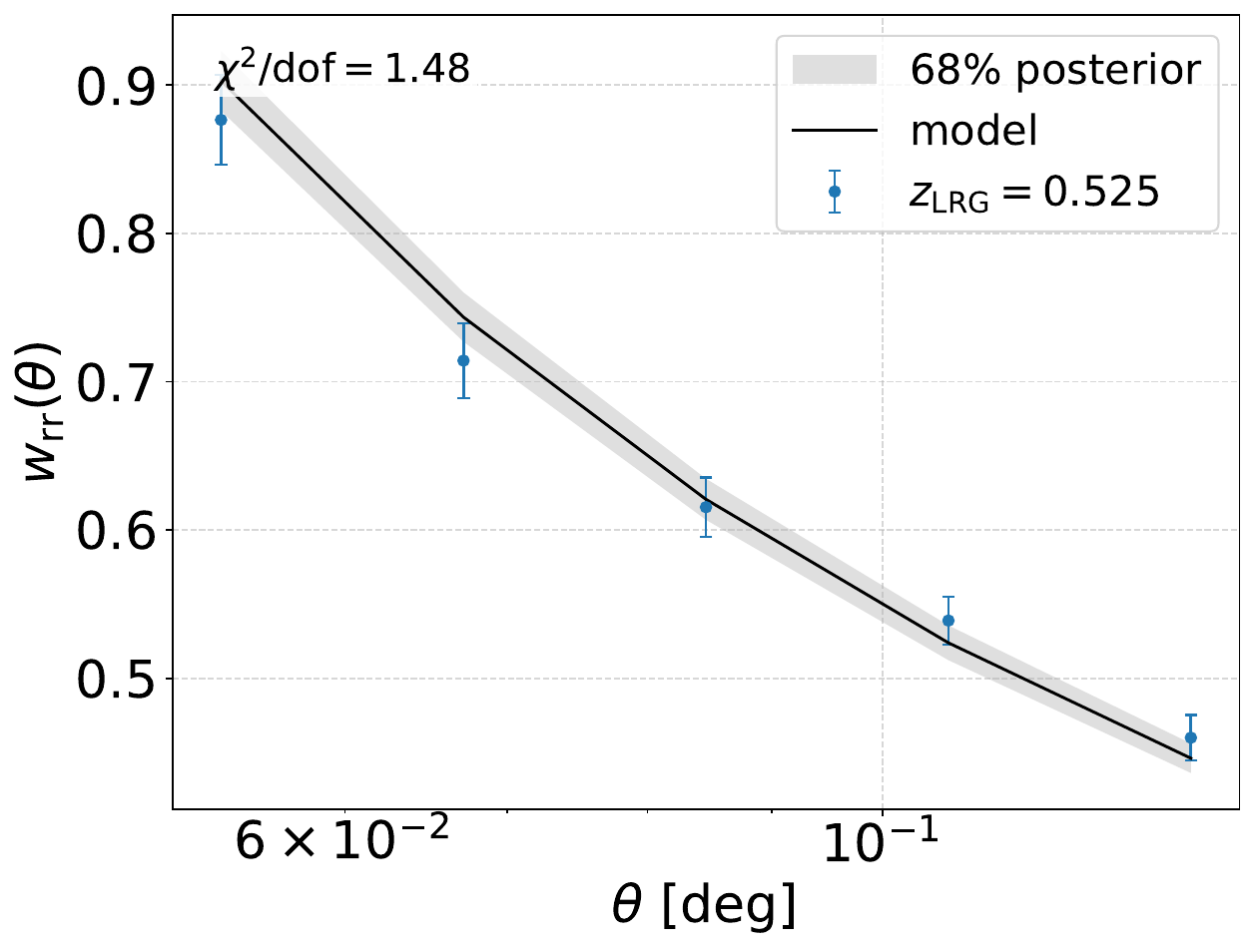}
\end{subfigure}\hfill
\begin{subfigure}[t]{0.32\textwidth}
  \centering
  \includegraphics[width=\linewidth]{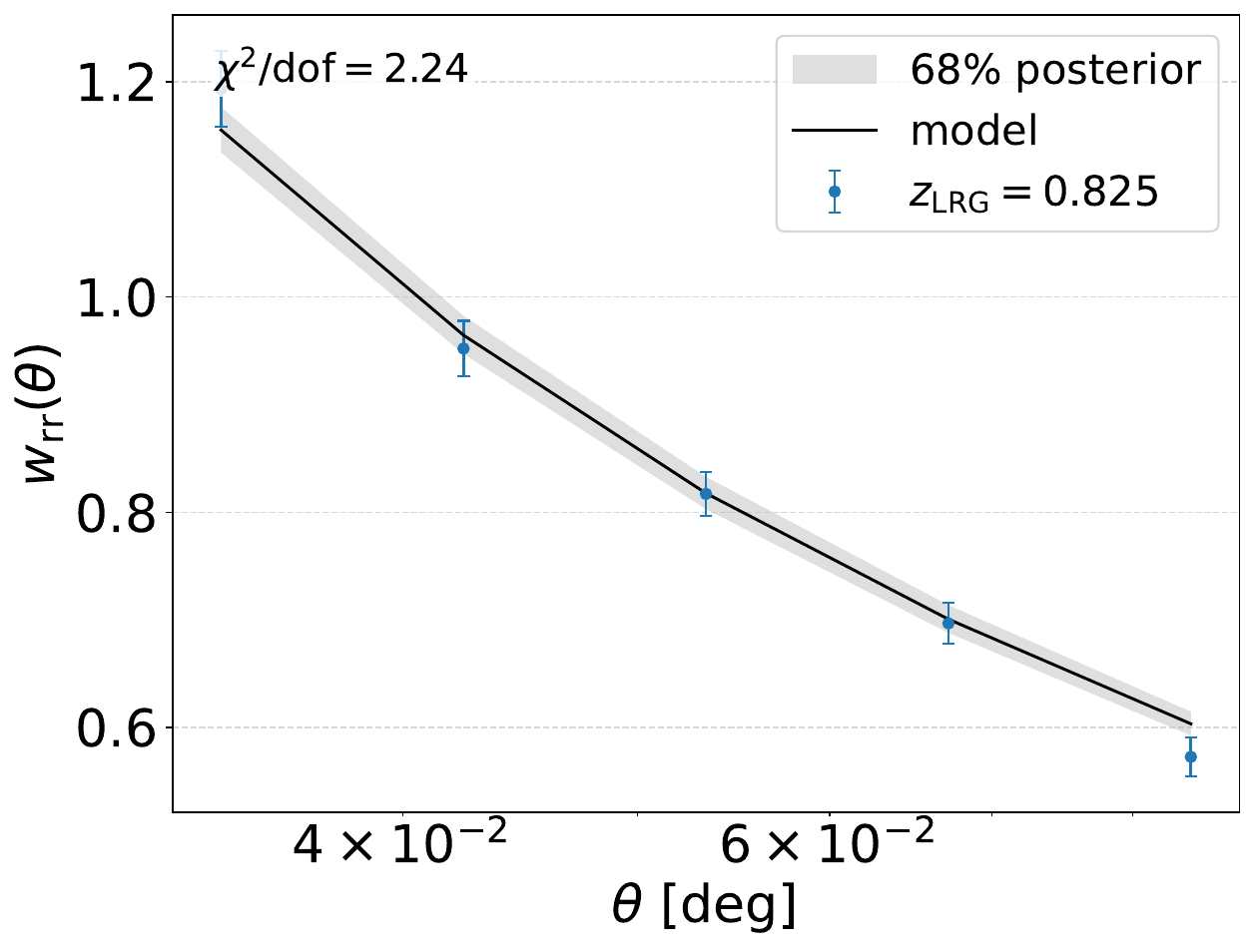}
\end{subfigure}\hfill
\begin{subfigure}[t]{0.32\textwidth}
  \centering
  \includegraphics[width=\linewidth]{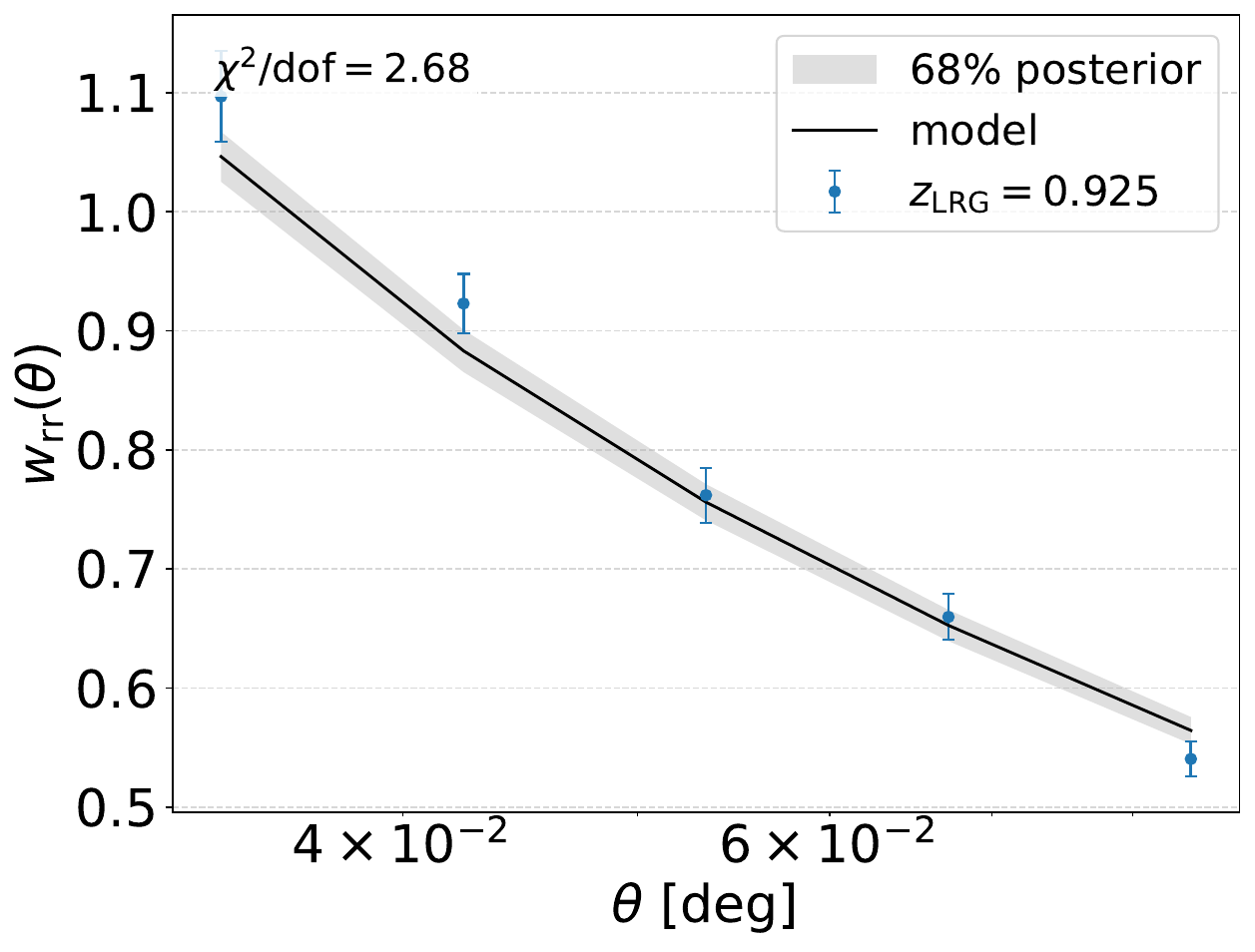}
\end{subfigure}

\begin{subfigure}[t]{0.32\textwidth}
  \centering
  \includegraphics[width=\linewidth]{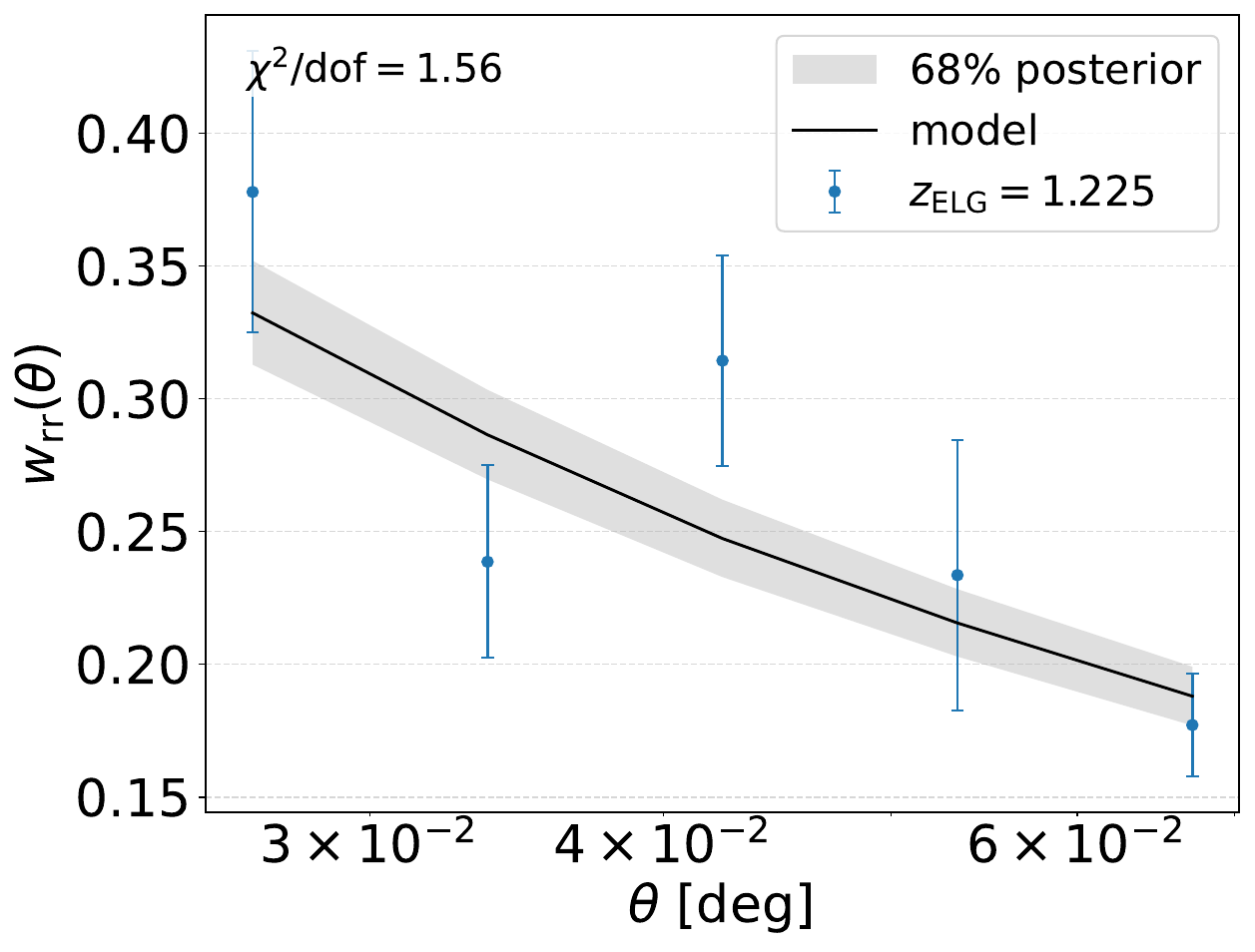}
\end{subfigure}\hfill
\begin{subfigure}[t]{0.32\textwidth}
  \centering
  \includegraphics[width=\linewidth]{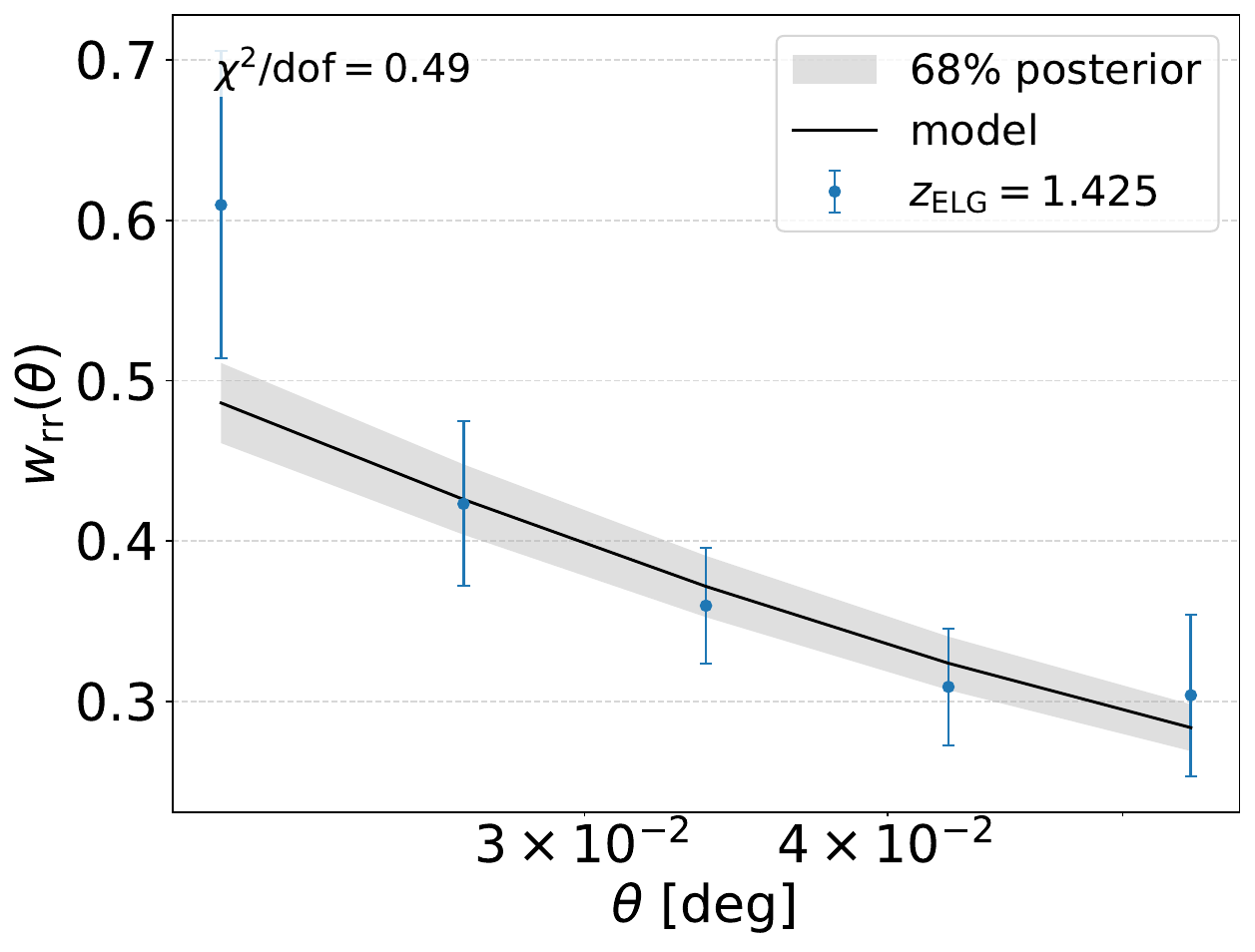}
\end{subfigure}\hfill
\begin{subfigure}[t]{0.32\textwidth}
  \centering
  \includegraphics[width=\linewidth]{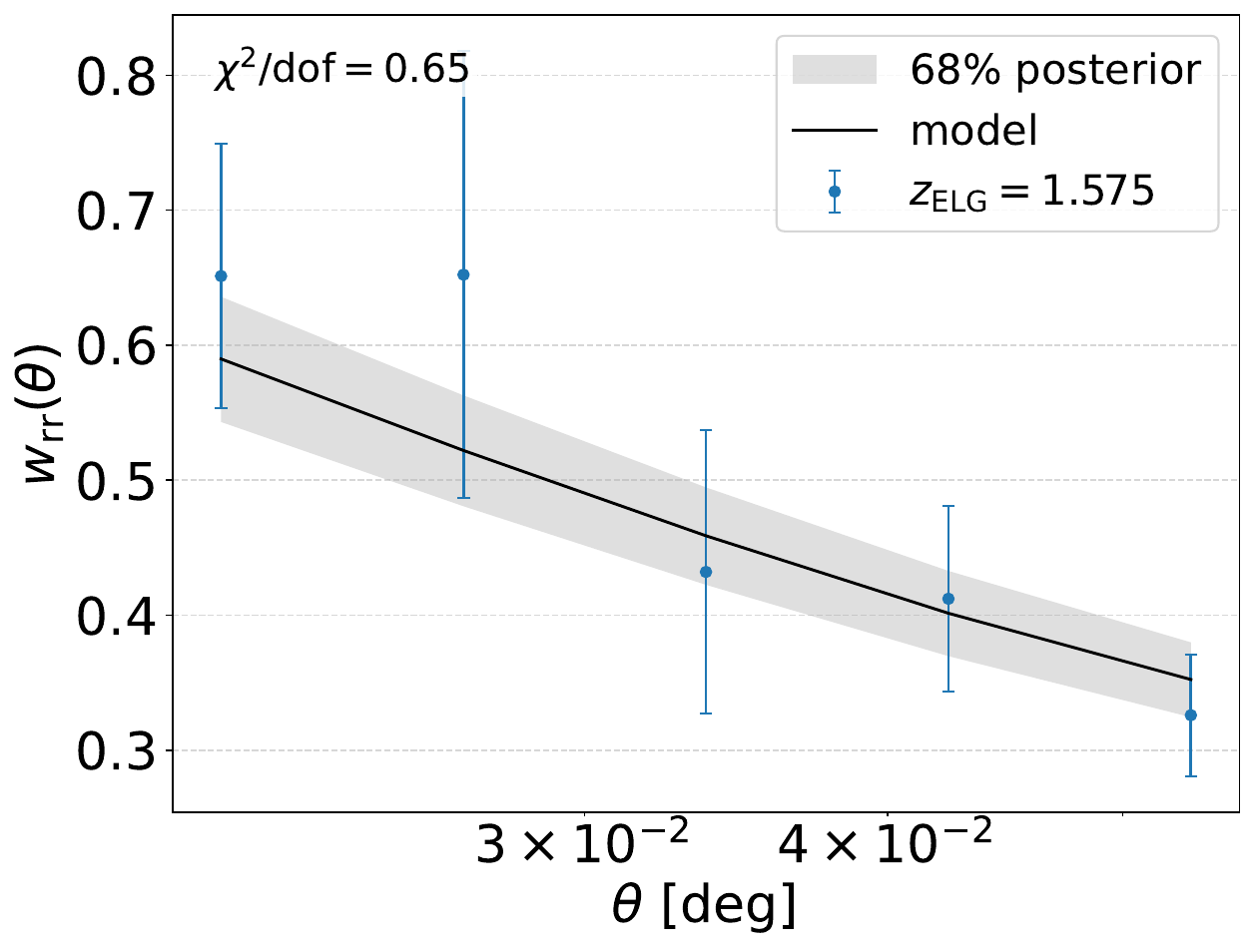}
\end{subfigure}
\caption{\rossi{Angular auto-correlation measurements $w_{rr}(\theta)$ for
the DESI-DR1 spectroscopic sample, shown for representative BGS
(top row), LRG (middle row), and ELG (bottom row) redshift bins.
The points show the measurements with jackknife-derived uncertainties,
the solid curves show the MAP model $A_{rr}w_m(\theta)$, and the
shaded regions indicate the 68\% posterior uncertainty.
The reduced $\chi^2$ of each fit, evaluated using the full covariance
matrix, is also shown in each panel.}}
\label{fig:wrr-desi-panels}
\end{figure*}

We fit the angular auto-correlation function $w_{rr}(\theta)$ in each narrow redshift bin using the one-parameter model defined in Sec.~\ref{sec:formalism}, $w_{rr}(\theta) = A_{rr} \, w_m(\theta)$, where $w_m(\theta)$ is the theoretical template computed from the matter power spectrum, defined by Eq.~\ref{eq:wll} with $b_\mathrm{r} = 1$.  The fit is performed by minimizing the $\chi^2$ function,
\begin{equation}
\chi^2(A_{rr}) = \left[ \vec{w}_{rr} - A_{rr} \vec{w}_m \right]^T \cdot \mathbf{C}^{-1} \cdot \left[ \vec{w}_{rr} - A_{rr} \vec{w}_m \right]
\end{equation}
where $\mathbf{C}^{-1}$ is the inverse covariance matrix computed using jackknife re-sampling.  We use a flat prior $A_{rr} \in (0, 20)$. The posterior on $A_{rr}$ is sampled using an MCMC sampler, and we determine both the ``maximum \textit{a posteriori}'' (MAP) estimate and the posterior mean and standard deviation.

\rossi{In Fig.~\ref{fig:wrr-desi-panels} we show representative auto-correlation measurements $w_{rr}(\theta)$ for the three DESI reference populations: BGS at $\bar z=0.125,0.275,0.375$, LRG at $\bar z=0.525,0.825,0.925$, and ELG at $\bar z=1.225,1.425,1.575$. These selections highlight the clear evolution of clustering amplitude with redshift and tracer type. The data points show the measurements with jackknife-derived uncertainties, while the solid curves show the MAP model $A_{rr}w_m(\theta)$. The shaded regions indicate the 68\% posterior uncertainty on the model.  The posterior mean and MAP values of $A_{rr}$ differ negligibly compared with the posterior uncertainty for these examples, and we therefore show only the MAP model.}

\rossi{We note that the measurements are strongly correlated across angular scales, such that the best-fit normalization obtained using the full covariance matrix need not coincide with the visually-expected normalization obtained by treating the data points as independent.  We verified that this behaviour is not associated with the prior boundary, and that repeating the analysis using only the diagonal covariance changes the individual fitted amplitudes modestly, but has a negligible impact on the recovered $b_\mathrm{u}(z_\mathrm{r})p_\mathrm{u}(z_\mathrm{r})$ distributions.}

\rossi{The corresponding $\chi^2/{\rm dof}$ values are also reported in each panel of Fig.~\ref{fig:wrr-desi-panels}.  Some BGS auto-correlation fits have large $\chi^2/{\rm dof}$, which could indicate an issue with the accuracy of the jack-knife inverse covariance in some cases where the off-diagonal elements are large.  However, this issue does not have a significant impact on the derived source redshift distributions, or the errors on those distributions, which are dominated by the cross-correlation measurement.}
%

\subsection{Cross-correlation amplitude \texorpdfstring{$A_{ur}$}{Aur}}\label{subsec:cross}

\begin{figure*}[t]
\centering

  \includegraphics[width=0.32\textwidth]{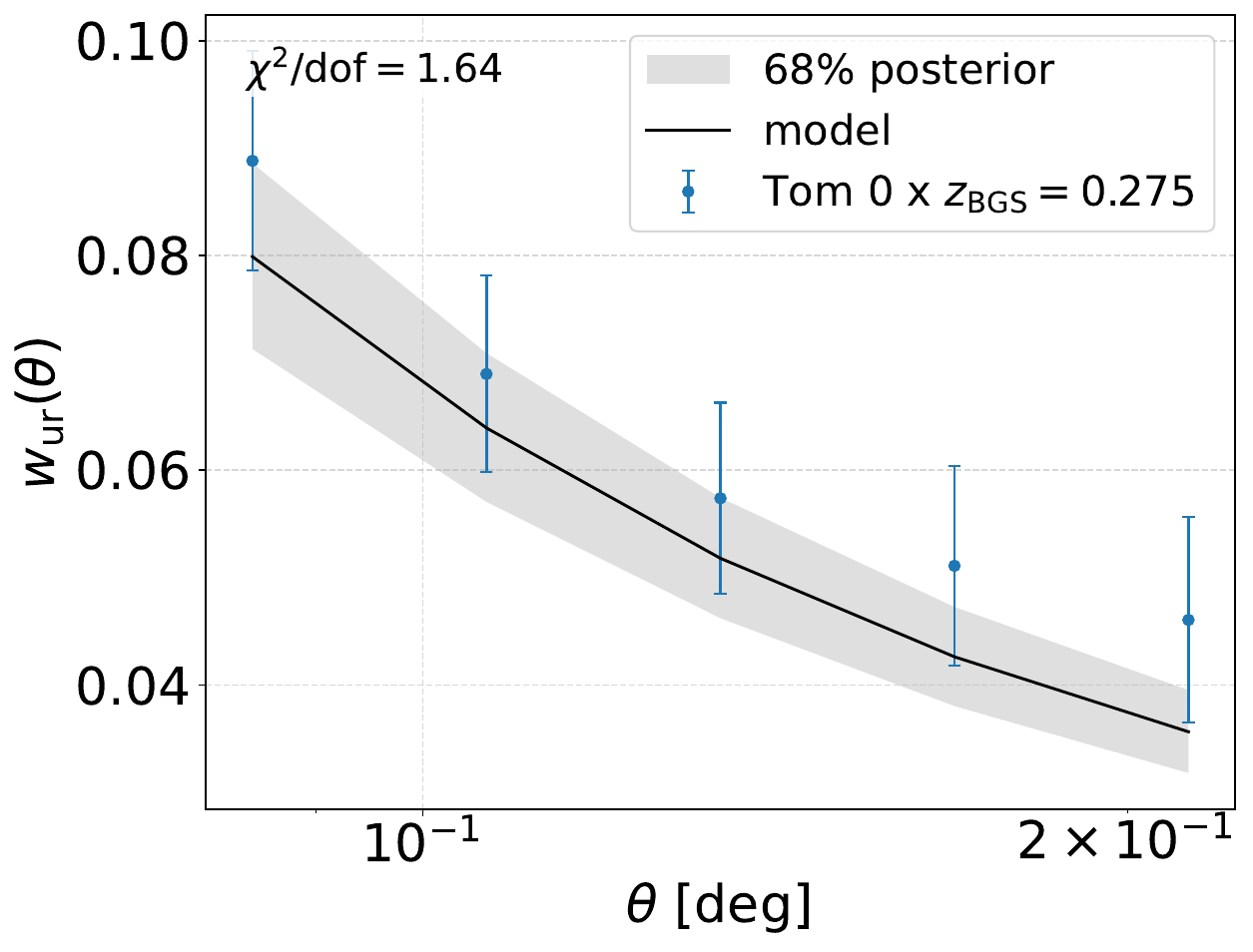}\hfill
  \includegraphics[width=0.32\textwidth]{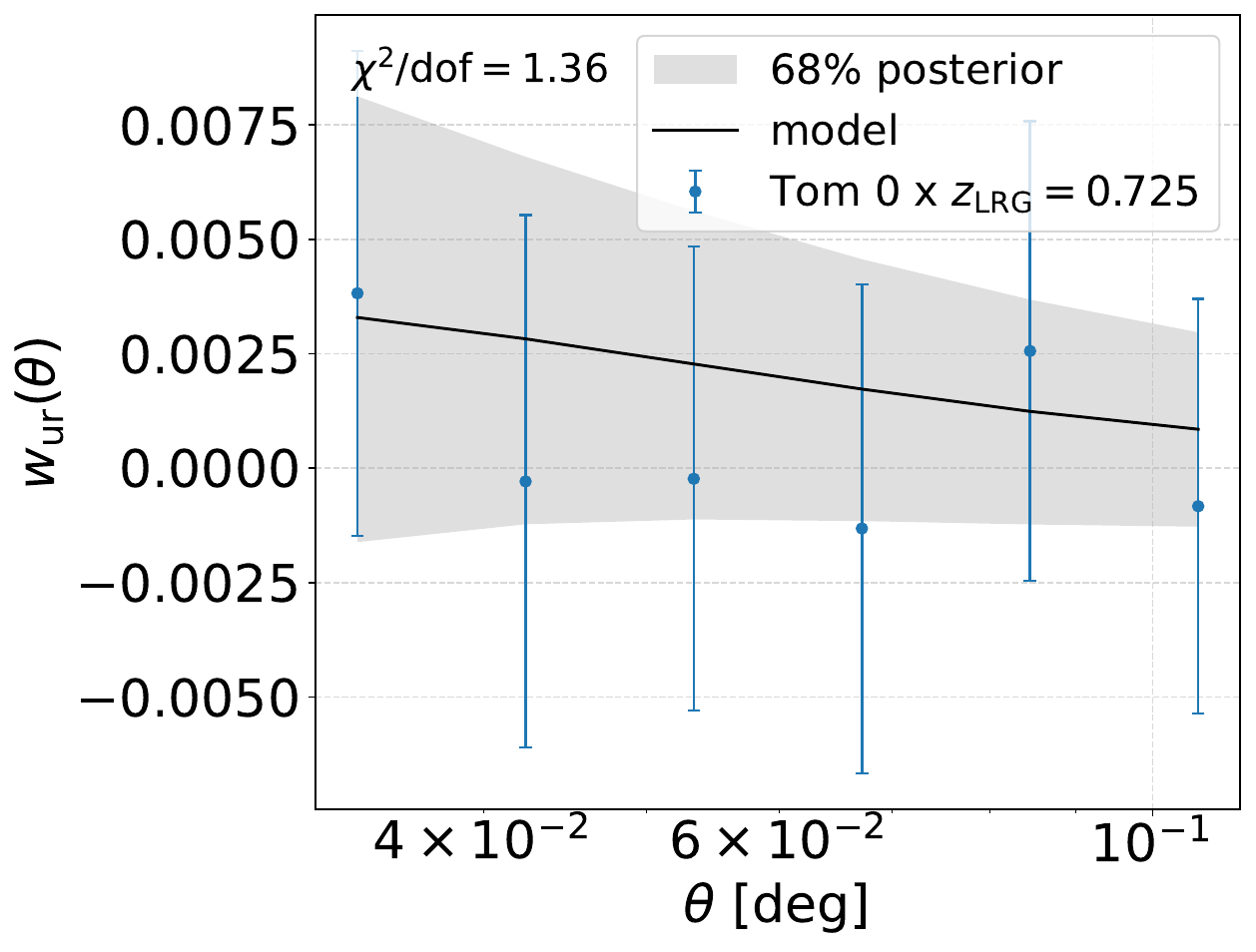}\hfill
  \includegraphics[width=0.32\textwidth]{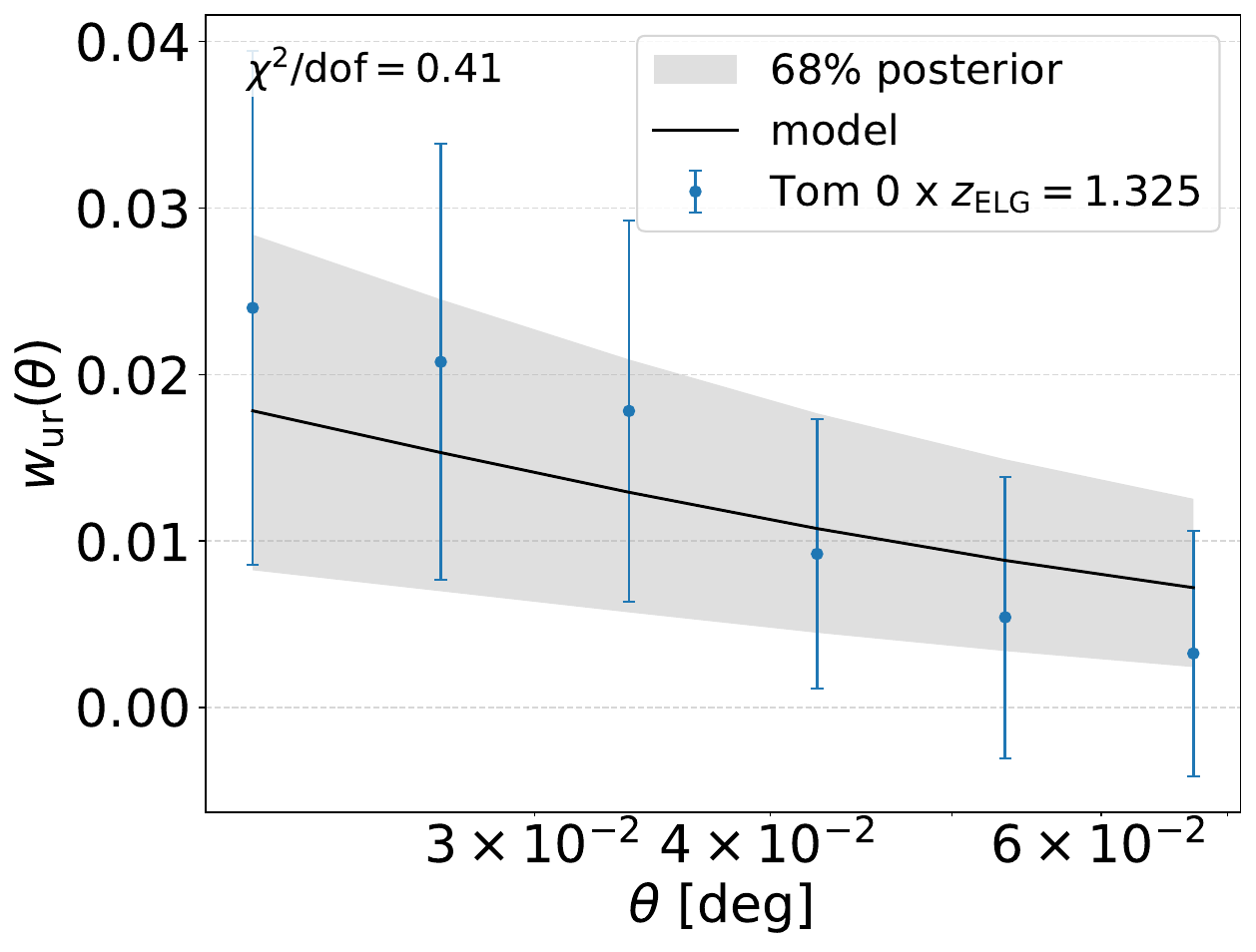}

  \includegraphics[width=0.32\textwidth]{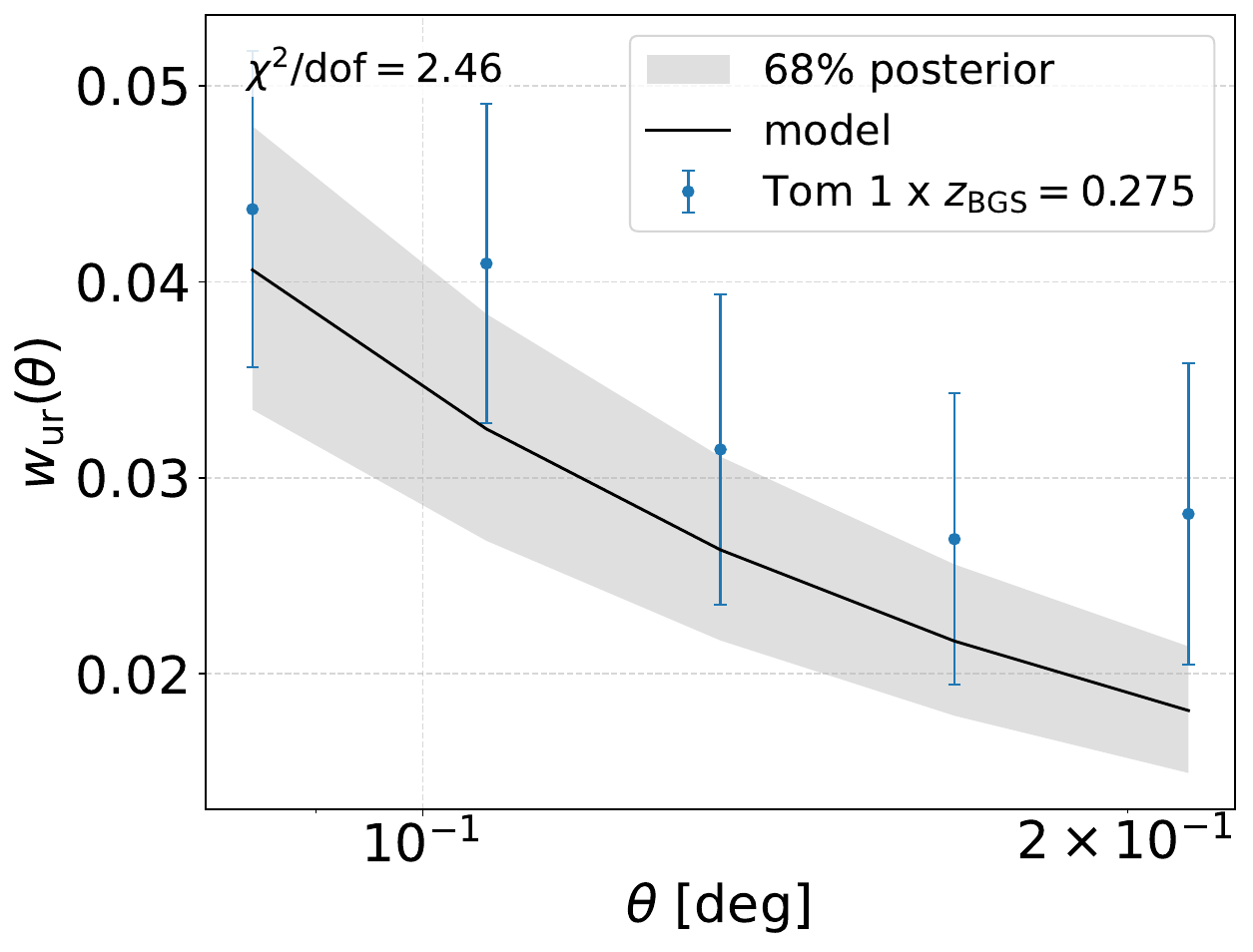}\hfill
  \includegraphics[width=0.32\textwidth]{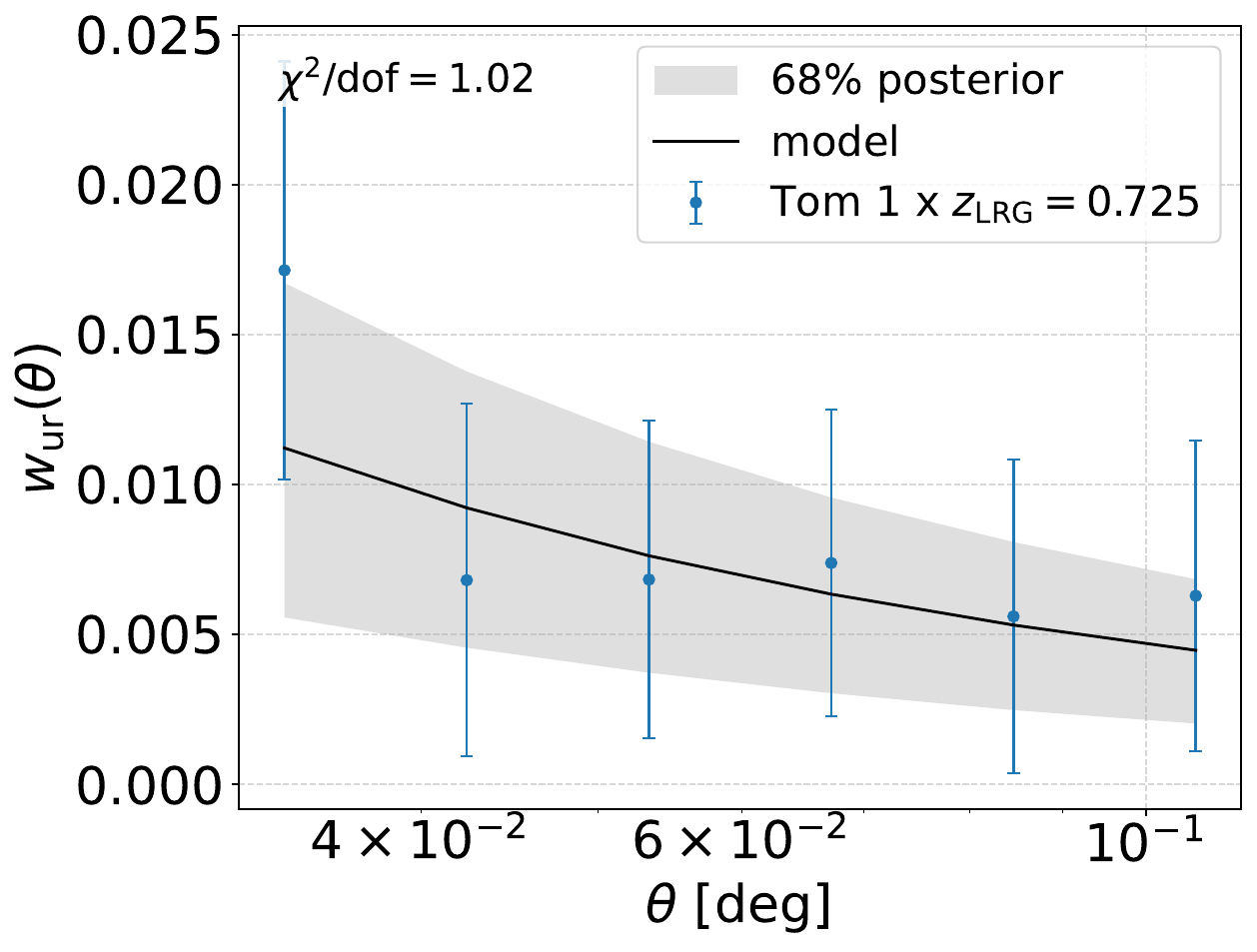}\hfill
  \includegraphics[width=0.32\textwidth]{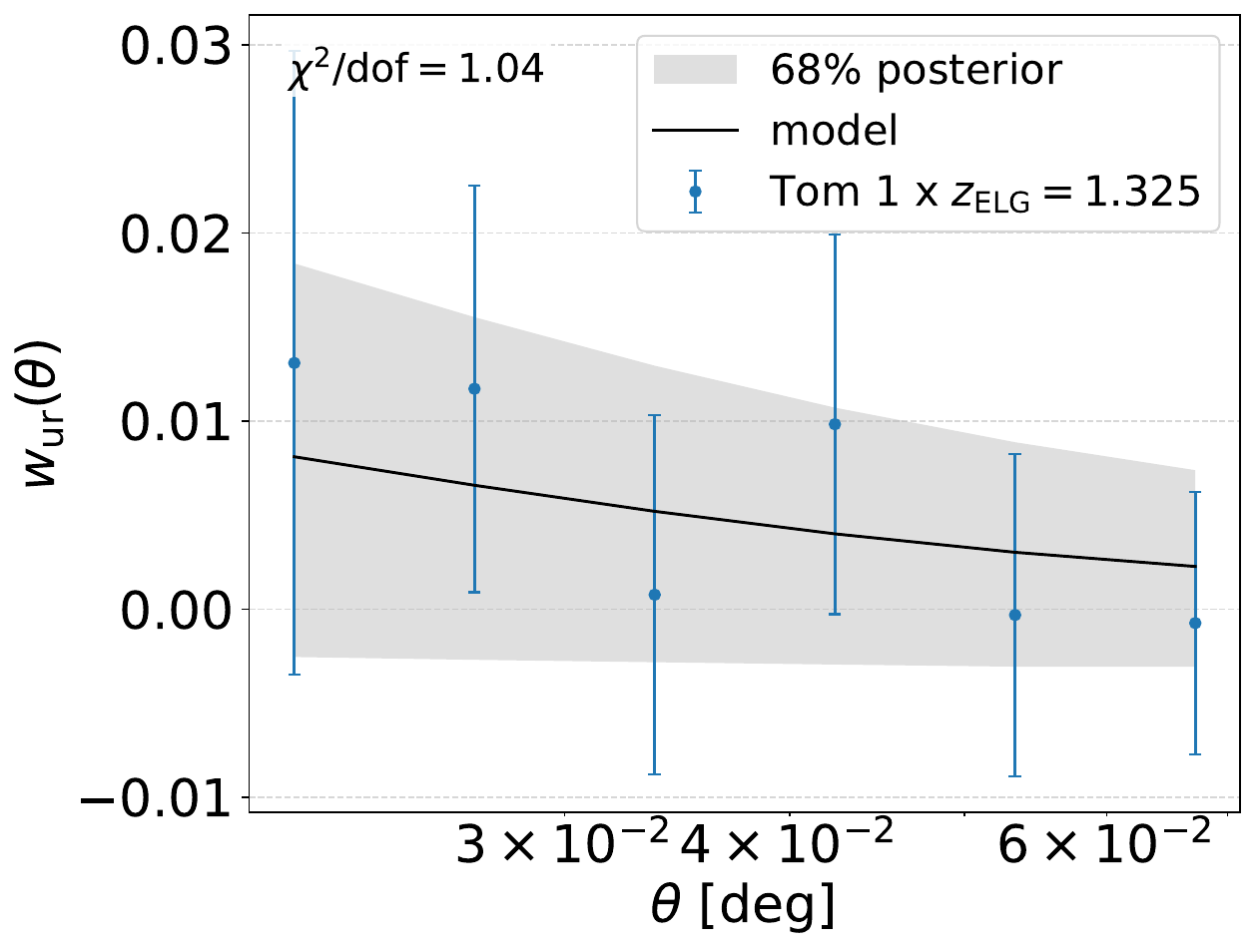}

  \includegraphics[width=0.32\textwidth]{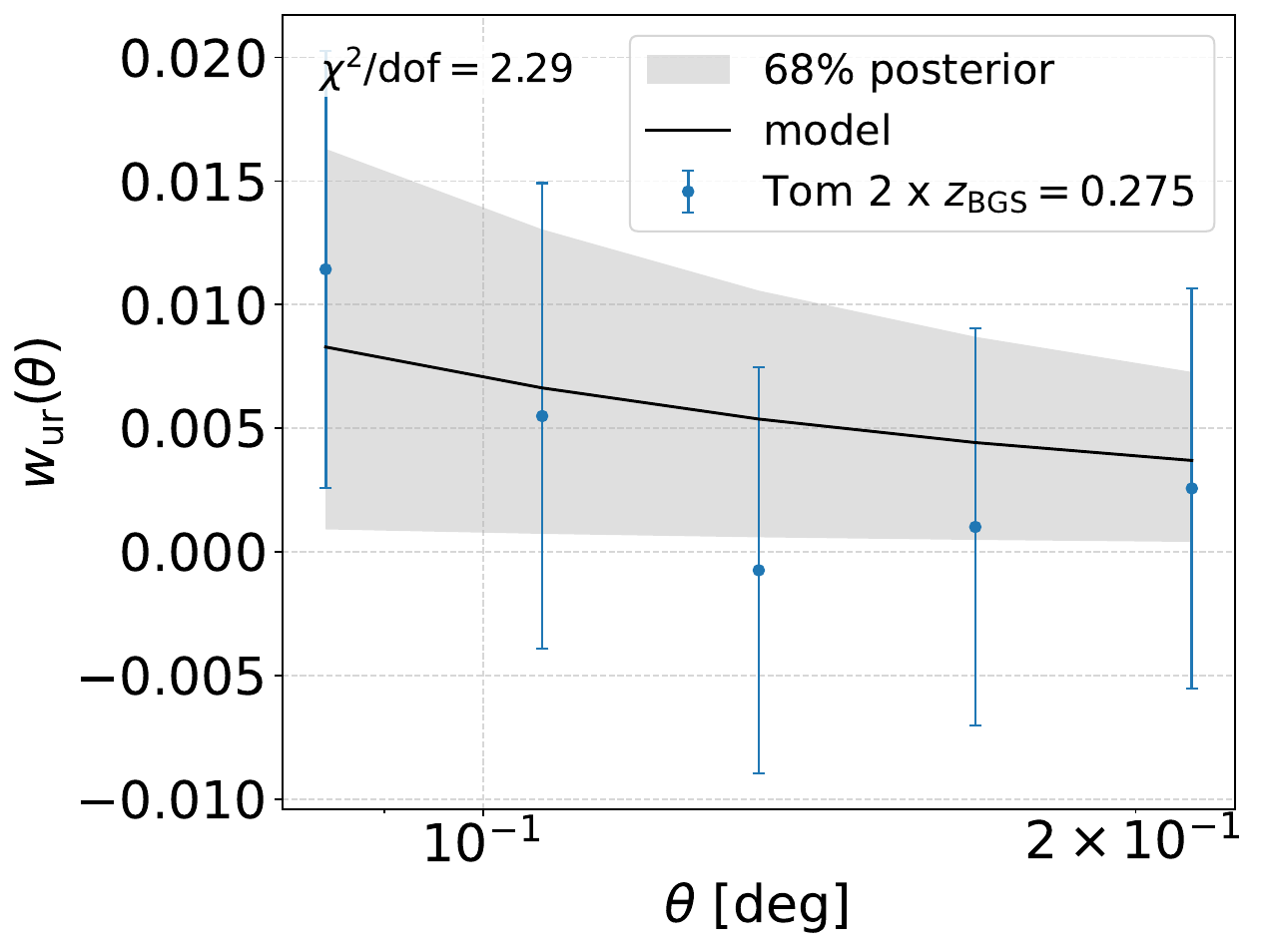}\hfill
  \includegraphics[width=0.32\textwidth]{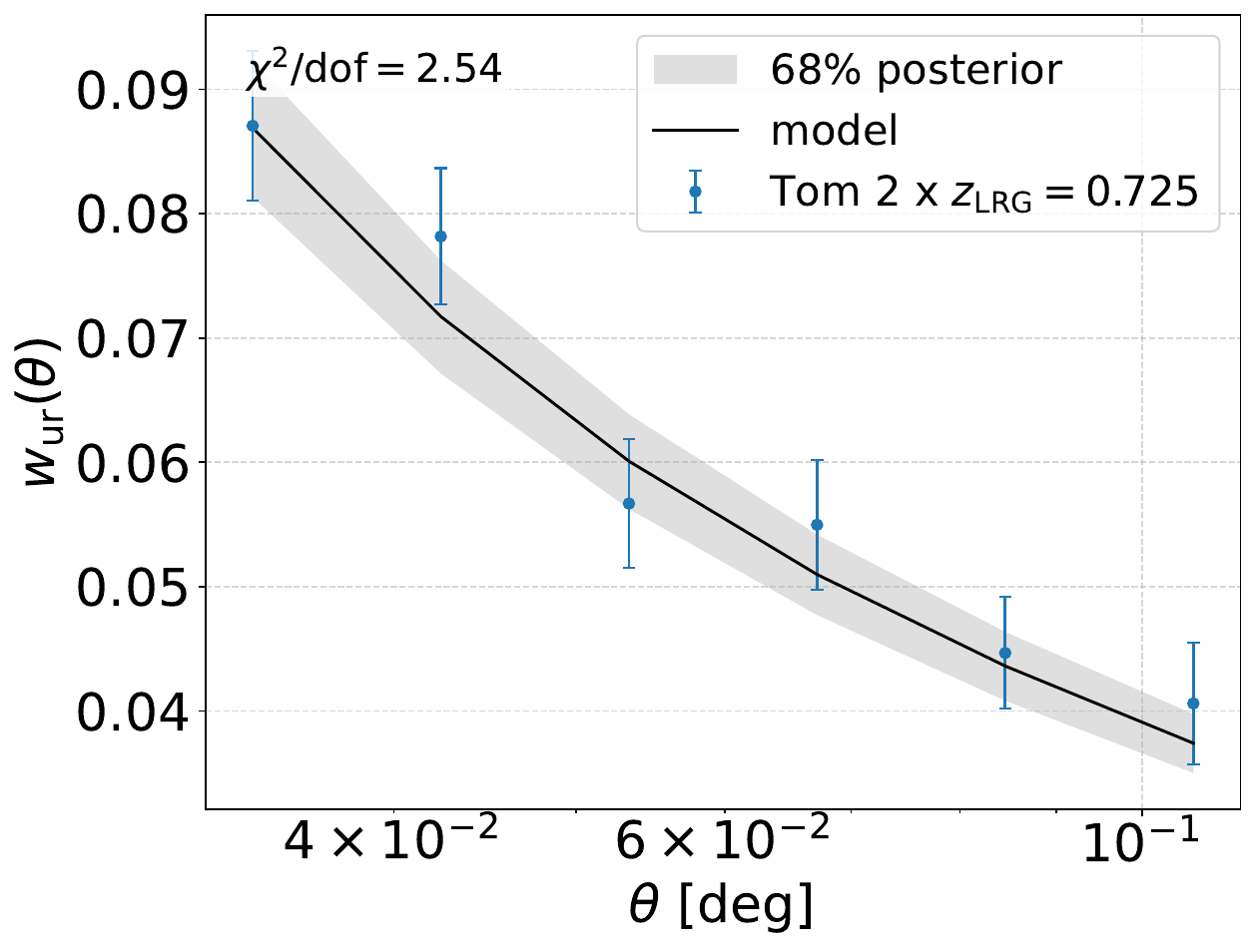}\hfill
  \includegraphics[width=0.32\textwidth]{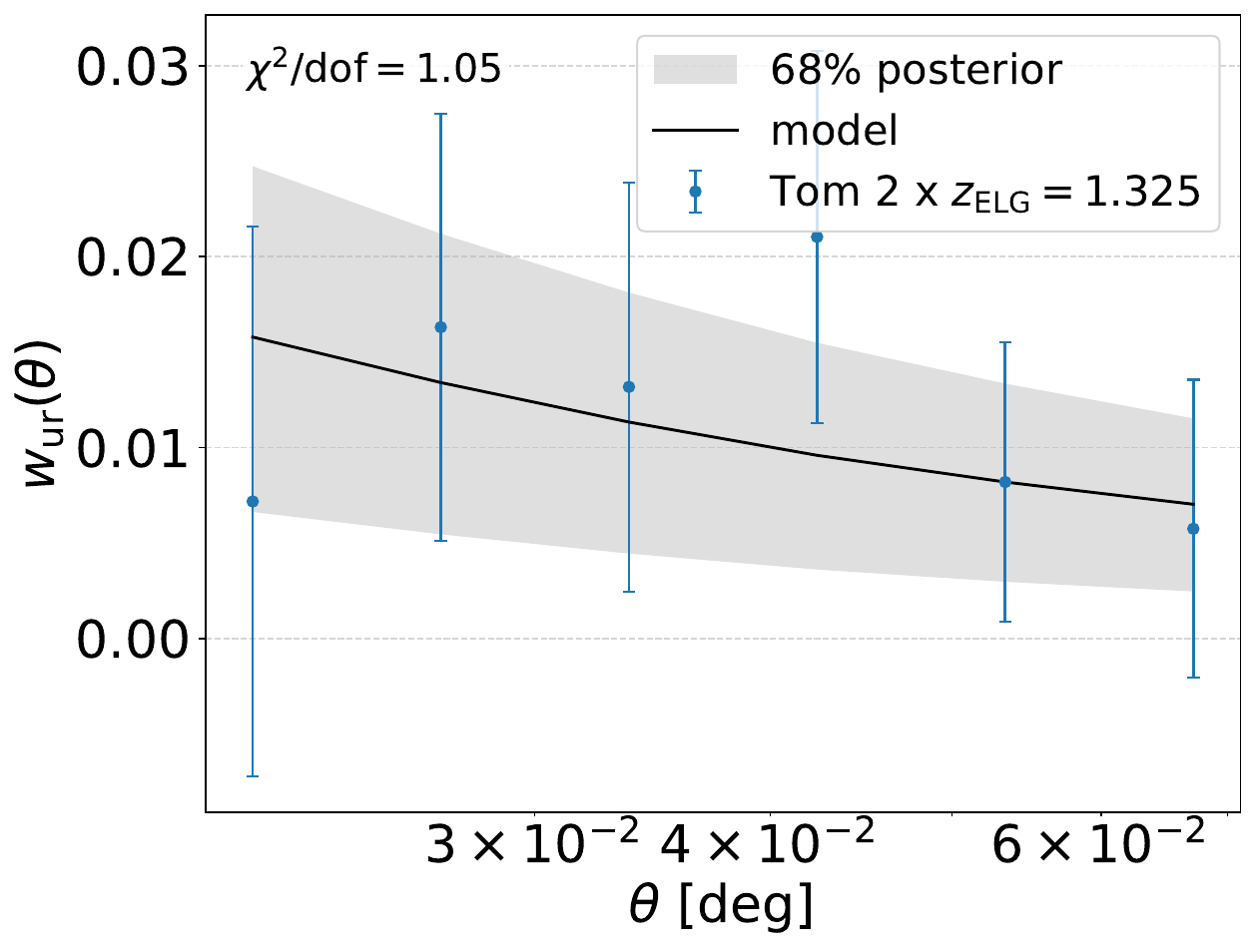}

  \includegraphics[width=0.32\textwidth]{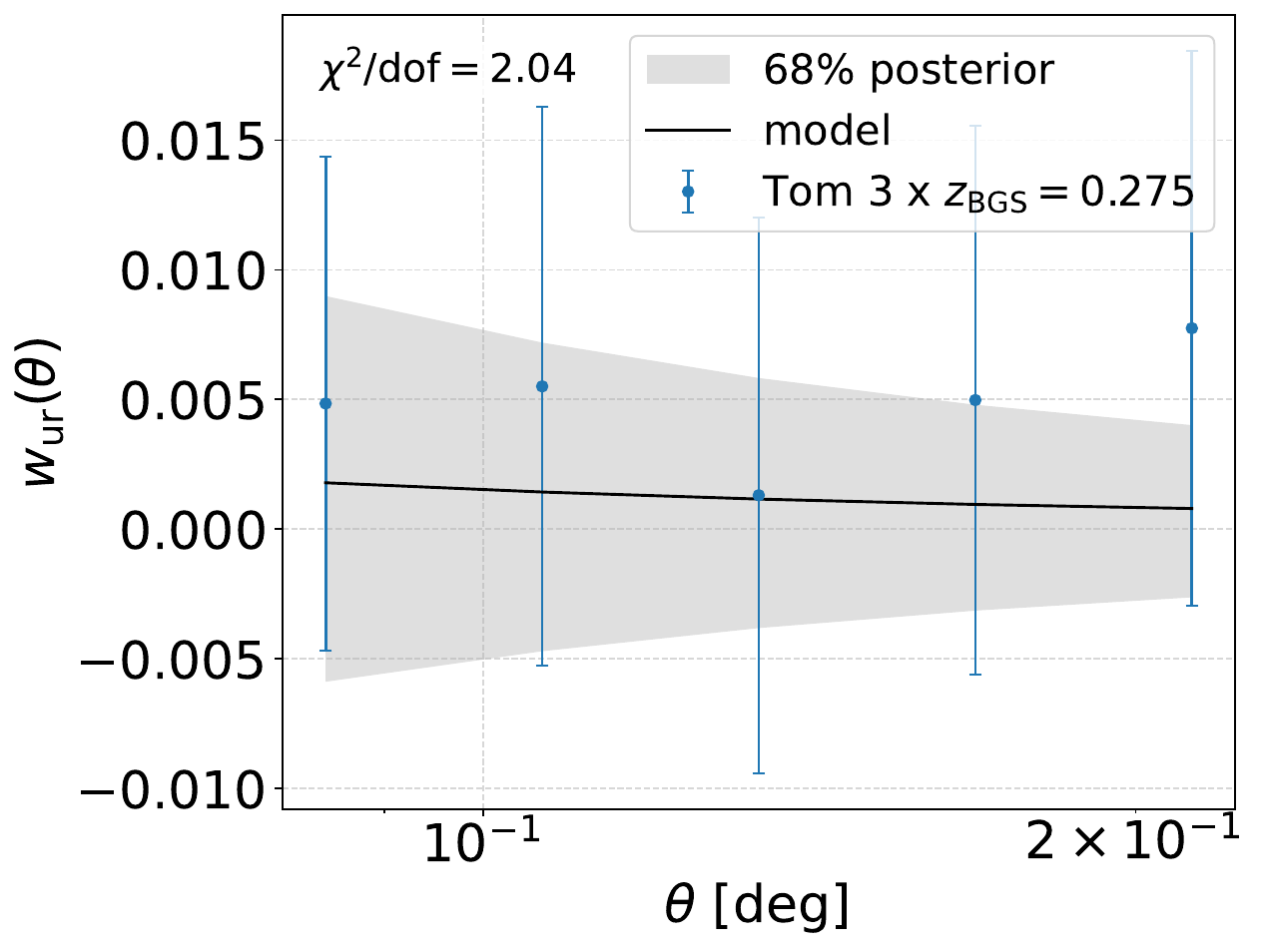}\hfill
  \includegraphics[width=0.32\textwidth]{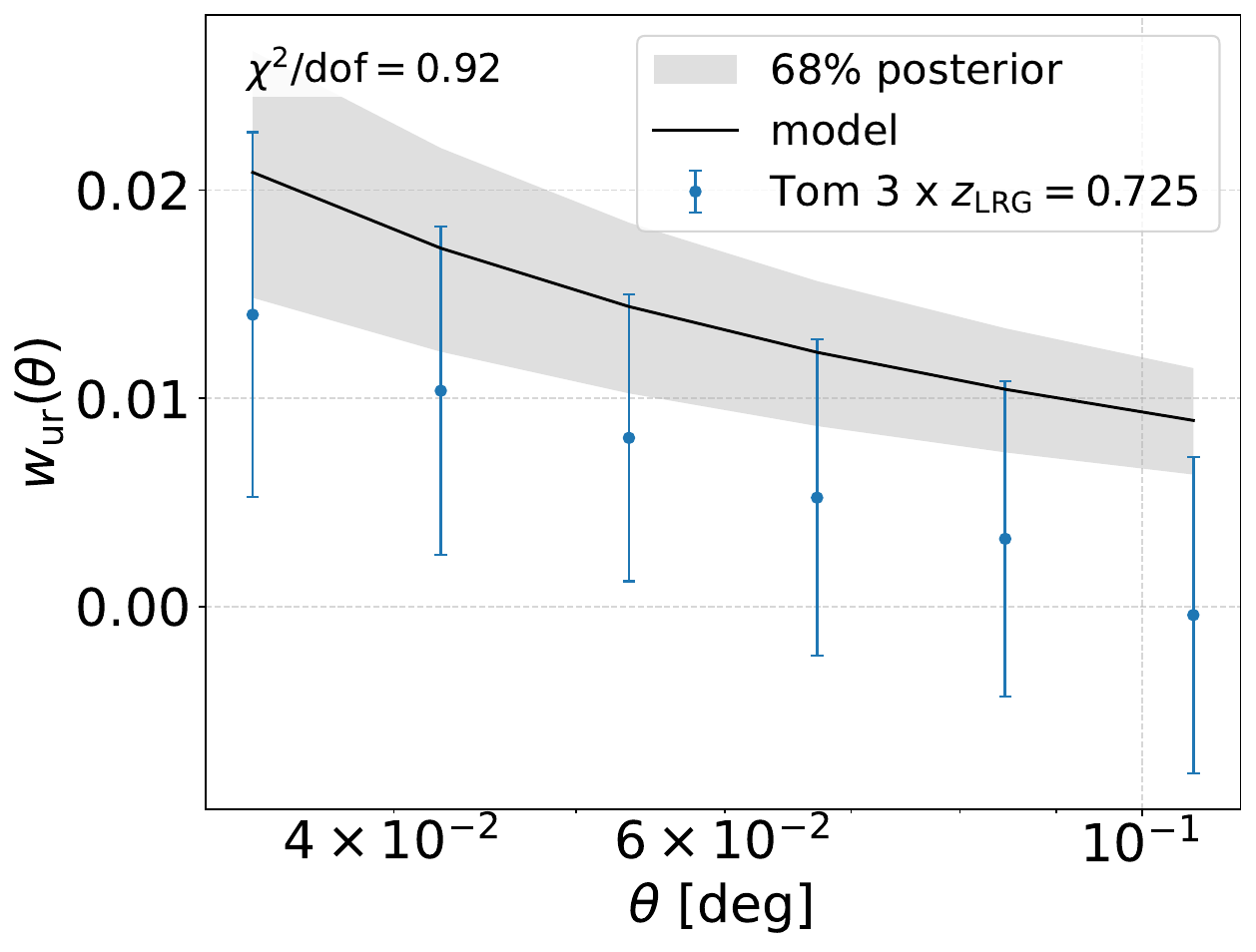}\hfill
  \includegraphics[width=0.32\textwidth]{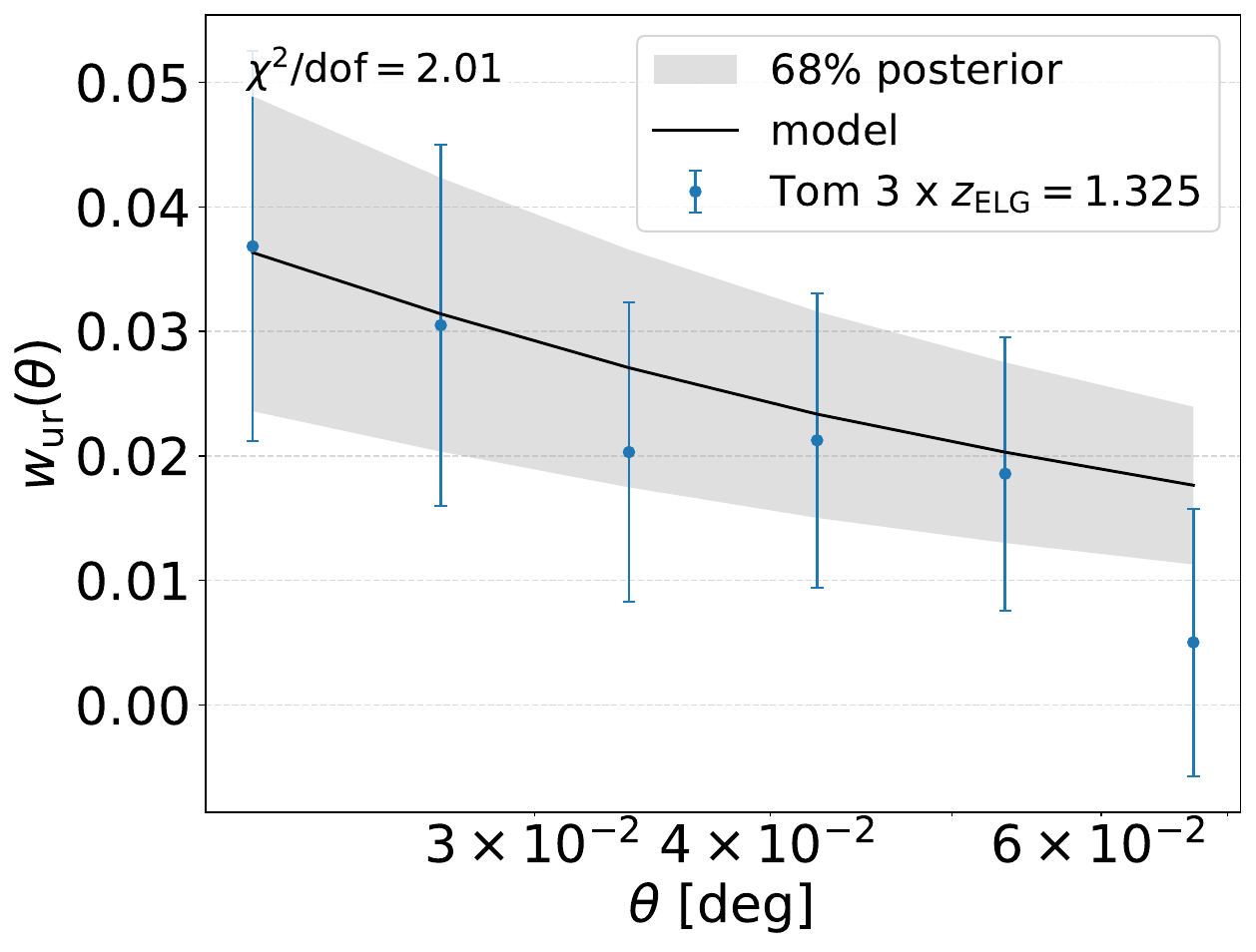}

%
\caption{\rossi{Angular cross-correlation measurements $w_{ur}(\theta)$
between DES-Y3 source galaxies and the DESI spectroscopic reference
sample. The rows show the four DES-Y3 tomographic bins and the columns show representative BGS ($\bar z=0.275$), LRG
($\bar z=0.725$), and ELG ($\bar z=1.325$) reference bins.
The points show the measurements with jackknife-derived uncertainties, the black curves show the total MAP model
$A_{ur}w_m+p\,w^{\rm mag}_{ru}+q\,w^{\rm mag}_{ur}$, and the shaded regions show its 68\% posterior uncertainty. The corresponding $\chi^2/{\rm dof}$ is reported in each panel.}}
\label{fig:wur-desy3xdesi-by-t}
\end{figure*}

\begin{figure*}[t]
  \centering
  \includegraphics[width=0.32\textwidth]{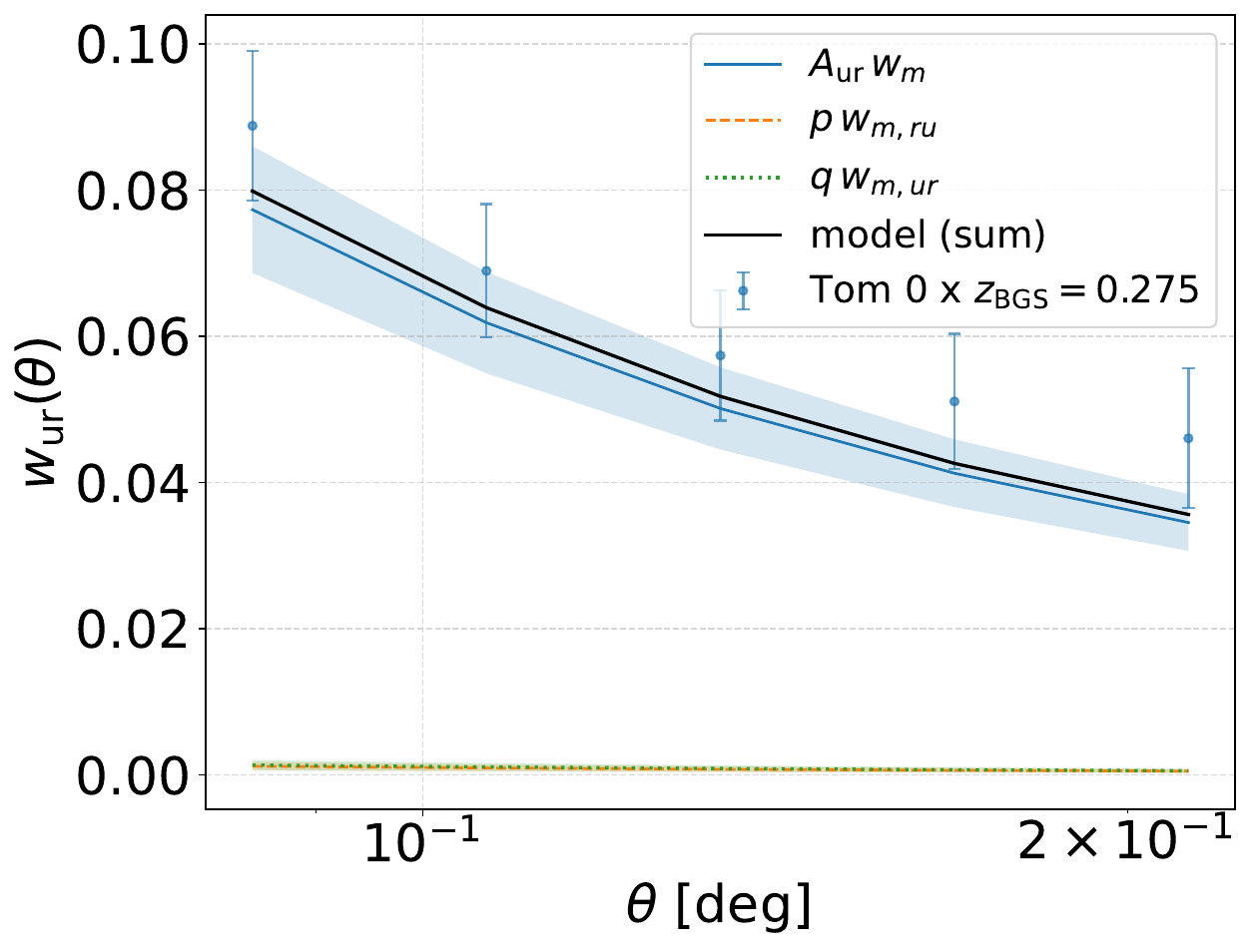}\hfill
  \includegraphics[width=0.32\textwidth]{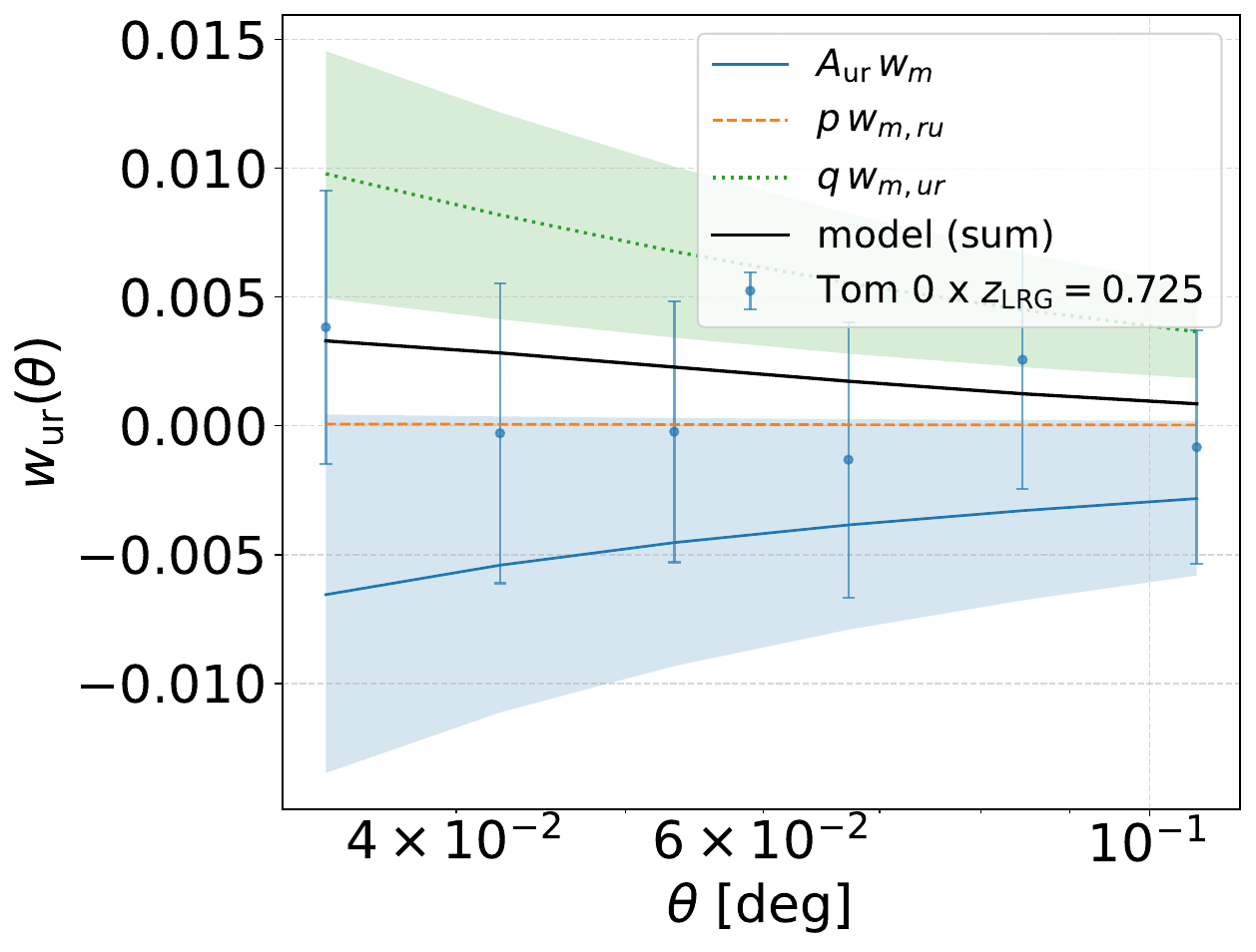}\hfill
  \includegraphics[width=0.32\textwidth]{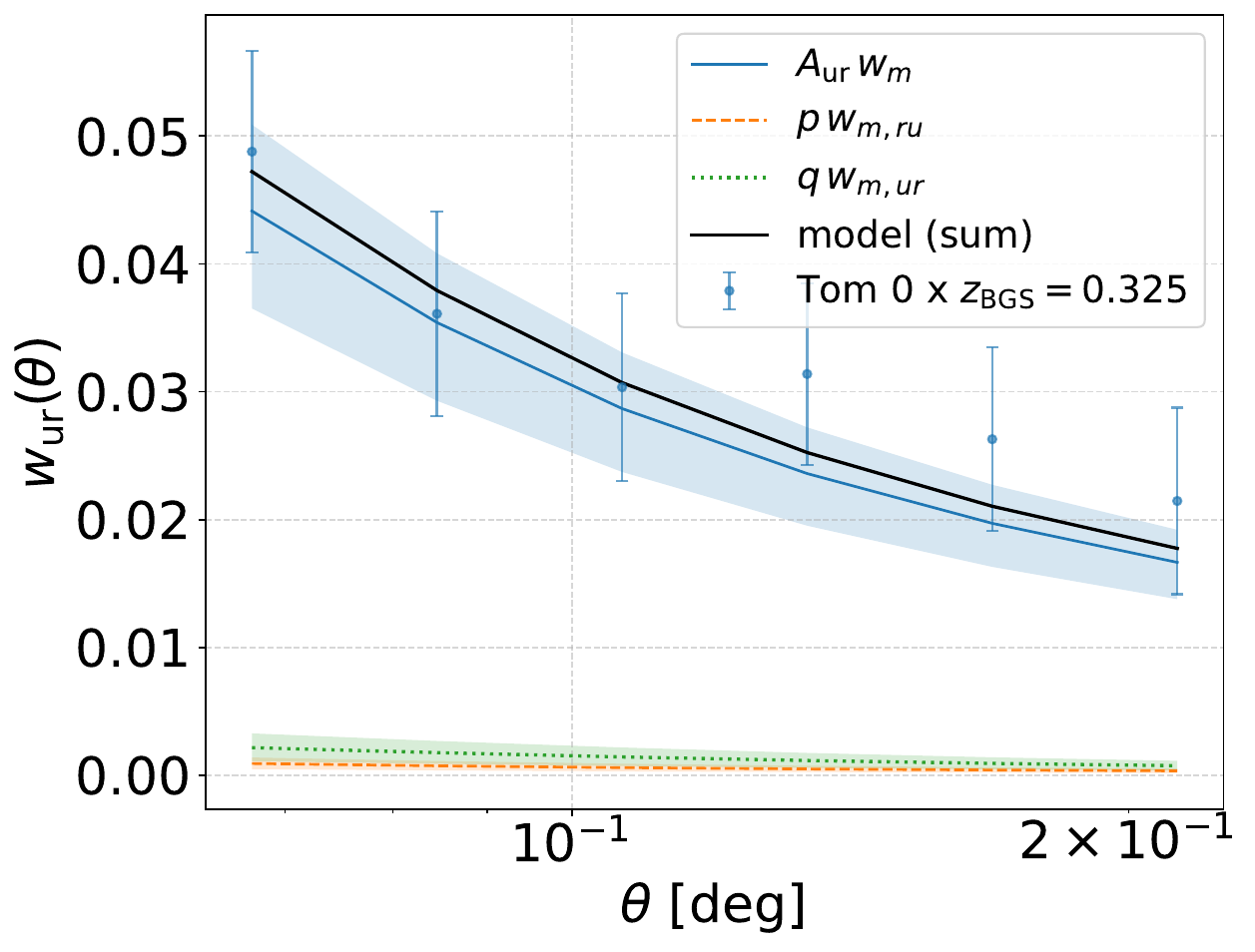}

  \includegraphics[width=0.32\textwidth]{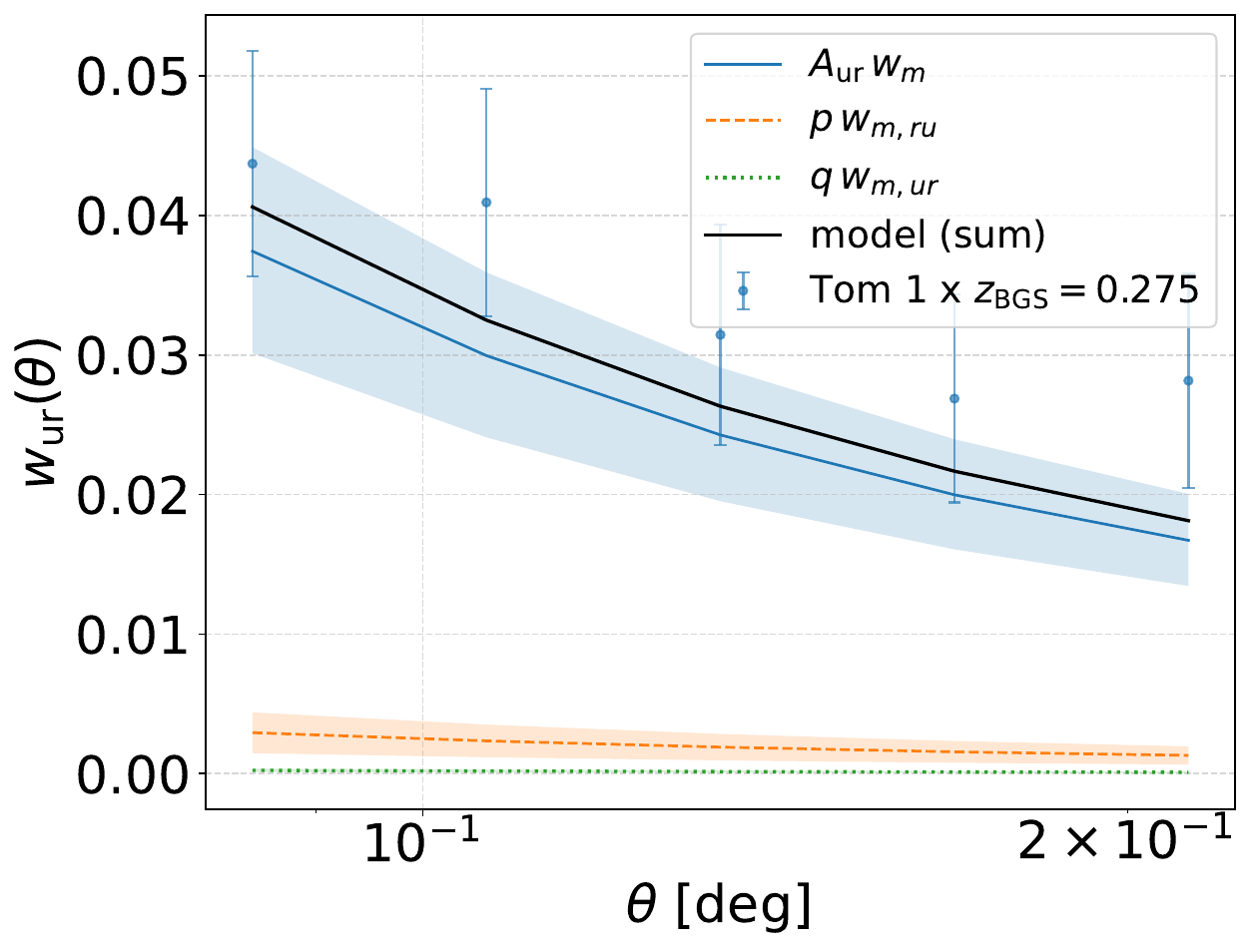}\hfill
  \includegraphics[width=0.32\textwidth]{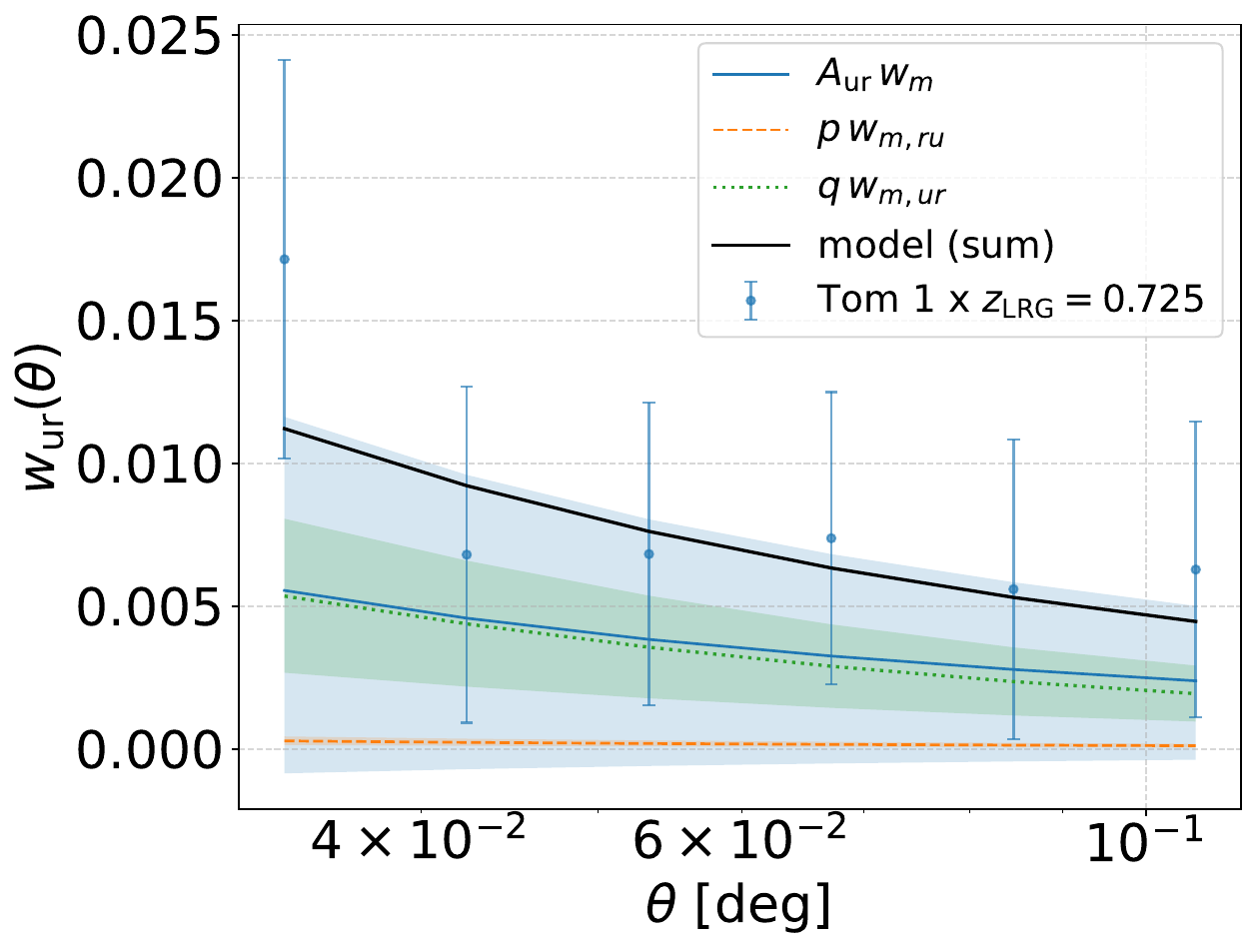}\hfill
  \includegraphics[width=0.32\textwidth]{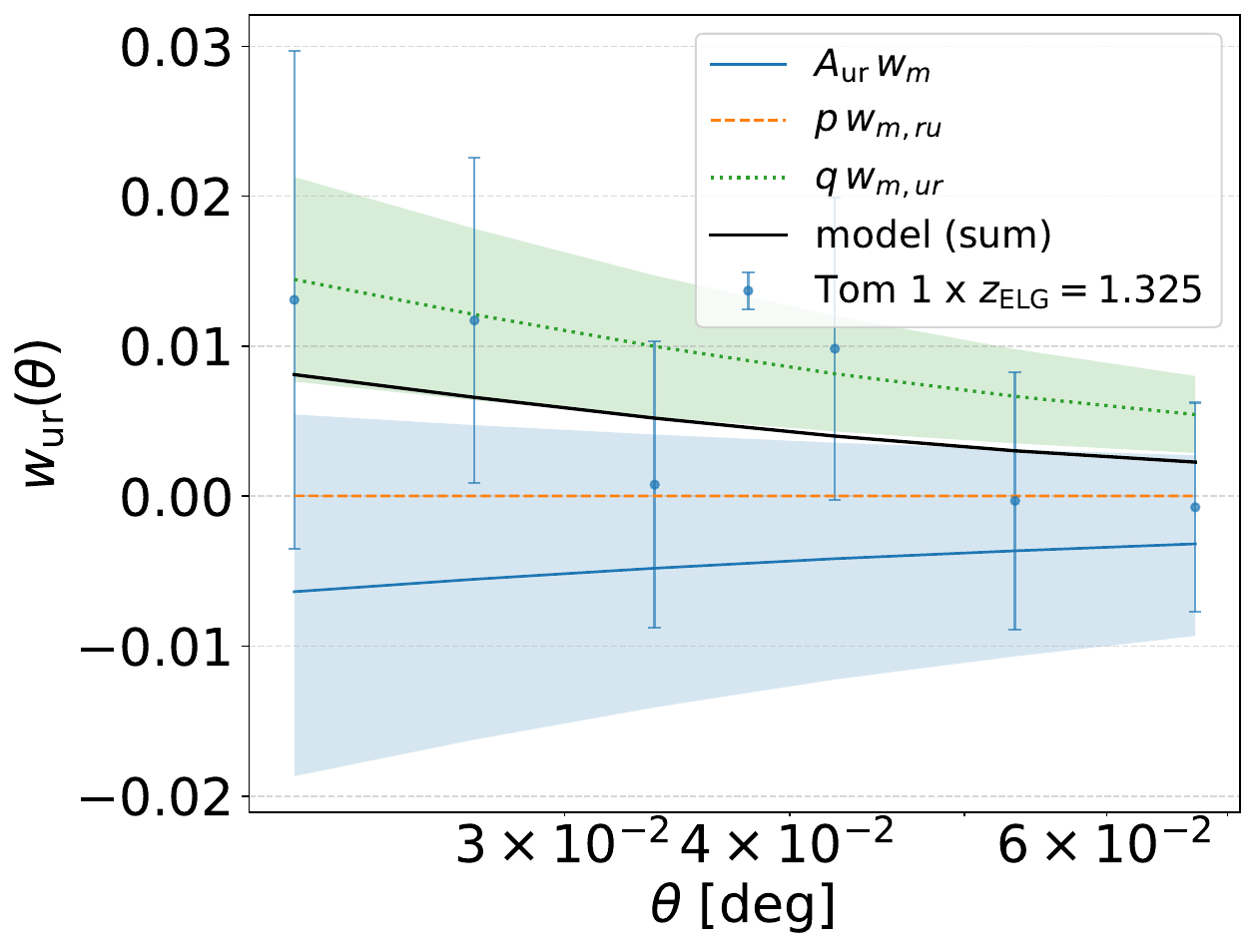}

  \includegraphics[width=0.32\textwidth]{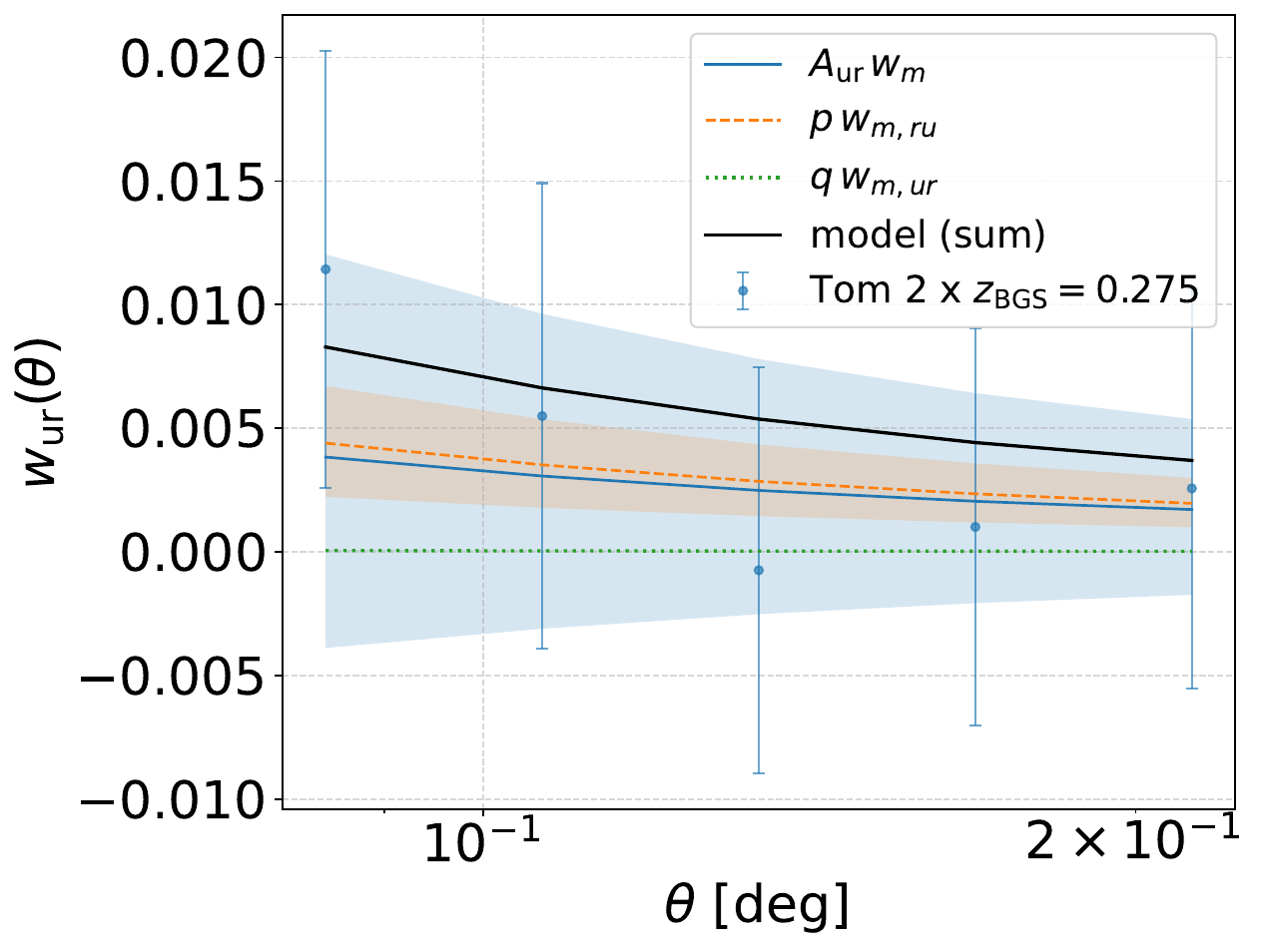}\hfill
  \includegraphics[width=0.32\textwidth]{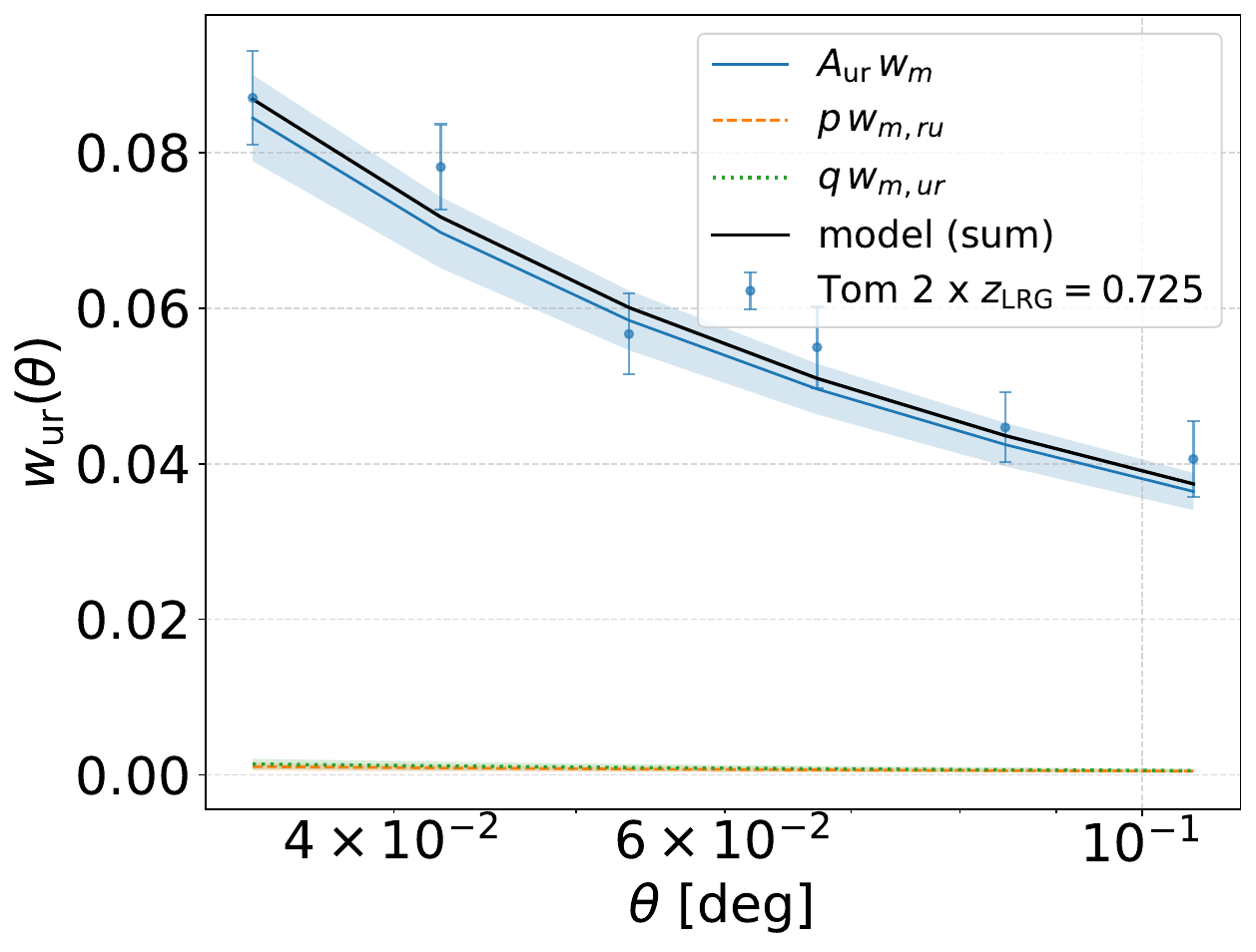}\hfill
  \includegraphics[width=0.32\textwidth]{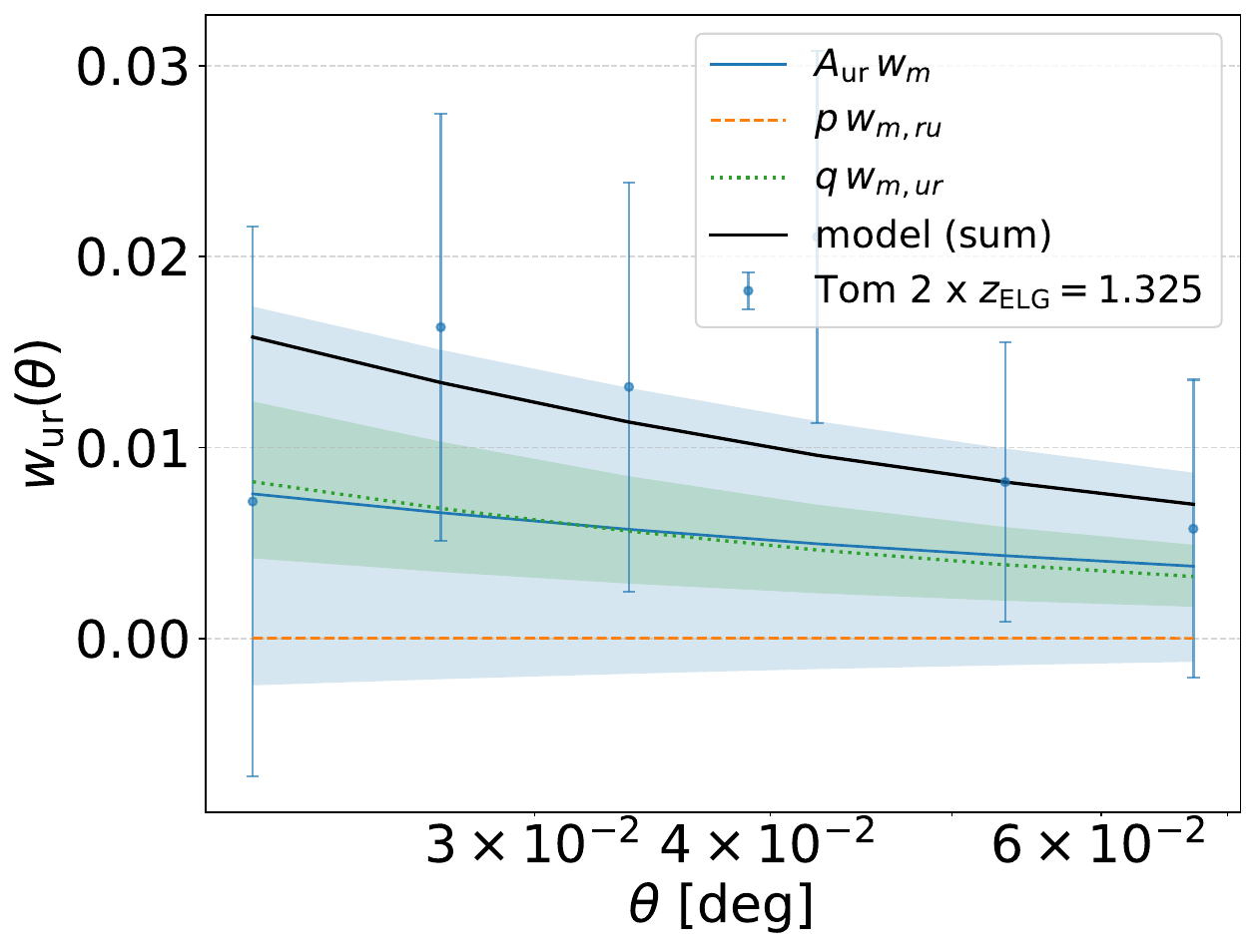}

  \includegraphics[width=0.32\textwidth]{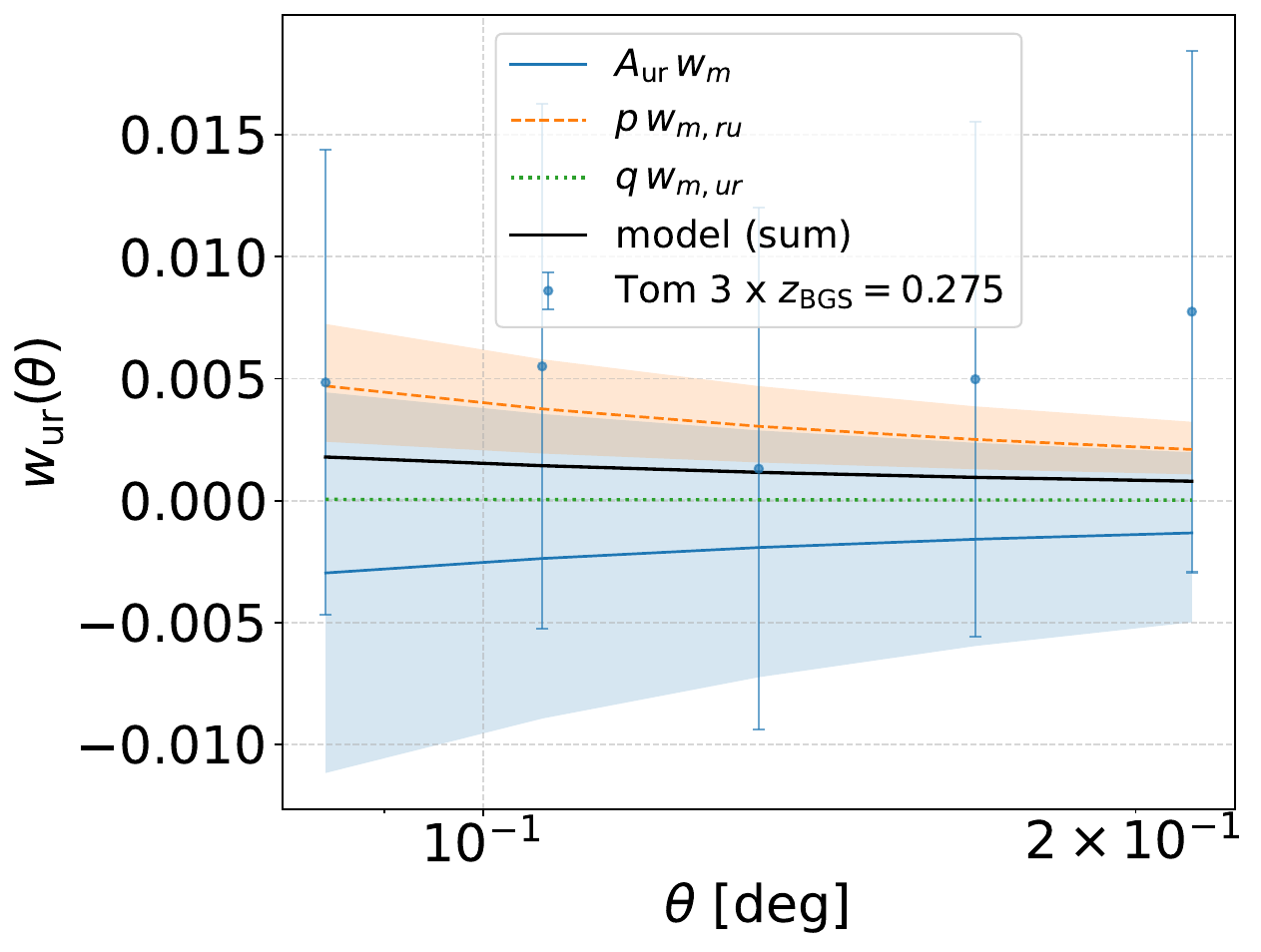}\hfill
  \includegraphics[width=0.32\textwidth]{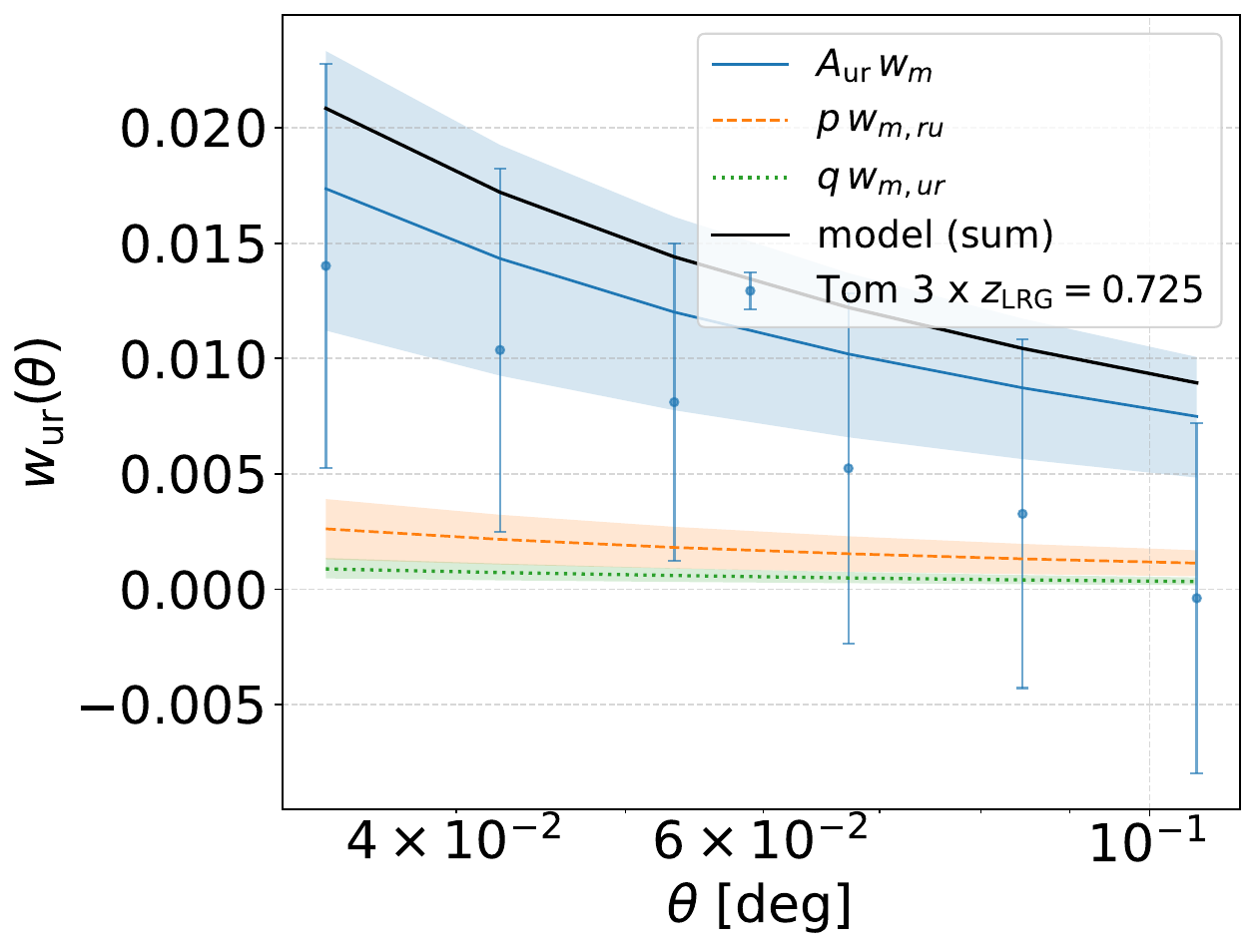}\hfill
  \includegraphics[width=0.32\textwidth]{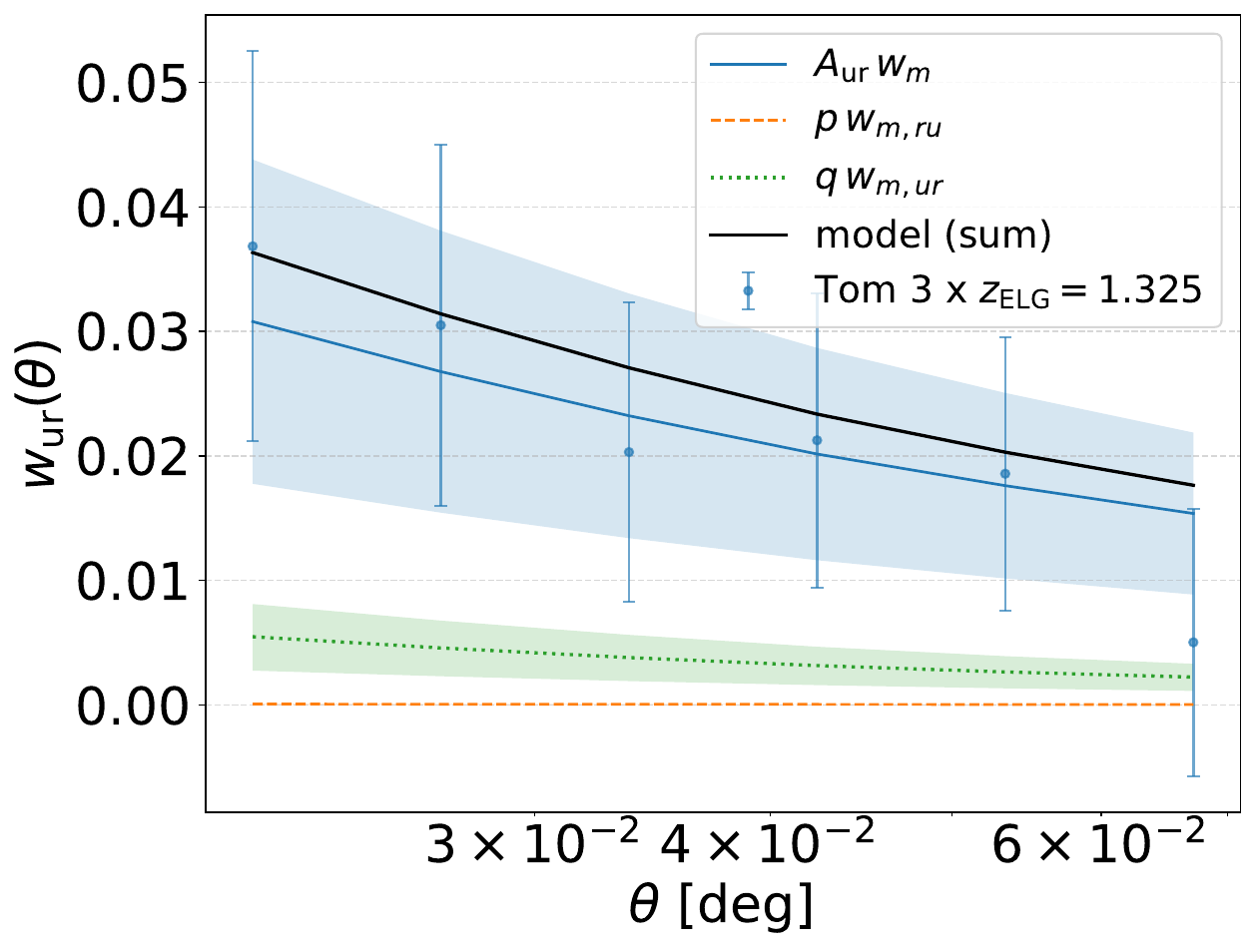}

\caption{\rossi{Component decomposition of the angular cross-correlation
$w_{ur}(\theta)$ between DES-Y3 source galaxies and the DESI
spectroscopic reference sample. The rows correspond to the four
DES-Y3 tomographic bins, while the columns show representative BGS ($\bar z=0.275$), LRG ($\bar z=0.725$), and ELG
($\bar z=1.325$) reference bins. The data points show the measured correlations with jackknife uncertainties. The curves show the clustering
component $A_{ur}w_m(\theta)$, the two magnification contributions
$p\,w_{ru}^{\rm mag}(\theta)$ and $q\,w_{ur}^{\rm mag}(\theta)$,
and their sum. The shaded regions show the 68\% posterior uncertainty
of each individual component. The posterior uncertainty of the total
model is shown separately in Fig.~\ref{fig:wur-desy3xdesi-by-t}.}}
\label{fig:wur-components-desy3xdesi}
\end{figure*}

The cross-correlation function $w_{ur}(\theta)$ in each narrow redshift bin includes contributions from both clustering and magnification, and is modelled according to Eq.~\ref{eq:wcross}.
The fit hence has three parameters: the clustering amplitude $A_{ur}$, and two magnification nuisance parameters, $p$ and $q$, which are products of the number-count slopes and galaxy bias factors (see Sec.~\ref{sec:magnification}).
We adopt Gaussian priors on $p$ and $q$, centred at 4.0 with standard deviation 2.0, \rossi{representing a broad order-of-magnitude expectation for the product of galaxy bias and magnification response, while allowing substantial variation around this value}, and use a flat prior for $A_{ur}$ over the range $(-10, 10)$.  \rossi{We allow negative $A_{ur}$ values so that noise-dominated measurements can fluctuate around zero without introducing a positivity bias.}
The $\chi^2$ function for this model is,
\begin{equation}
\chi^2(A_{ur}, p, q) = \Delta\vec{w}^T \cdot \mathbf{C}^{-1} \cdot \Delta\vec{w} ,
\end{equation}
where,
\begin{equation}
\Delta\vec{w} = \vec{w}_{ur} - A_{ur} \, \vec{w}_m - p \, \vec{w}^{\mathrm{mag}}_{ru} - q \, \vec{w}^{\mathrm{mag}}_{ur} .
\end{equation}
We sample the posterior distribution using MCMC, and determine the MAP and mean estimates of all parameters along with the associated errors.

Fig.~\ref{fig:wur-desy3xdesi-by-t} shows example cross-correlation measurements $w_{ur}(\theta)$, and corresponding best-fit models, for the DES-Y3 $\times$ DESI case (we find consistent trends for KIDS-1000, HSC-Y1 and HSC-Y3).  Each panel corresponds to a specific combination of tomographic source and spectroscopic reference redshift bin.  The blue points show the measured angular cross-correlations with jackknife-derived uncertainties. The solid black curve indicates the best-fit model computed from the MAP parameters $(A_{ur}, p, q)$.  
\rossi{The shaded
region shows the 68\% posterior uncertainty of the total model,
obtained from the joint MCMC samples. We also report
$\chi^2/{\rm dof}$ for each fit, evaluated from the likelihood using
the full covariance matrix.  In general the $\chi^2$ values indicate an acceptable goodness-of-fit, considering the small number of data points.}
Similarly as for $A_{rr}$, the off-diagonal structure in the covariance implies that the best-fitting normalisation does not necessarily coincide with the visual least-squares fit.  The variation of $w_{ur}(\theta)$ across bins traces the redshift distribution of the unknown sample, together with the evolution of the clustering amplitude with redshift and the scale dependence of the magnification contributions $p$ and $q$, illustrating where each term dominates in the cross-correlation signal.  We separately plot these different components in Fig.~\ref{fig:wur-components-desy3xdesi}.

\subsection{Clustering-\texorpdfstring{$z$}{z} amplitude \texorpdfstring{$b_\mathrm{u} p_\mathrm{u}$}{bupu}}

We used the fitted correlation amplitudes to derive the clustering-$z$ signal $b_\mathrm{u}(z_\mathrm{r}) p_\mathrm{u}(z_\mathrm{r})$ via Eq.~\ref{eq:bupu}, where $z_\mathrm{r}$ denotes a  narrow redshift bin of the spectroscopic reference sample.  The error in this signal is computed by propagating the error in $A_{rr}$ and $A_{ur}$,
\begin{equation}
\sigma_{b_\mathrm{u} p_\mathrm{u}} = \sqrt{
\left( \frac{\sigma_{A_{ur}}}{\sqrt{A_{rr}} \, \Delta z} \right)^2 +
\left( \frac{A_{ur} \, \sigma_{A_{rr}}}{2 \, A_{rr}^{3/2} \, \Delta z} \right)^2 },
\end{equation}
where $\sigma_{A_{ur}}$ and $\sigma_{A_{rr}}$ are the uncertainties in the respective fitted amplitudes determined by the individual MCMC fits described above. This method fully accounts for the parameter degeneracies and provides our most reliable estimate of the error in $b_\mathrm{u}(z_\mathrm{r}) p_\mathrm{u}(z_\mathrm{r})$.  We also verified that the posterior errors obtained via our MCMC fitting procedure are consistent with the scatter observed in individual jackknife samples and across different mock realizations, as discussed in the next section.

\section{Results}
\label{sec:results}

\subsection{Validation using mocks}
\label{sec:mock_source_bias}

\begin{figure}
\centering
\includegraphics[width=\columnwidth]{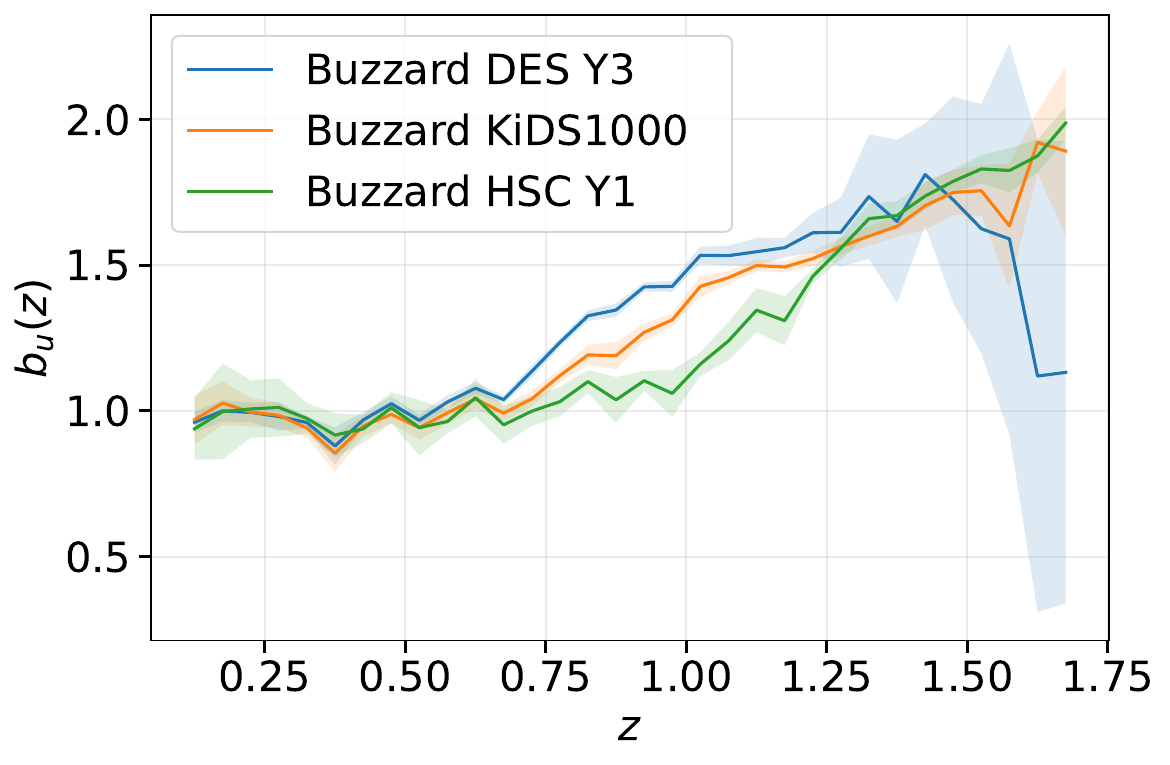}
 \caption{The evolution of the source bias $b_\mathrm{u}(z_\mathrm{r})$ across surveys, as determined from the \textsc{Buzzard} mock catalogues for each imaging survey.  The blue curve corresponds to DES-Y3, the orange curve to KiDS-1000, and the green curve to HSC-Y1.  For each survey, the solid line represents the average over the ten mock regions, and the shaded band indicates the region-to-region scatter.}
\label{fig:sourcebias}
\end{figure} 

\begin{figure*}
\centering
\begin{subfigure}{0.48\textwidth}
    \centering
    \includegraphics[width=\linewidth]{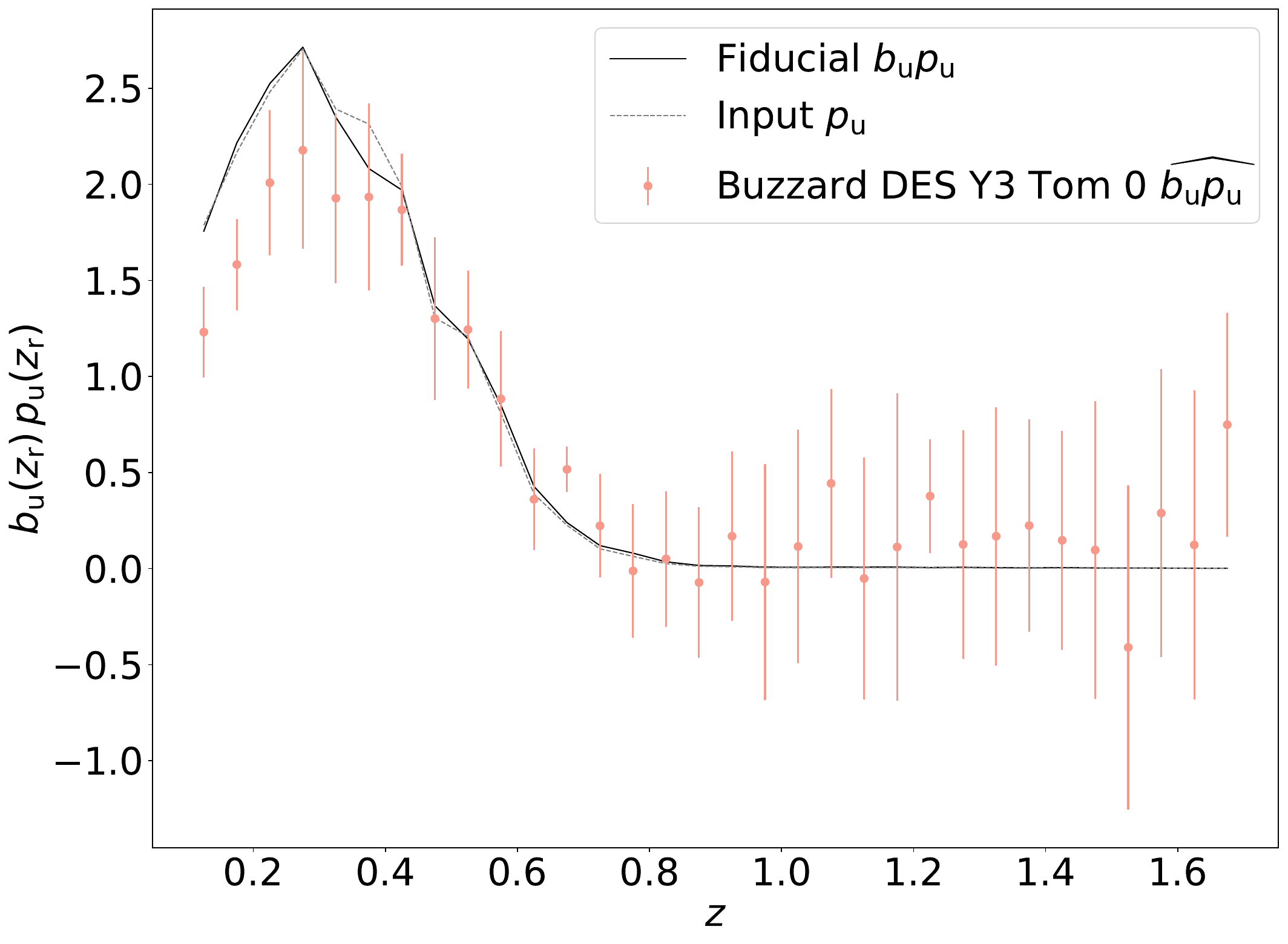}
\end{subfigure}\hfill
\begin{subfigure}{0.48\textwidth}
    \centering
    \includegraphics[width=\linewidth]{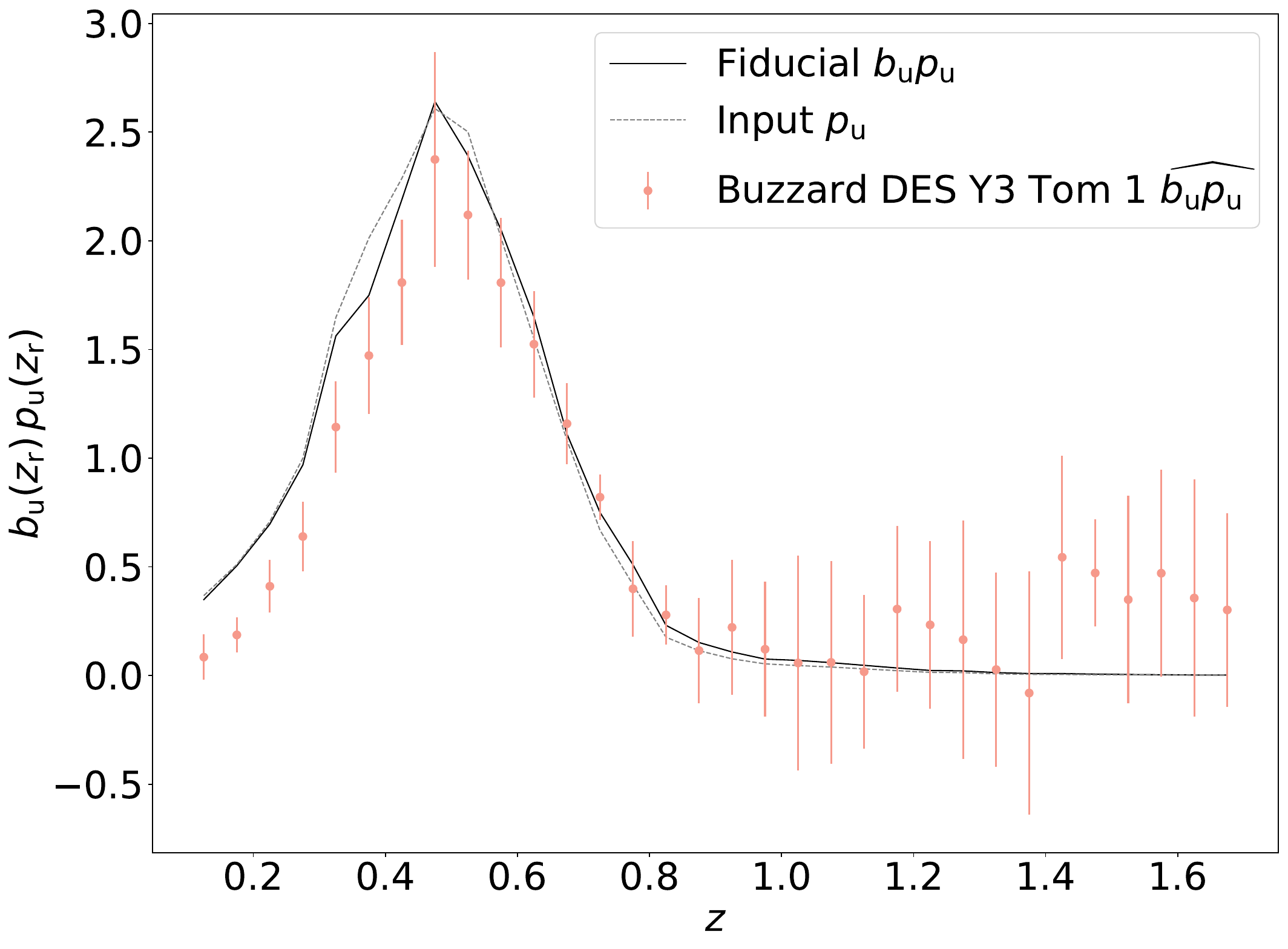}
\end{subfigure}
\vspace{0.5em}
\begin{subfigure}{0.48\textwidth}
    \centering
    \includegraphics[width=\linewidth]{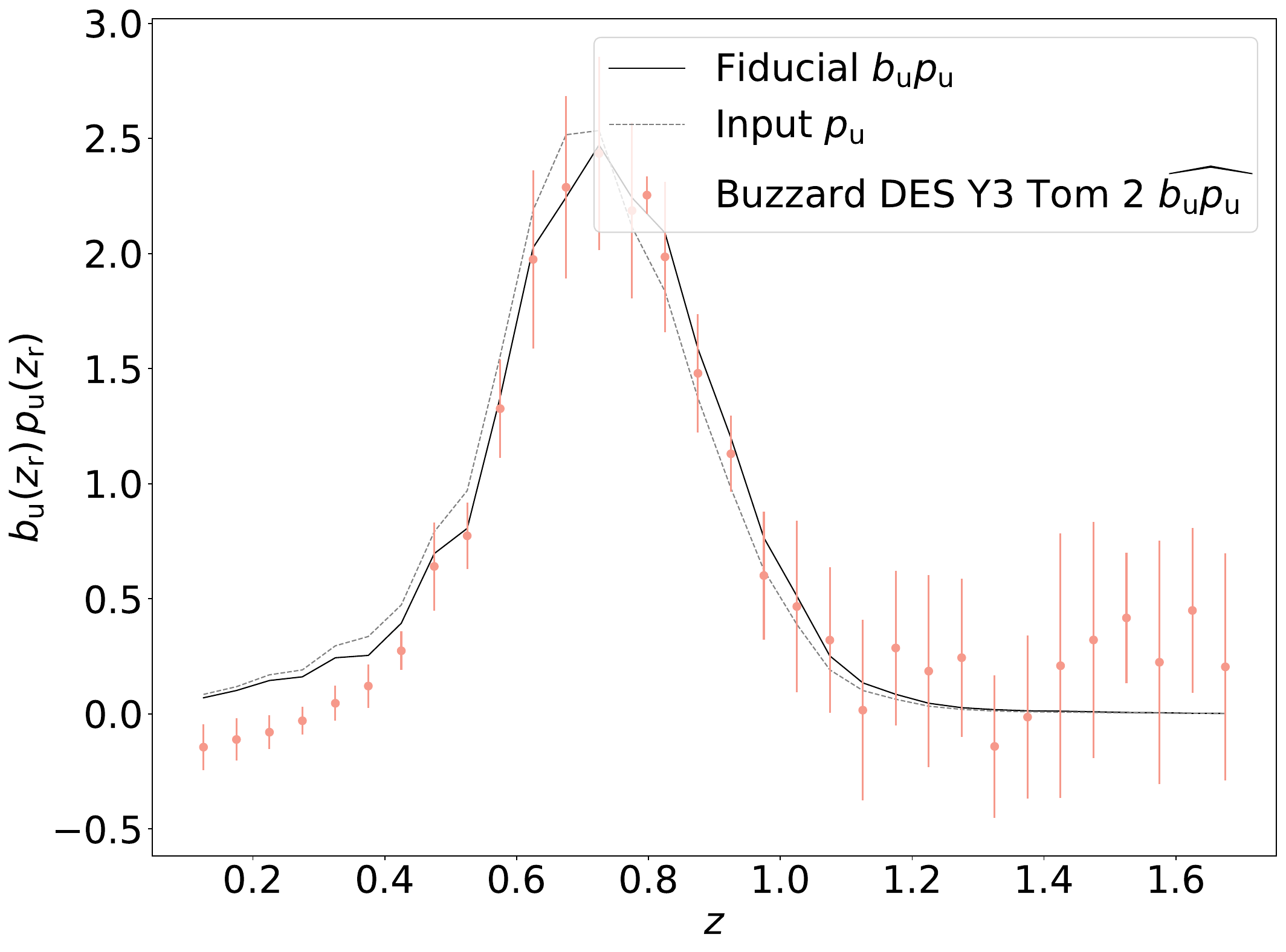}
\end{subfigure}\hfill
\begin{subfigure}{0.48\textwidth}
    \centering
    \includegraphics[width=\linewidth]{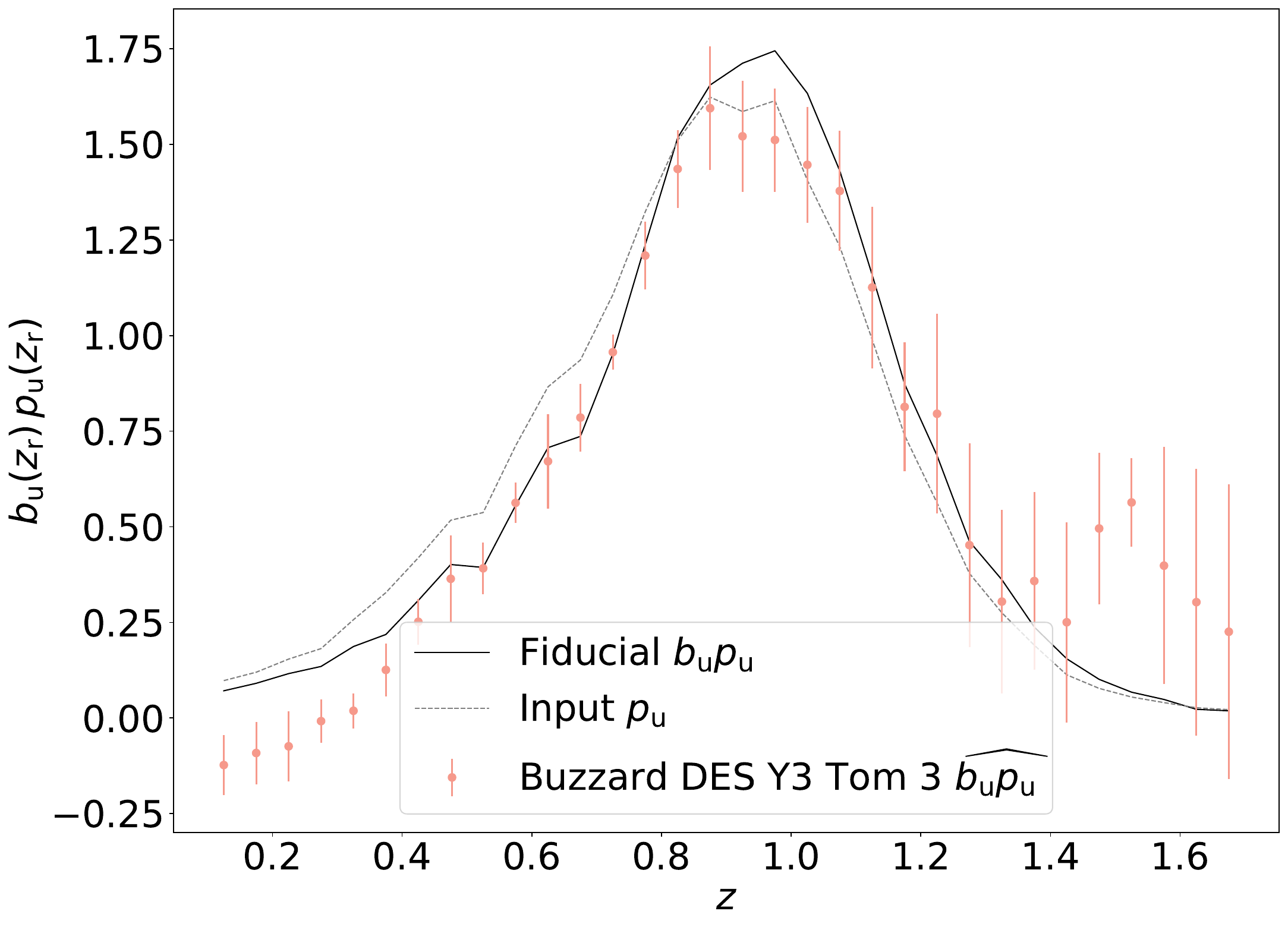}
\end{subfigure}
\caption{Clustering-$z$ validation using mock catalogues for DESI and DES-Y3. The panels show the recovery of the source redshift distribution in the four DES-Y3 tomographic bins. The solid black line shows the \rossi{fiducial bias-weighted redshift distribution} $b_\mathrm{u}(z_\mathrm{r})\,p_\mathrm{u}(z_\mathrm{r})$ from the simulation, while the dashed black line shows the \rossi{input source redshift distribution} $p_\mathrm{u}(z_\mathrm{r})$. \rossi{The red points show the clustering-based estimate of $b_\mathrm{u}(z_\mathrm{r})\,p_\mathrm{u}(z_\mathrm{r})$, measured from $w_{ur}(\theta)$ in discrete redshift slices, with error bars
indicating the mock-to-mock scatter.}
\rossi{The quantitative recovery of the mean redshift is reported in
Table~\ref{tab:mock_mean_z_stats}.}}
?
\label{fig:clustz-desy3-mock}
\end{figure*}

To validate our clustering redshift methodology and associated uncertainty estimates, we performed a series of tests on realistic mock catalogues constructed for each imaging survey using the \textsc{Buzzard} simulations, as described in Sec.~\ref{sec:mocks}. We analyse ten independent realizations of the DES-Y3, KiDS-1000 and HSC-Y1 source samples, allowing us to assess the scatter in the inferred amplitudes $A_{rr}$ and $A_{ur}$ across the realizations. We also estimated uncertainties using jackknife resampling within each mock region. In the regime considered, the contribution of $A_{rr}$ to the total uncertainty is subdominant relative to $A_{ur}$.

\rossi{We find good agreement between the uncertainties obtained from
jackknife resampling, the MCMC posterior including marginalisation
over $p$ and $q$, and the mock-to-mock scatter. In Fig.~5 the
error bars show the mock-to-mock scatter, which is consistent with
the posterior uncertainty estimated for the individual mock regions.
}


We then used the mocks to test the recovery of the input source redshift distribution. Because the clustering redshift estimator measures the bias-weighted quantity $b_\mathrm{u}(z_\mathrm{r}) \, p_\mathrm{u}(z_\mathrm{r})$, rather than the redshift distribution $p_\mathrm{u}(z_\mathrm{r})$ directly, we also determined the effective bias $b_\mathrm{u}(z_\mathrm{r})$ for the mock source samples from their angular auto-correlations in narrow redshift bins, which is possible for the mock sources because their true redshifts are known.  Following the methods described in Sec.~\ref{sec:fits}, for each mock region and redshift bin, we computed the source auto-correlation function measurements $w_{uu}(\theta)$ with \textsc{treecorr}, using jackknife resampling.  We then fit the measured auto-correlation at each redshift $\bar z$ with a single-parameter amplitude model, $w_{uu}(\theta) = A_{uu}\,w_m(\theta)$, where $w_m(\theta)$ is the matter correlation function projected at $\bar z$ evaluated using Eq.~\ref{eq:wll}, using the same fitting range $1.5 < R < 5 \,h^{-1}\mathrm{Mpc}$.  The amplitude $A_{uu}$ was obtained by minimizing $\chi^2(A_{uu}) = \Delta \vec{w}^\mathrm{T} \mathbf{C}^{-1} \Delta \vec{w}$, with $\Delta \vec{w} = \vec{w}_{uu} - A_{uu} \vec{w}_m$ and $\mathbf{C}$ is the jackknife covariance restricted to the angular fitting range, with flat priors $A_{uu}\in(0,20)$.  For each survey, the source bias was then determined as,
\begin{equation}
b_\mathrm{u}(z_\mathrm{r}) = \langle A_{uu}^{1/2}(z_\mathrm{r}) \rangle_\mathrm{reg},
\end{equation}
averaged across all mock regions.

Fig.~\ref{fig:sourcebias} shows the mean source-galaxy clustering bias $b_\mathrm{u}(z_\mathrm{r})$ inferred from the mock catalogues for each imaging survey (blue: DES-Y3, orange: KiDS-1000, and green: HSC-Y1).  The shaded regions represent the standard deviation obtained as the scatter across the mock regions.  The rise of $b_\mathrm{u}(z_\mathrm{r})$ with redshift reflects the increasing typical halo mass of galaxies selected at higher redshift.  
\rossi{At low redshift, $b_\mathrm{u}(z_\mathrm{r})$ evolves only weakly, such that the source galaxy bias correction has little impact on the shape of the recovered low-$z$ distributions.}
Although the amplitude of $b_\mathrm{u}(z_\mathrm{r})$ differs somewhat across surveys due to differing selection functions and depths, the trends are smooth and consistent with expectations from galaxy--halo connection models, \rossi{as $b_\mathrm{u}(z_\mathrm{r})$ is expected to evolve with redshift due to the evolution of halo bias and changes in the selected galaxy population \citep[e.g.][]{2010ApJ...724..878T, 2018ARA&A..56..435W}.}

We incorporated this empirically-derived $b_\mathrm{u}(z_\mathrm{r})$ into the construction of the normalized source redshift distribution,
\begin{equation}
p_\mathrm{u}(z_\mathrm{r}) \propto \frac{1}{b_\mathrm{u}(z_\mathrm{r})}\,\widehat{b_\mathrm{u} p_\mathrm{u}}(z_\mathrm{r}),
\end{equation}
and compared the resulting $p_\mathrm{u}(z_\mathrm{r})$ obtained with and without source bias correction on the mocks.  We obtained a smooth bias evolution function by interpolating between the discrete $b_\mathrm{u}(z_\mathrm{r})$ values.  
We note that \rossi{data-driven estimates of $b_\mathrm{u}(z_\mathrm{r})$, inferred directly from source clustering rather than calibrated from simulations,} can also be applied \citep{2020A&A...642A.200V, 2022MNRAS.513.5517C, 2025A&A...702A.155D}, although we do not consider these in the current study.

Fig.~\ref{fig:clustz-desy3-mock} shows the result for the DES-Y3 mock sources (results for KiDS-1000 and HSC-Y1 are similar). The solid black curve denotes the \rossi{fiducial bias-weighted distribution} $b_\mathrm{u}(z_\mathrm{r}) \, p_\mathrm{u}(z_\mathrm{r})$, while the dashed black curve shows the underlying \rossi{input} source redshift distribution $p_\mathrm{u}(z_\mathrm{r})$. 
The red points show the clustering-based estimate of
$b_\mathrm{u}(z_\mathrm{r})\,p_\mathrm{u}(z_\mathrm{r})$,
with error bars indicating the mock-to-mock scatter.
\rossi{The dashed and solid black distributions remain broadly similar across the redshift range, indicating that source-bias evolution does not dominate the recovered source redshift distribution. Its effect is nevertheless non-negligible: across the mock samples, bias weighting shifts the fiducial mean redshift by up to $\sim0.04$, as quantified in Table~\ref{tab:mock_mean_z_stats}.}


We quantify these results by comparing the mean redshifts of the recovered bias-weighted distributions and the fiducial \textsc{Buzzard} inputs.  Table~\ref{tab:mock_mean_z_stats} lists, for each survey and tomographic bin, the means of the fiducial bias-weighted distribution $\bar z_{\rm fid}^{(b_\mathrm{u} p_\mathrm{u})}$, the underlying source distribution $\bar z_{\rm fid}^{(p_\mathrm{u})}$ (i.e.\ with $b_\mathrm{u}=1$), and the recovered mean $\bar z_{\rm meas}$ from the clustering-$z$ measurements with its propagated uncertainty.

We adopt the same procedure as for the data, computing the means within a restricted redshift window $z \in [z_{\min}, z_{\max}]$ defined from the fiducial bias-weighted template: $z_{\min}$ is fixed to the analysis lower bound, while $z_{\max} = \min(z_{\rm max,grid}, z_{\rm peak} + 0.6)$, where $z_{\rm peak}$ is the peak of the fiducial $b_\mathrm{u}(z_\mathrm{r})p_\mathrm{u}(z_\mathrm{r})$ model for each tomographic bin. 
\rossi{Relative to the fiducial bias-weighted distribution
$b_\mathrm{u} \, p_\mathrm{u}$, the recovered means are within
$1.5\sigma$ in all bins except the highest-redshift HSC-Y1 bin
($1.81\sigma$). Relative instead to the underlying unweighted
$p_\mathrm{u}(z)$, larger shifts occur for DES-Y3 tomographic bin 3
($2.14\sigma$), KiDS-1000 bin 4 ($1.75\sigma$), and HSC-Y1 bins 2
and 3 ($1.50\sigma$ and $2.07\sigma$, respectively). This difference
reflects the effect of source-bias evolution: bias weighting changes the fiducial mean redshift by up to
$\sim0.04$ across the mock samples, and the recovered means are
typically within the quoted statistical uncertainties of the
fiducial bias-weighted values.}

\begin{table*}
    \centering
 \begin{tabular}{lccccc}
\hline\hline

Survey, tomographic bin &
$\bar z_{\rm fid}^{(b_\mathrm{u} p_\mathrm{u})}$ &
$\bar z_{\rm fid}^{(p_\mathrm{u})}$ &
$\bar z_{\rm meas} \pm \sigma(\bar z_{\rm meas})$ &
$\sigma_{\Delta z}^{(b_\mathrm{u} p_\mathrm{u})}$ &
$\sigma_{\Delta z}^{(p_\mathrm{u})}$ \\
\hline
Buzzard DES-Y3, tom 0    & 0.335 & 0.334 & $0.358 \pm 0.037$ &  0.62 &  0.67 \\
Buzzard DES-Y3, tom 1    & 0.489 & 0.478 & $0.522 \pm 0.049$ &  0.67 &  0.89 \\
Buzzard DES-Y3, tom 2    & 0.723 & 0.700 & $0.765 \pm 0.064$ &  0.66 &  1.02 \\
Buzzard DES-Y3, tom 3    & 0.901 & 0.858 & $0.979 \pm 0.057$ &  1.38 &  2.14 \\
Buzzard KiDS-1000, tom 0 & 0.250 & 0.251 & $0.251 \pm 0.047$ &  0.03 & -0.01 \\
Buzzard KiDS-1000, tom 1 & 0.393 & 0.392 & $0.407 \pm 0.071$ &  0.20 &  0.21 \\
Buzzard KiDS-1000, tom 2 & 0.527 & 0.522 & $0.554 \pm 0.066$ &  0.41 &  0.48 \\
Buzzard KiDS-1000, tom 3 & 0.770 & 0.755 & $0.828 \pm 0.074$ &  0.78 &  0.99 \\
Buzzard KiDS-1000, tom 4 & 0.964 & 0.937 & $1.052 \pm 0.066$ &  1.33 &  1.75 \\
Buzzard HSC-Y1, tom 0    & 0.440 & 0.439 & $0.405 \pm 0.108$ & -0.32 & -0.32 \\
Buzzard HSC-Y1, tom 1    & 0.758 & 0.752 & $0.837 \pm 0.096$ &  0.82 &  0.88 \\
Buzzard HSC-Y1, tom 2    & 1.068 & 1.045 & $1.185 \pm 0.093$ &  1.26 &  1.50 \\
Buzzard HSC-Y1, tom 3    & 1.311 & 1.291 & $1.447 \pm 0.075$ &  1.81 &  2.07 \\
\hline
    \end{tabular}
\caption{Validation of the recovery of the mean redshifts of the source distributions applying the clustering-$z$ pipeline to the \textsc{Buzzard} mock catalogues.  For each survey and tomographic bin we report: (i) the mean redshift of the fiducial bias-weighted distribution $\bar z_{\rm fid}^{(b_\mathrm{u} p_\mathrm{u})}$; (ii) the mean of the underlying source redshift distribution $\bar z_{\rm fid}^{(p_\mathrm{u})}$ (i.e.\ with $b_\mathrm{u}=1$); (iii) the mean redshift of the recovered clustering-$z$ estimator $\bar z_{\rm meas}$ and its propagated uncertainty $\sigma(\bar z_{\rm meas})$; and (iv) the normalized shifts $\sigma_{\Delta z}^{(b_\mathrm{u} p_\mathrm{u})} \equiv (\bar z_{\rm meas}-\bar z_{\rm fid}^{(b_\mathrm{u} p_\mathrm{u})}) / \sigma(\bar z_{\rm meas})$ and $\sigma_{\Delta z}^{(p_\mathrm{u})} \equiv (\bar z_{\rm meas}-\bar z_{\rm fid}^{(p_\mathrm{u})}) / \sigma(\bar z_{\rm meas})$. 
\rossi{Bias weighting changes the fiducial mean redshift by up to
$\sim0.04$ across the mock samples} ($|\bar z_{\rm fid}^{(b_\mathrm{u} p_\mathrm{u})}-\bar z_{\rm fid}^{(p_\mathrm{u})}| \lesssim 0.04$), and the recovered means are typically within a statistical error margin of the true values.}

\label{tab:mock_mean_z_stats}
\end{table*}

\subsection{Posterior estimates of \texorpdfstring{$b_\mathrm{u}(z_\mathrm{r})\,p_\mathrm{u}(z_\mathrm{r})$}{bupu}}
\label{sec:pdf_results}

\subsubsection{Results for different source samples}

\begin{figure*}[t]
\centering
\begin{subfigure}[t]{0.48\textwidth}\centering
 \includegraphics[width=\linewidth]{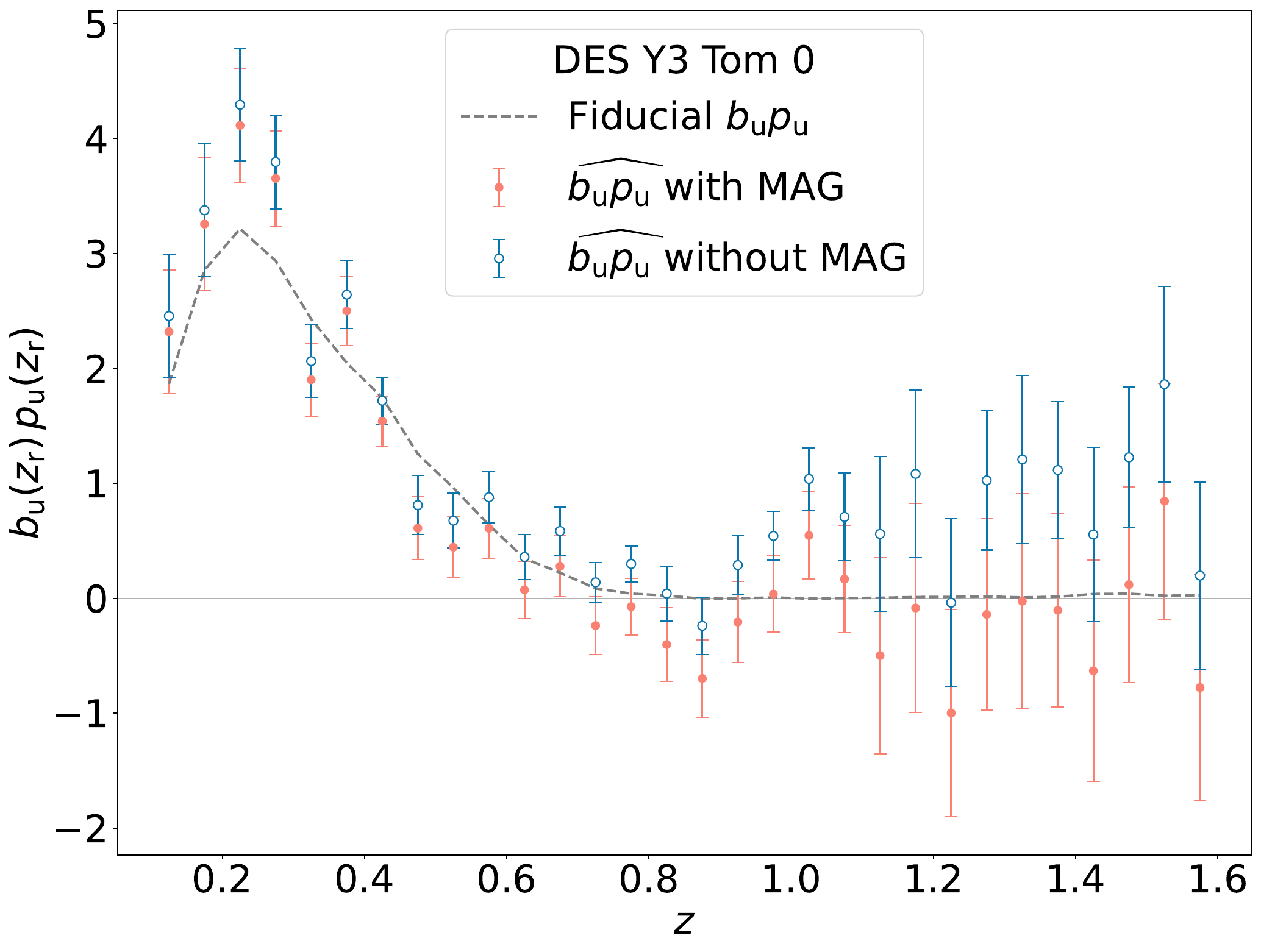}
 \end{subfigure}\hfill
\begin{subfigure}[t]{0.48\textwidth}\centering
\includegraphics[width=\linewidth]{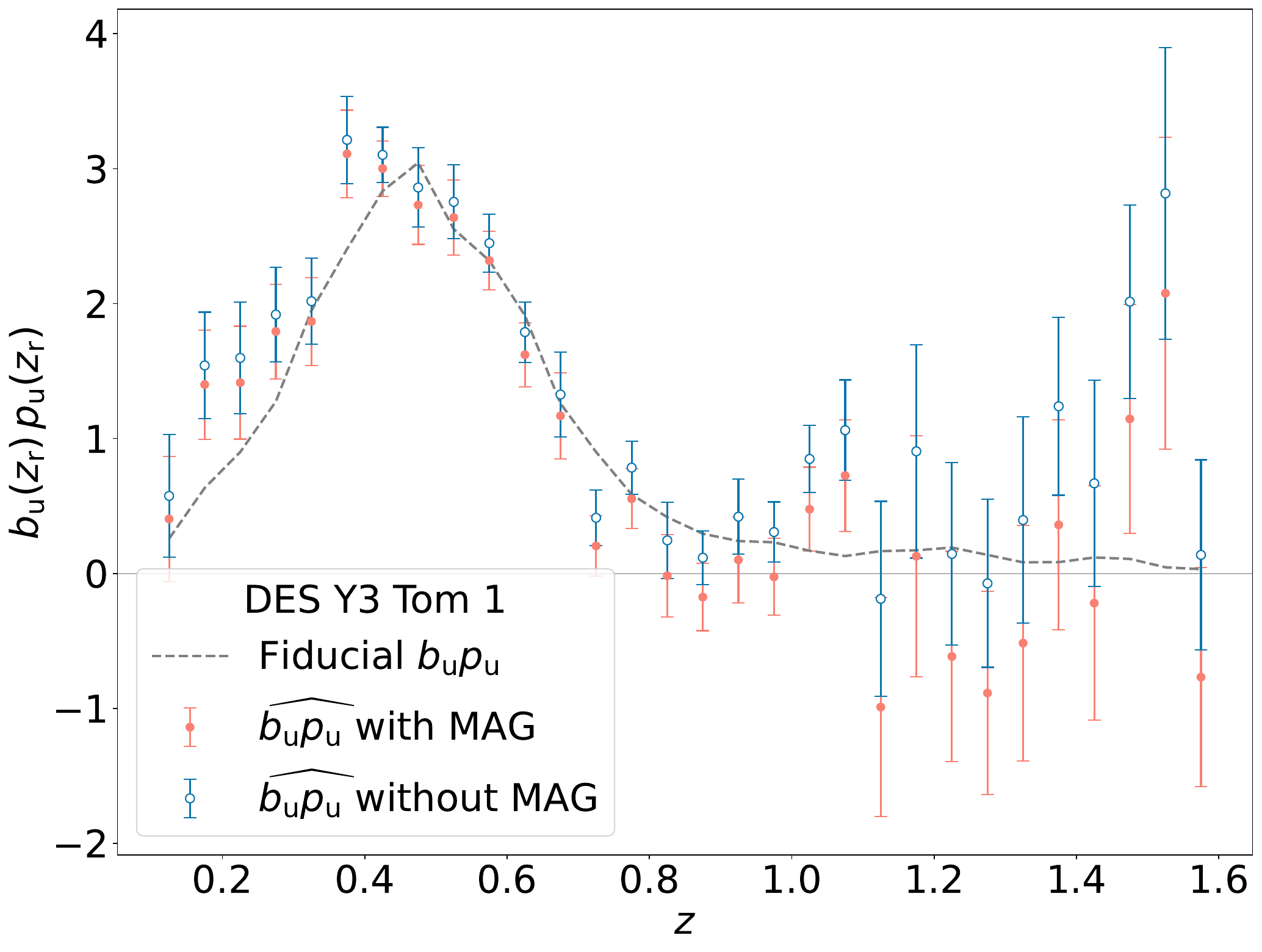}
 \end{subfigure}\\[0.6em]
\begin{subfigure}[t]{0.48\textwidth}\centering
\includegraphics[width=\linewidth]{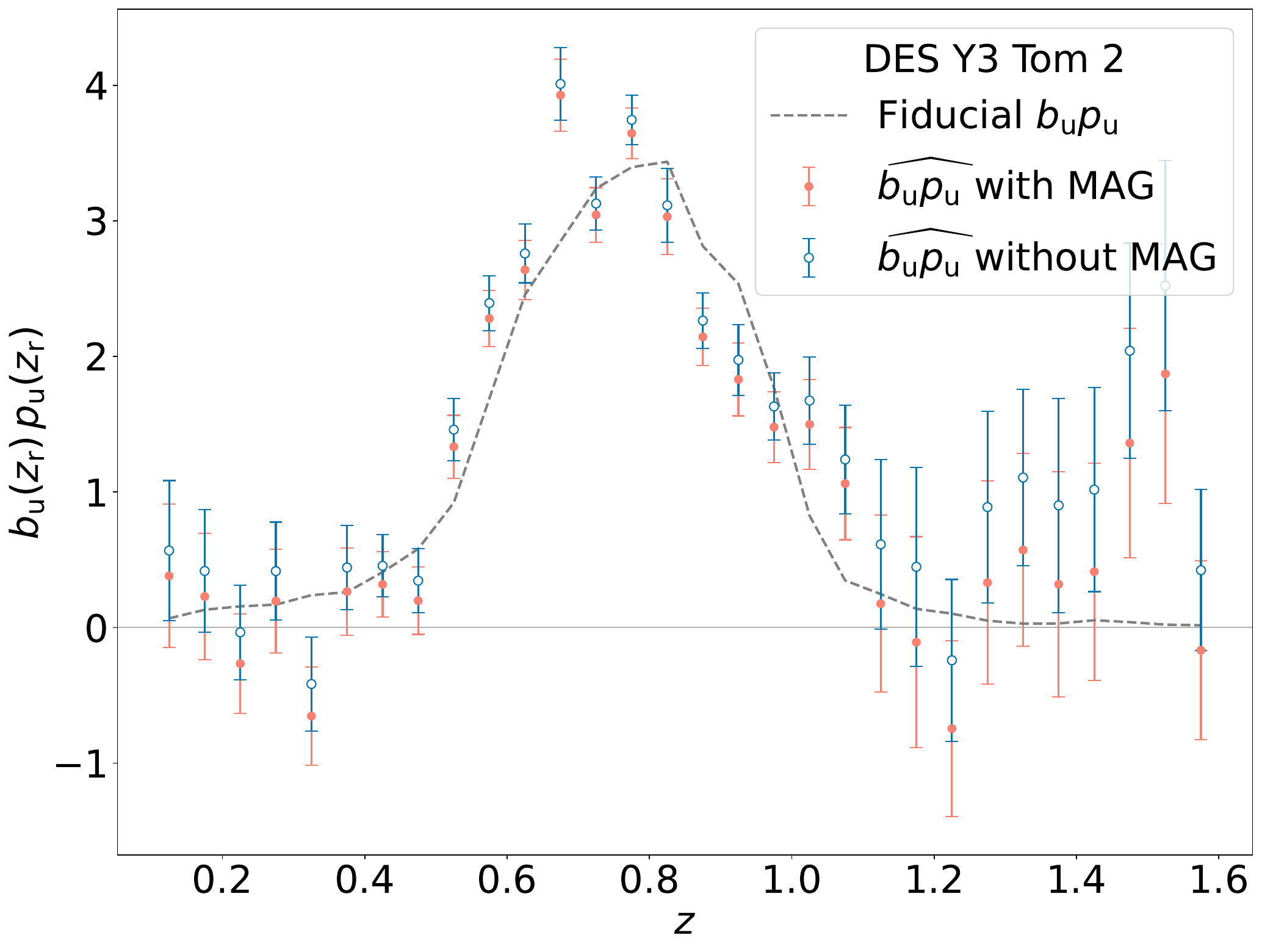}
 \end{subfigure}\hfill
\begin{subfigure}[t]{0.48\textwidth}\centering
\includegraphics[width=\linewidth]{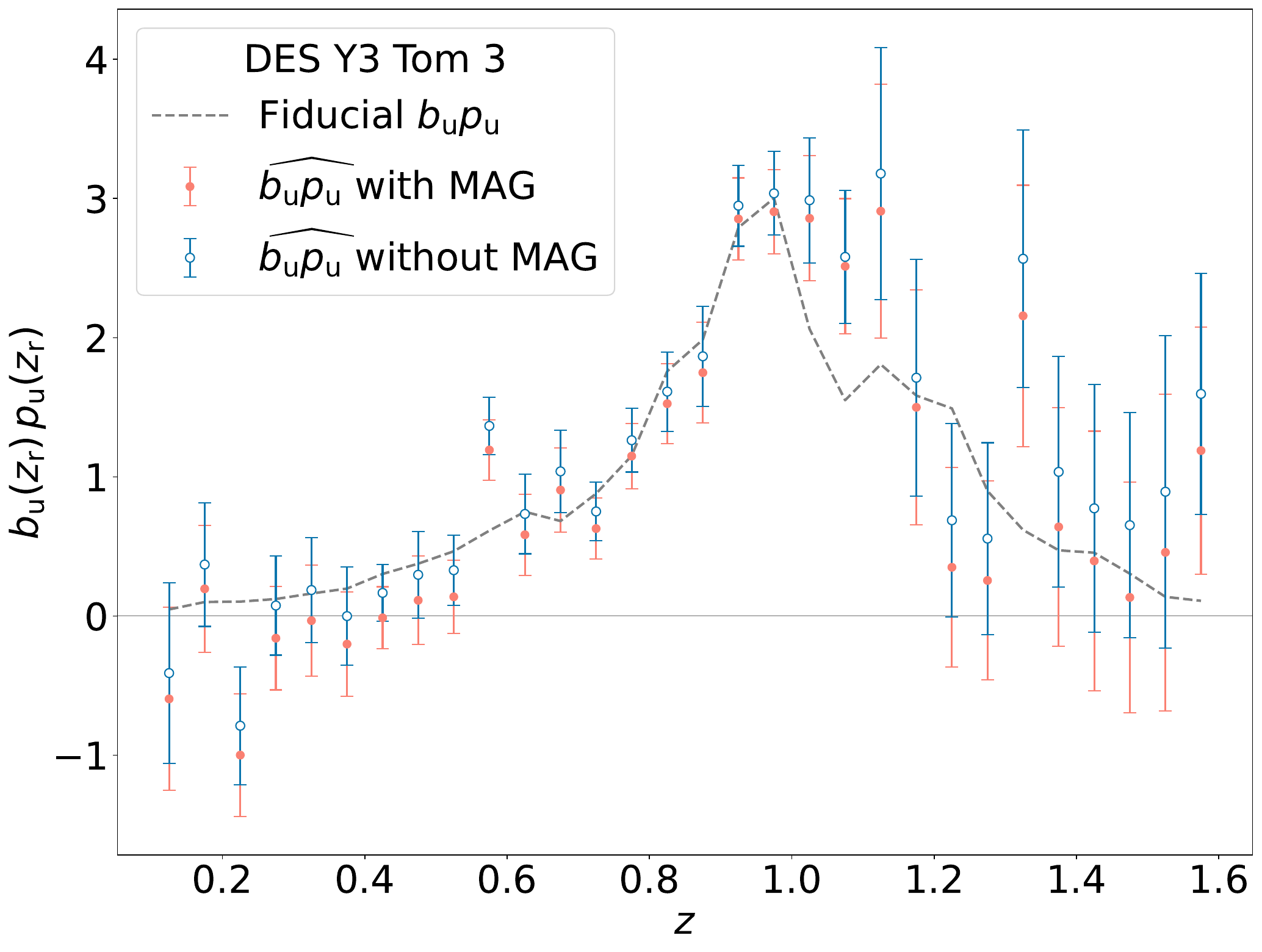}
 \end{subfigure}
\caption{Posterior estimates of the bias–weighted source redshift distribution $b_\mathrm{u}(z_\mathrm{r})\,p_\mathrm{u}(z_\mathrm{r})$ in each DES-Y3 tomographic bin.  The black dashed curve shows the survey's fiducial redshift distribution multiplied by the source bias $b_\mathrm{u}$ trend calibrated on the mocks, multiplied by a single free amplitude parameter $A$.   The coloured points show the estimator $b_\mathrm{u} \, p_\mathrm{u}(z_\mathrm{r})$ with $1\sigma$ uncertainties, where the red points correspond to the baseline analysis with magnification contributions included, and blue points show the result when magnification is excluded.}
\label{fig:post_desy3}
\end{figure*}

\begin{figure*}
\centering
\begin{subfigure}[t]{0.48\textwidth}\centering
\includegraphics[width=\linewidth]{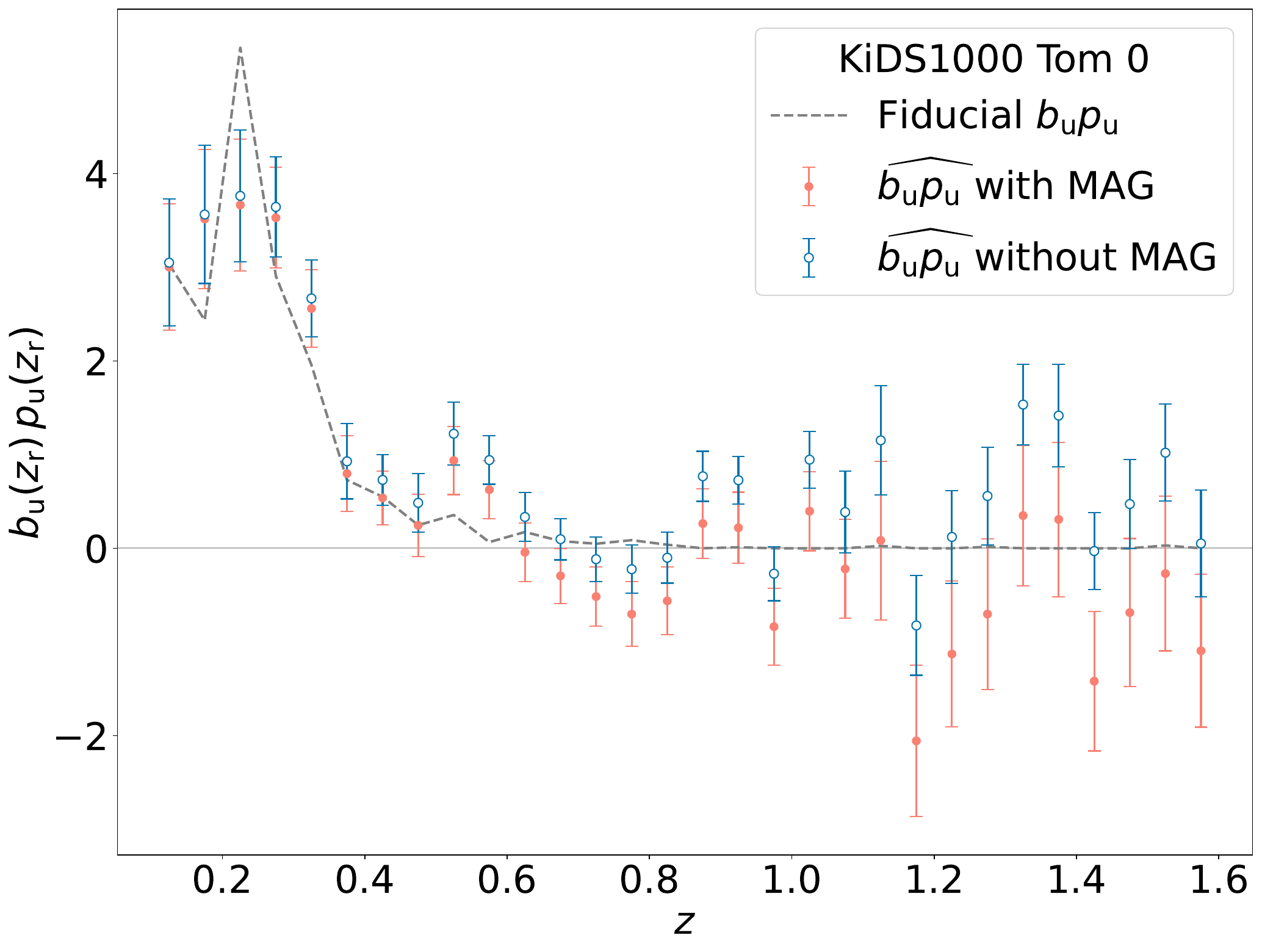}
 \end{subfigure}\hfill
\begin{subfigure}[t]{0.48\textwidth}\centering
\includegraphics[width=\linewidth]{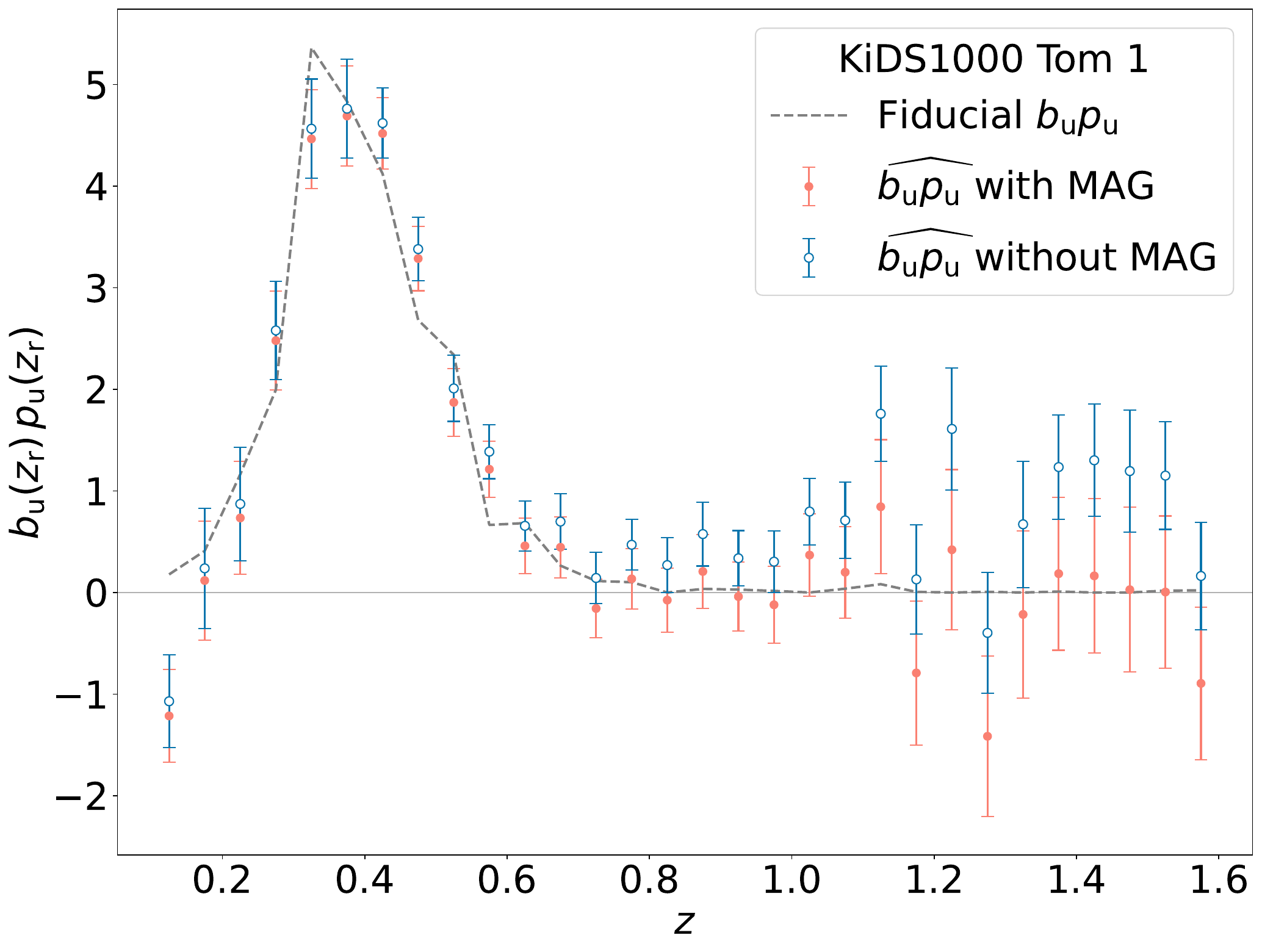}
 \end{subfigure}\\[0.6em]
\begin{subfigure}[t]{0.48\textwidth}\centering
\includegraphics[width=\linewidth]{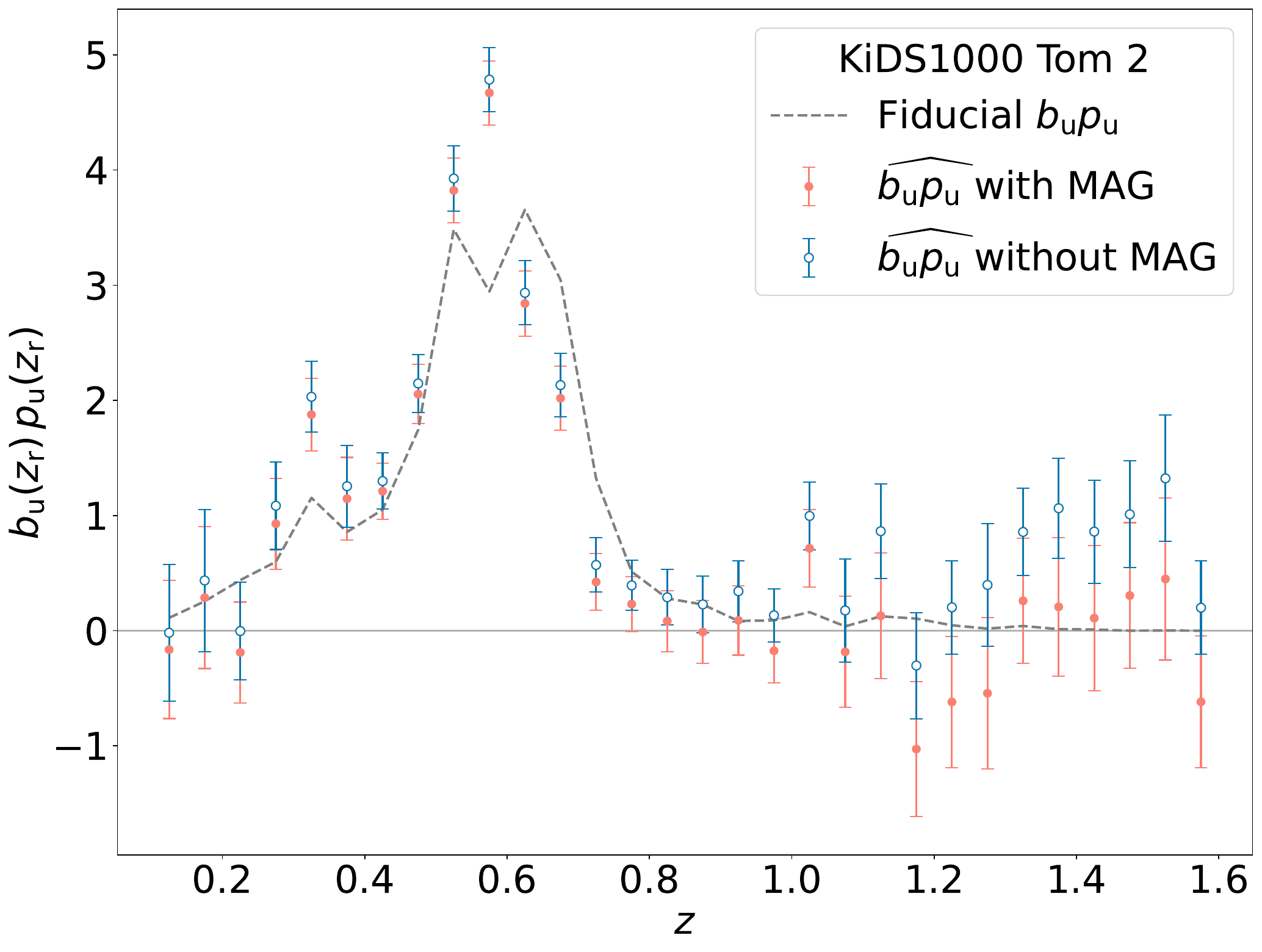}
 \end{subfigure}\hfill
\begin{subfigure}[t]{0.48\textwidth}\centering
\includegraphics[width=\linewidth]{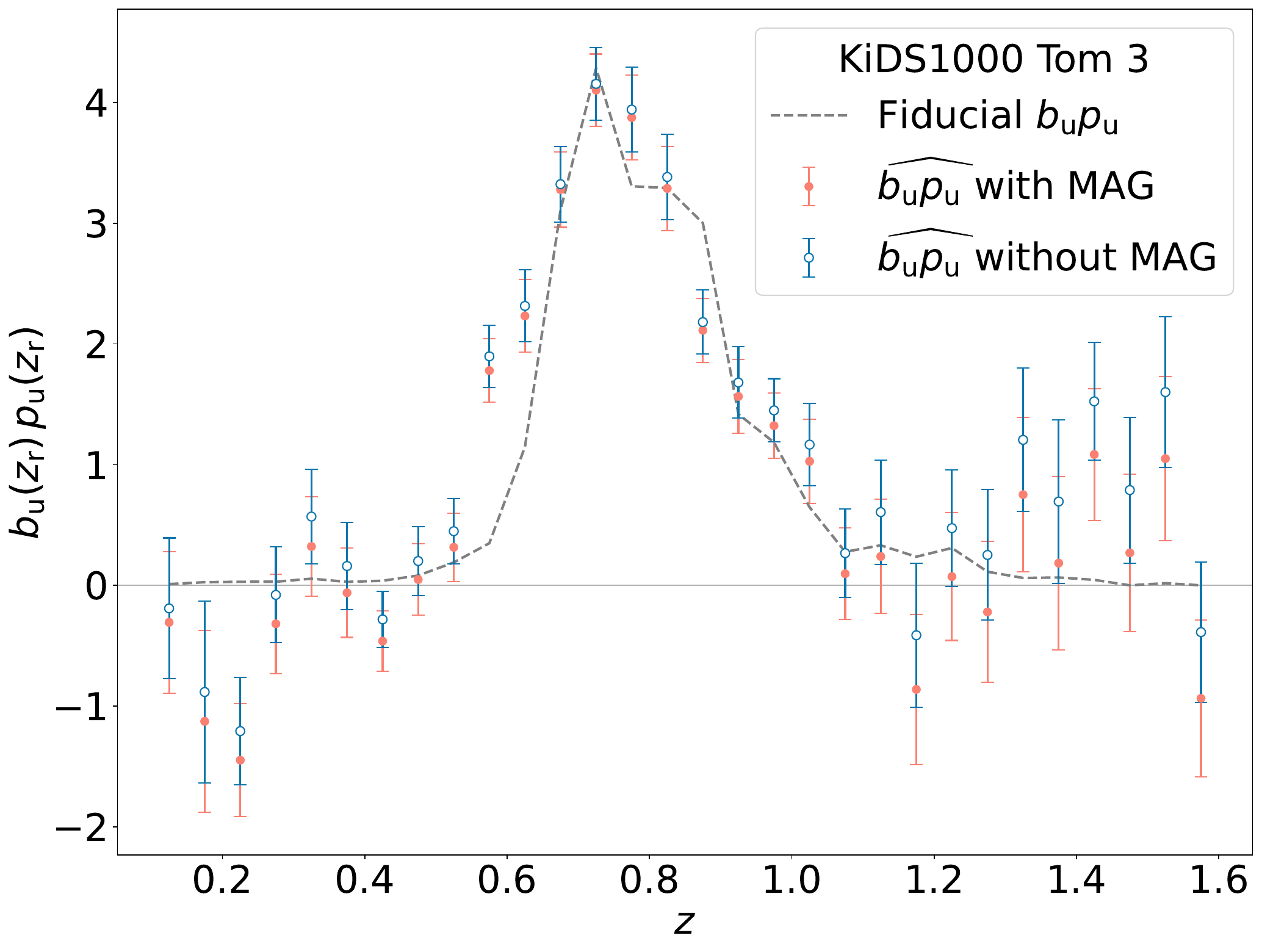}
 \end{subfigure}
\begin{subfigure}[t]{0.48\textwidth}\centering
\includegraphics[width=\linewidth]{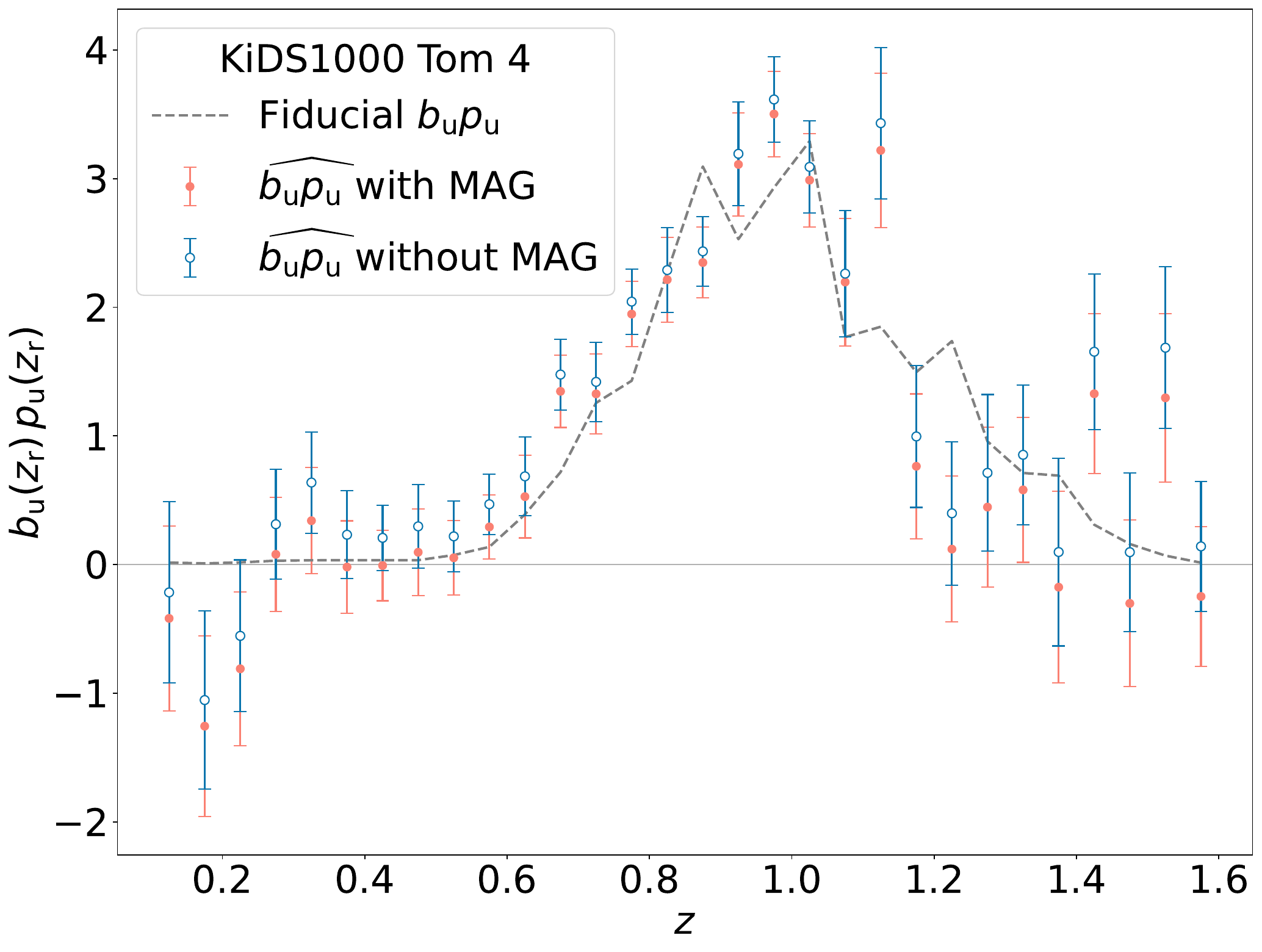}
 \end{subfigure}

\caption{Posterior estimates of the bias-weighted source redshift distribution $b_\mathrm{u}(z_\mathrm{r})\,p_\mathrm{u}(z_\mathrm{r})$ in each KiDS-1000 tomographic bin, with 1$\sigma$ uncertainties. The black dashed curve shows the survey's fiducial redshift distribution, and the red and blue points corresponding to analyses with and without magnification, respectively.}
\label{fig:post_kids}
\end{figure*}

\begin{figure*}
\centering
\begin{subfigure}[t]{0.48\textwidth}\centering
\includegraphics[width=\linewidth]{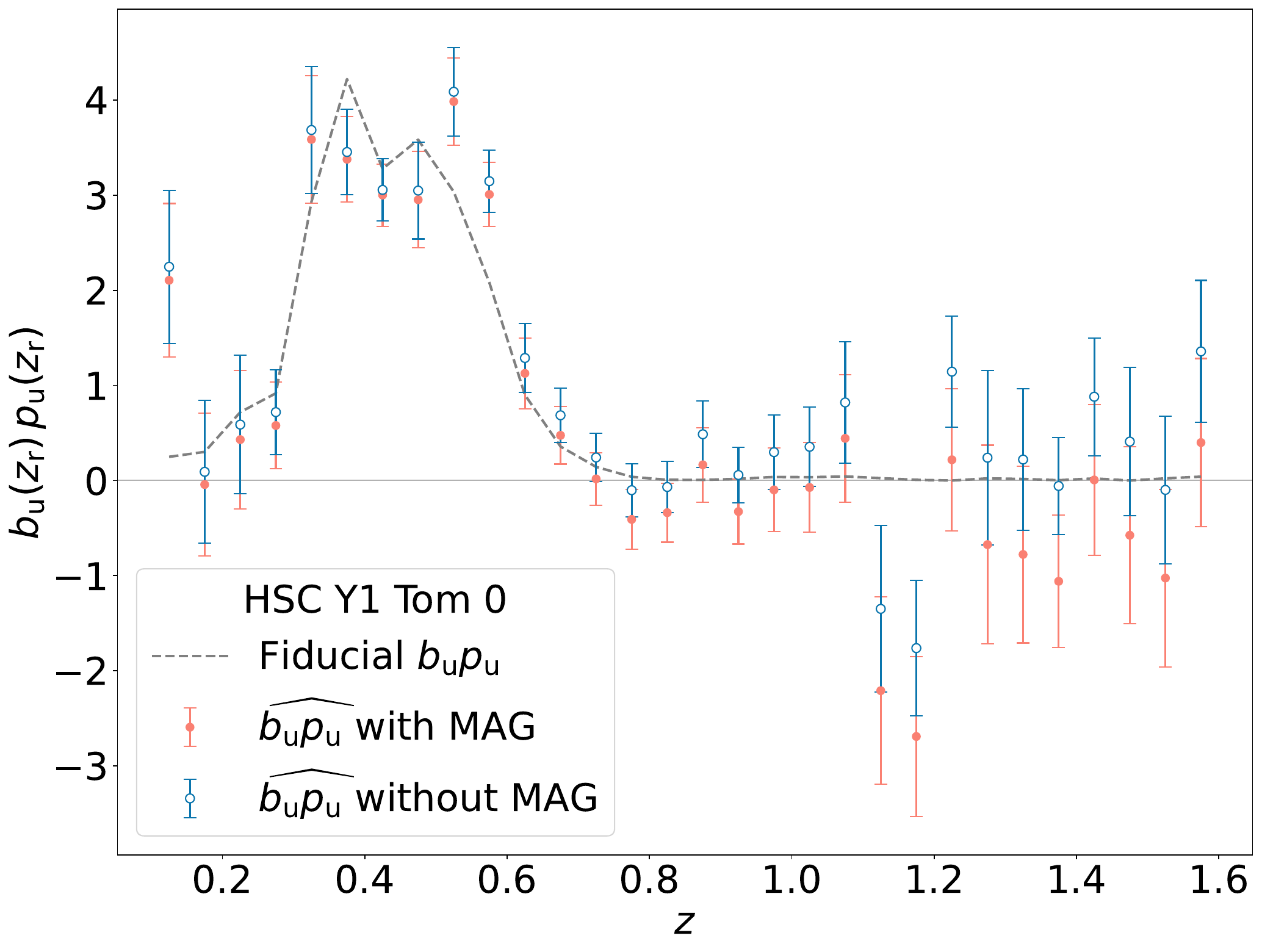}
 \end{subfigure}\hfill
\begin{subfigure}[t]{0.48\textwidth}\centering
\includegraphics[width=\linewidth]{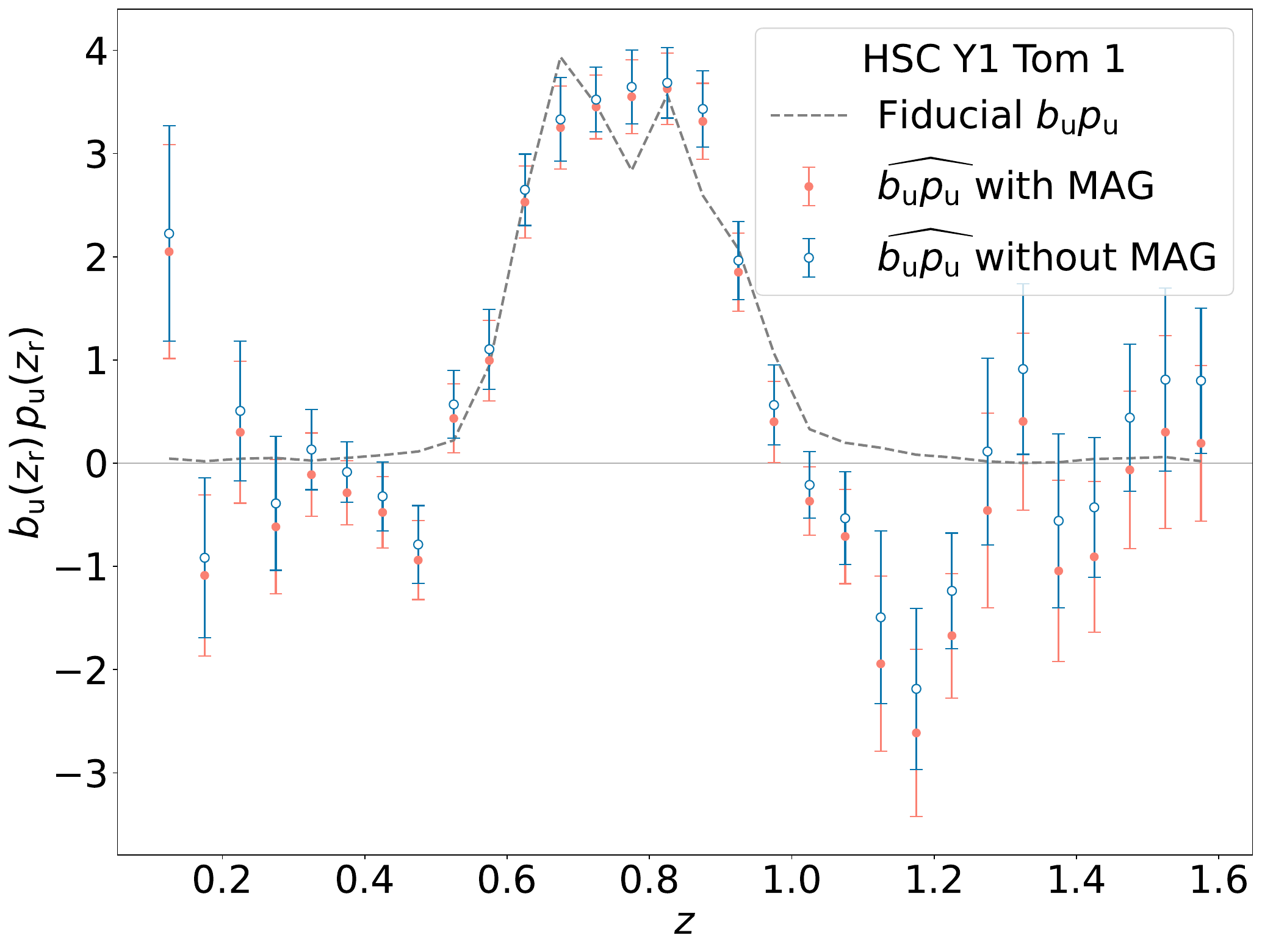}
 \end{subfigure}\\[0.6em]
\begin{subfigure}[t]{0.48\textwidth}\centering
\includegraphics[width=\linewidth]{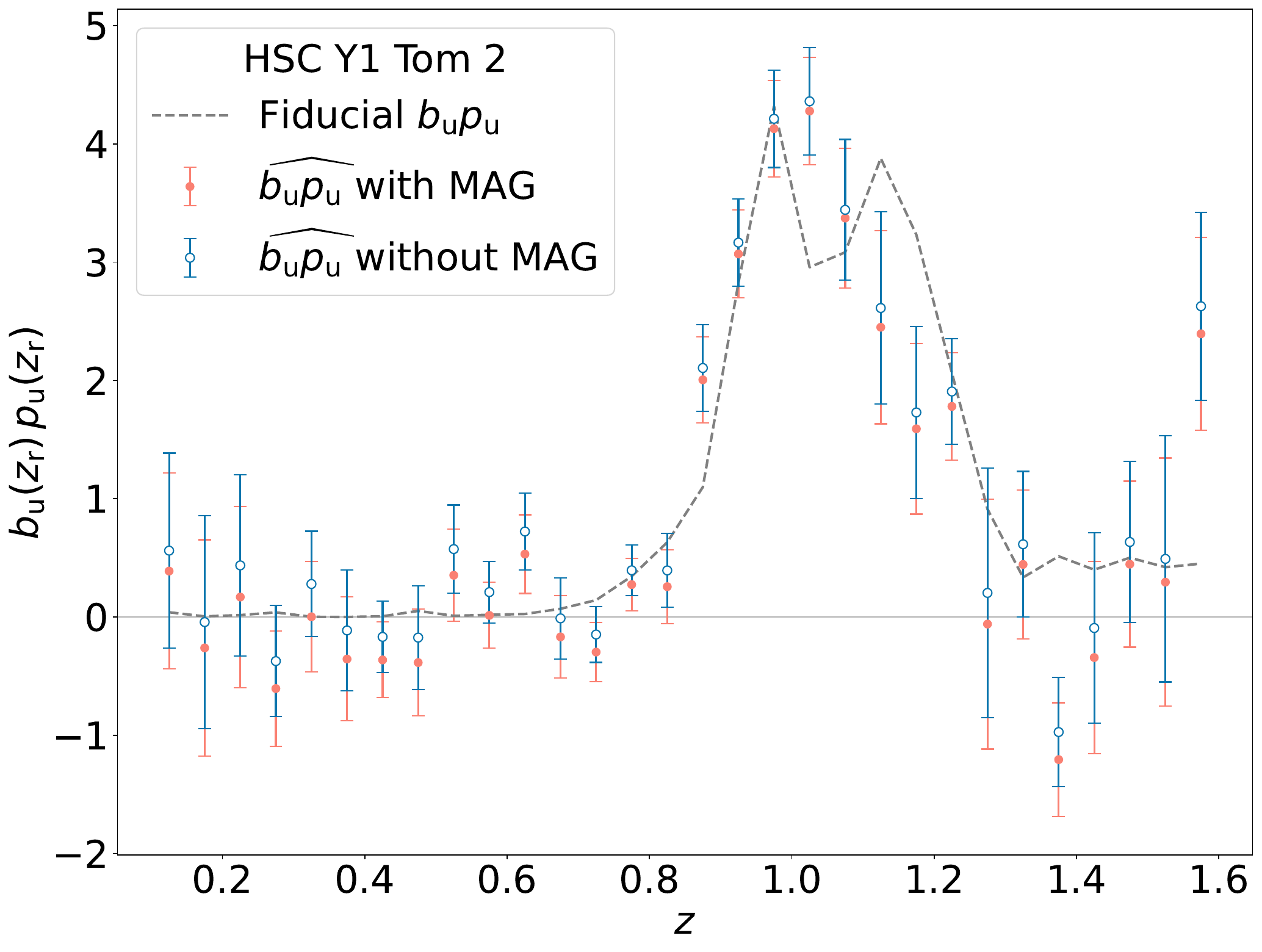}
 \end{subfigure}\hfill
\begin{subfigure}[t]{0.48\textwidth}\centering
\includegraphics[width=\linewidth]{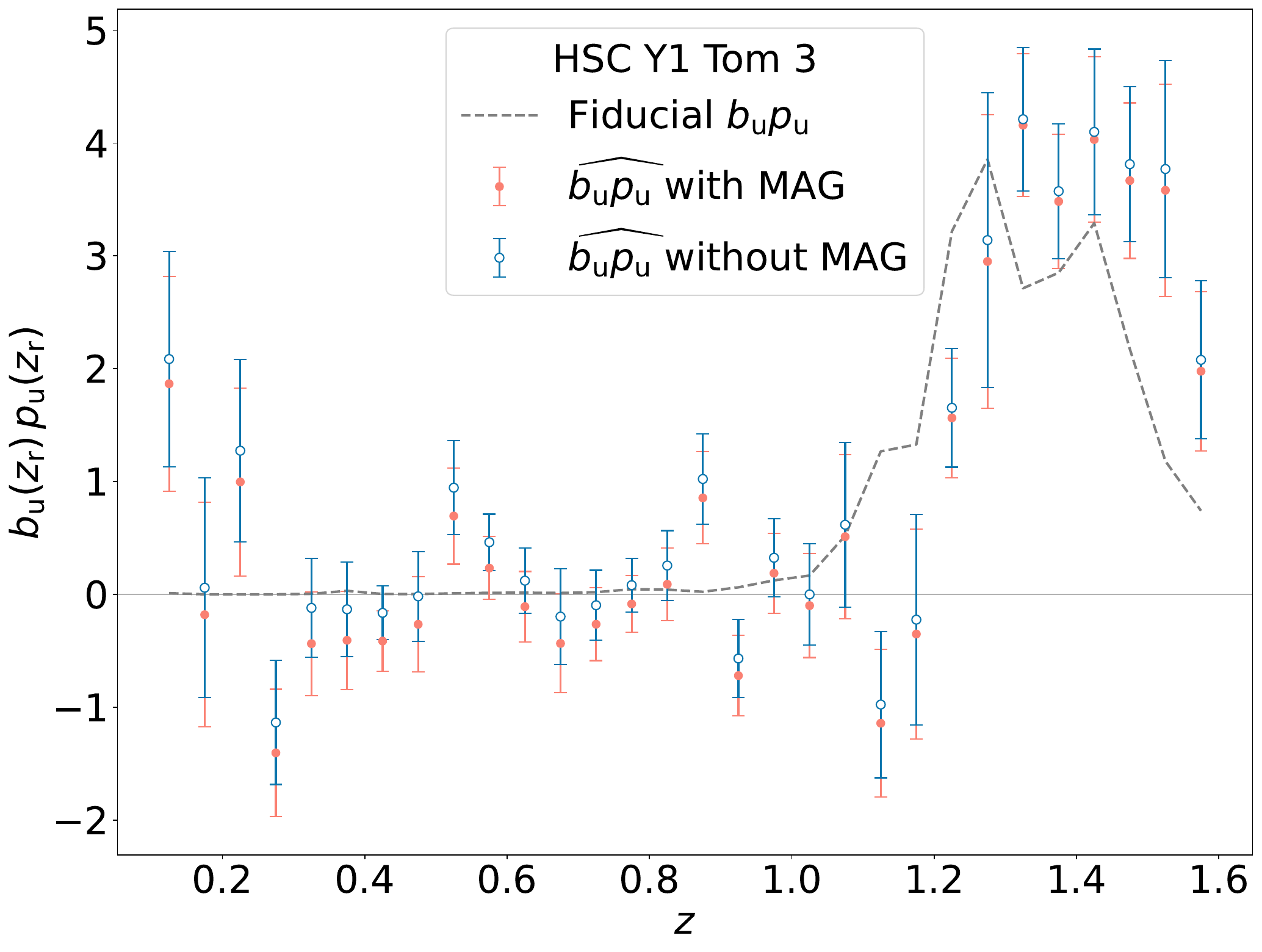}
 \end{subfigure}
\caption{Posterior estimates of the bias-weighted source redshift distribution $b_\mathrm{u}(z_\mathrm{r})\,p_\mathrm{u}(z_\mathrm{r})$ in each HSC-Y1 tomographic bin, with 1$\sigma$ uncertainties.  The black dashed curve shows the survey's fiducial redshift distribution, and the red and blue points corresponding to analyses with and without magnification, respectively.}
\label{fig:post_hscy1}
\end{figure*}

\begin{figure*}
\centering
\begin{subfigure}[t]{0.48\textwidth}\centering
\includegraphics[width=\linewidth]{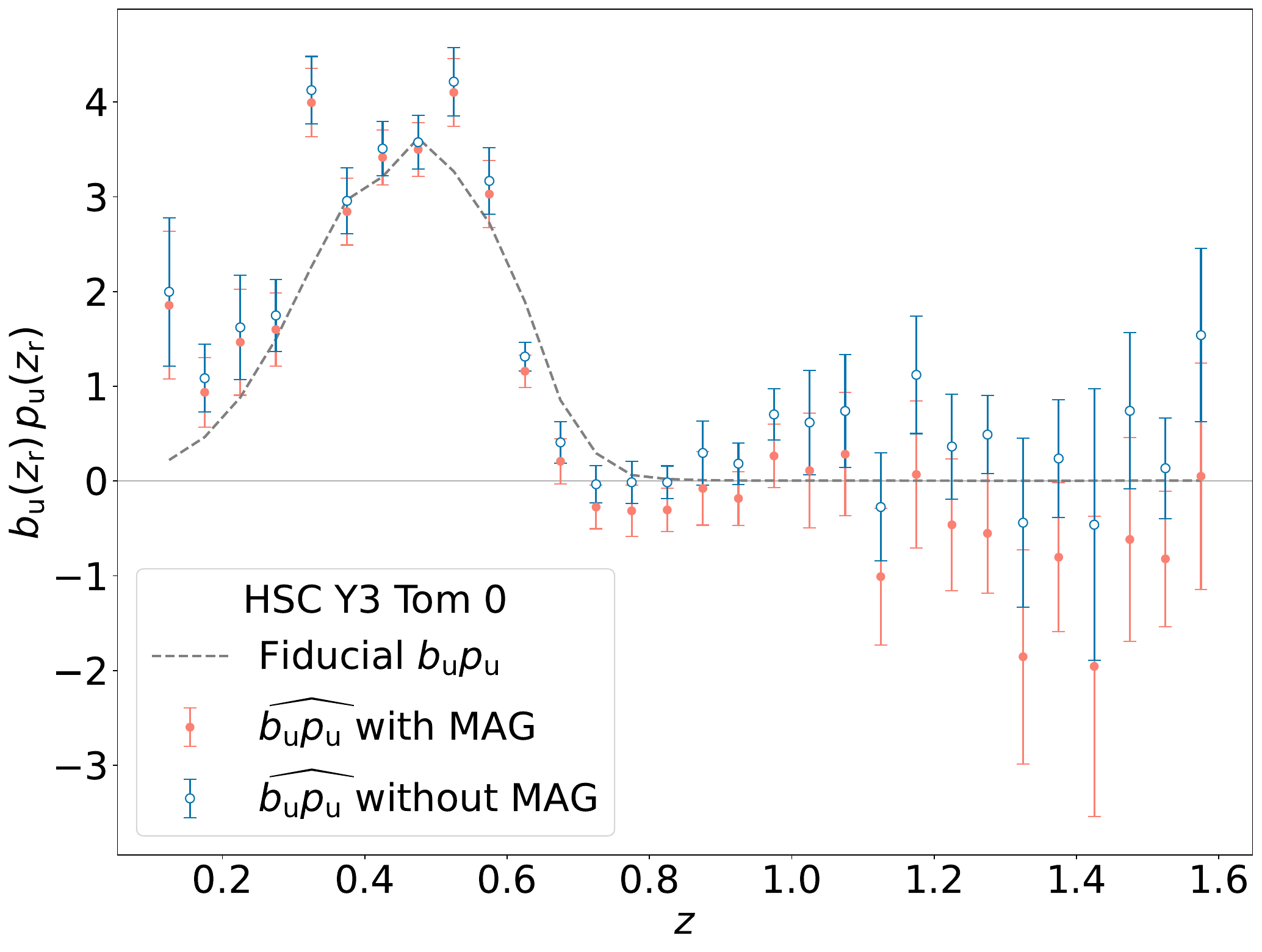}
 \end{subfigure}\hfill
\begin{subfigure}[t]{0.48\textwidth}\centering
\includegraphics[width=\linewidth]{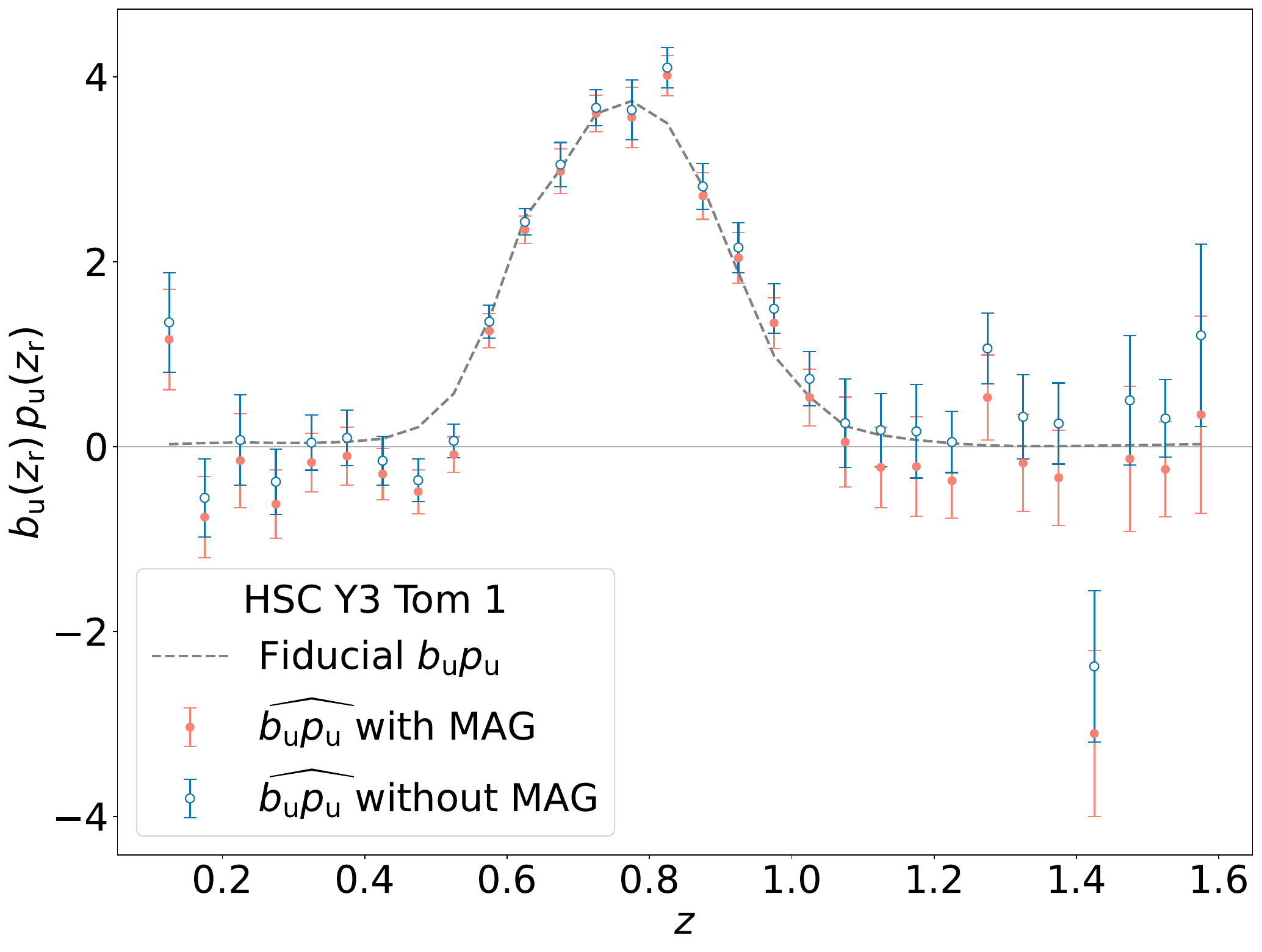}
 \end{subfigure}\\[0.6em]
\begin{subfigure}[t]{0.48\textwidth}\centering
\includegraphics[width=\linewidth]{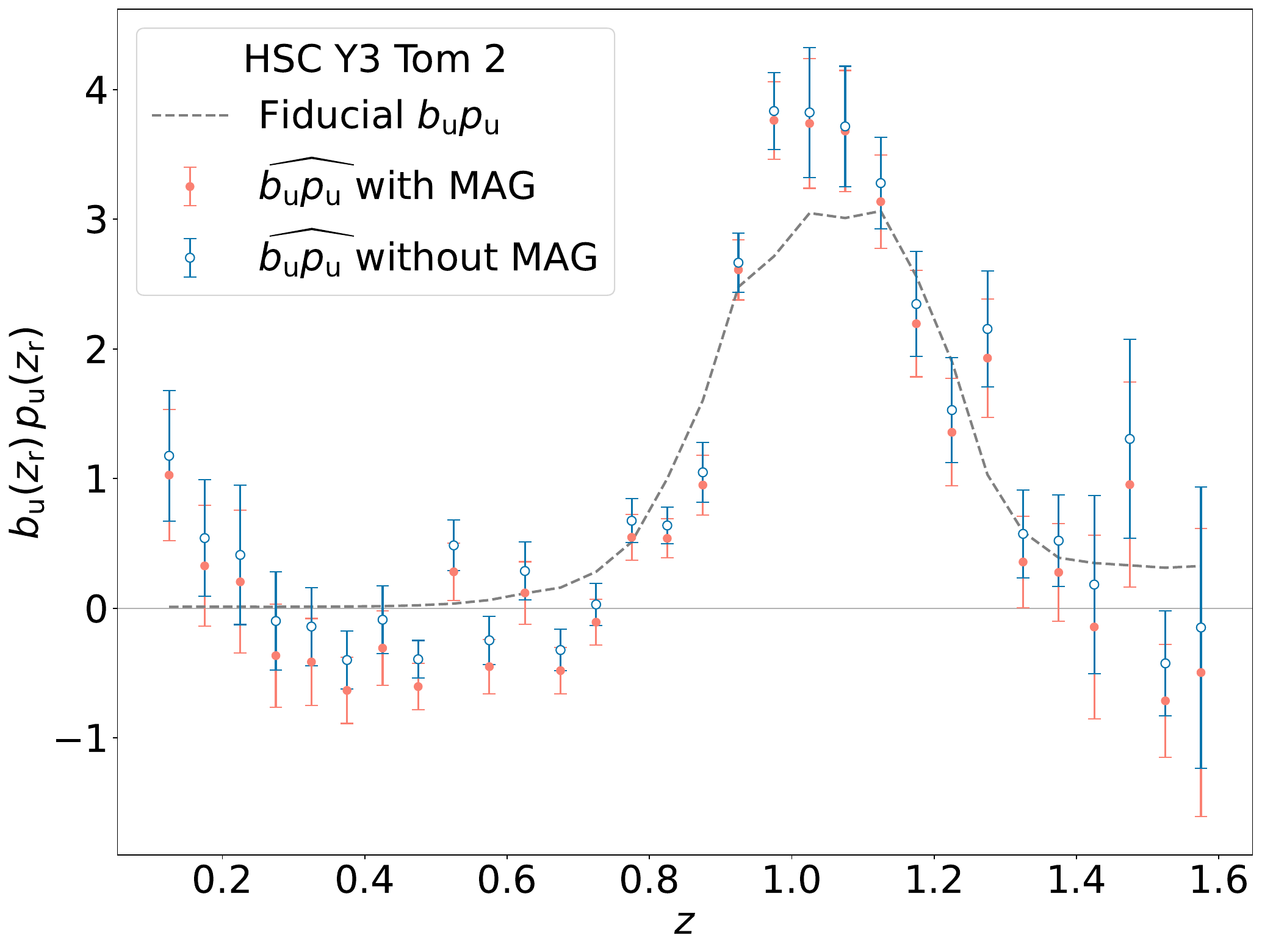}
 \end{subfigure}\hfill
\begin{subfigure}[t]{0.48\textwidth}\centering
\includegraphics[width=\linewidth]{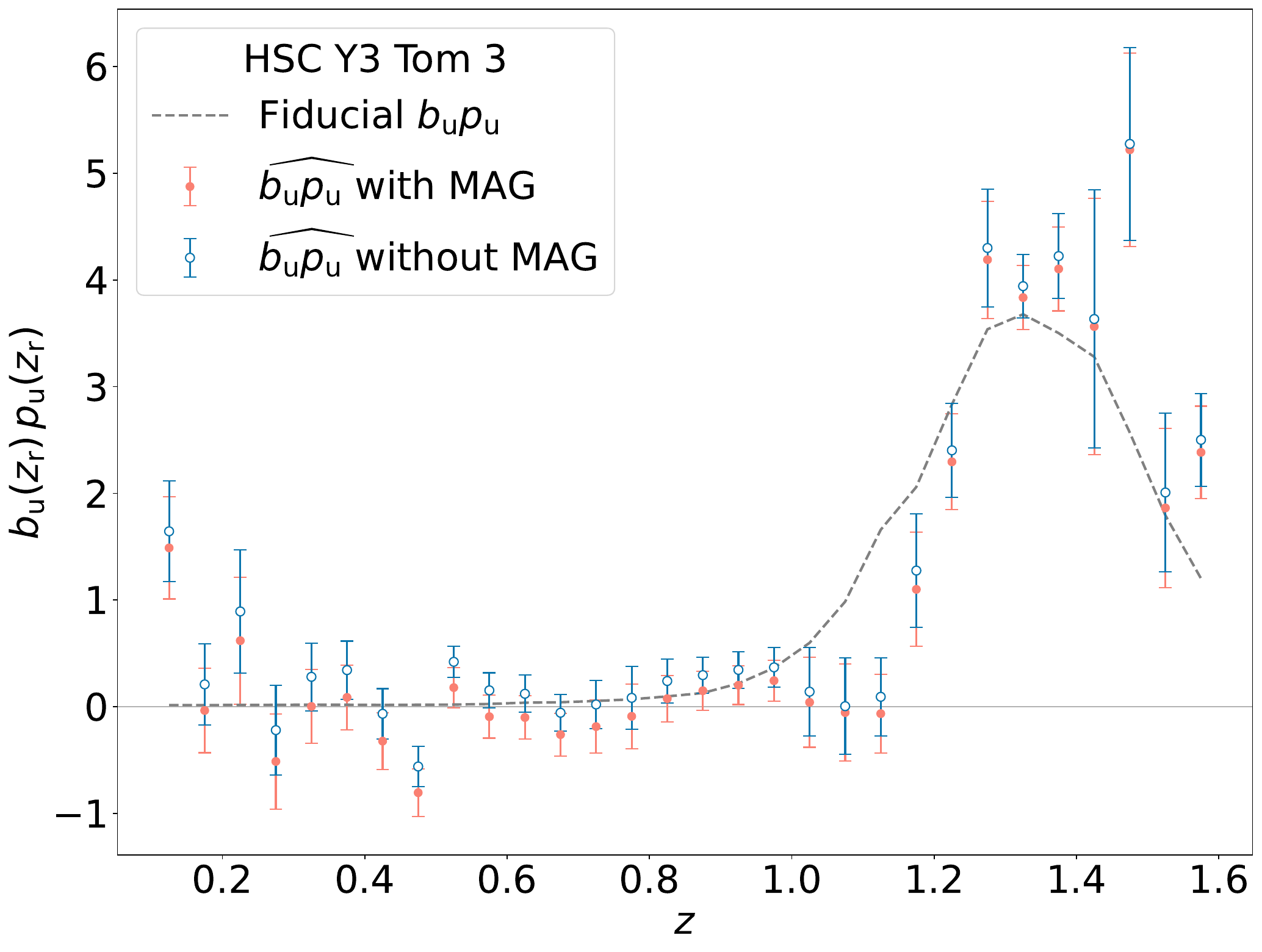}
 \end{subfigure}
\caption{Posterior estimates of the bias-weighted source redshift distribution $b_\mathrm{u}(z_\mathrm{r})\,p_\mathrm{u}(z_\mathrm{r})$ in each HSC-Y3 tomographic bin, with 1$\sigma$ uncertainties.  The black dashed curve shows the survey's fiducial redshift distribution, and the red and blue points corresponding to analyses with and without magnification, respectively.}
\label{fig:post_hscy3}
\end{figure*}

Having validated our analysis framework on the mocks, we now proceed to fit the DESI-DR1 and weak lensing survey correlations.  Figs.~\ref{fig:post_desy3}, \ref{fig:post_kids}, \ref{fig:post_hscy1} and \ref{fig:post_hscy3} show the posterior constraints on the product $b_\mathrm{u}(z_\mathrm{r})\,p_\mathrm{u}(z_\mathrm{r})$, obtained from clustering-$z$ analyses in tomographic bins of the source samples from DES-Y3, KiDS-1000, HSC-Y1 and HSC-Y3, respectively.

\rossi{For comparison, we use the fiducial source redshift distributions previously calibrated by the respective survey analyses, as described in Sec.~\ref{sec:datasets} \citep{2021MNRAS.505.4249M,2021A&A...647A.124H,2019PASJ...71...43H,2023MNRAS.524.5109R}. These include photometric-redshift calibration methods based on SOM and related direct-calibration techniques. The ``fiducial redshift distribution'' used in Figs.~\ref{fig:post_desy3}, \ref{fig:post_kids}, \ref{fig:post_hscy1} and \ref{fig:post_hscy3} refers to these externally calibrated survey redshift distributions. To compare them with the measured bias-weighted clustering-$z$ estimates, we construct
\begin{equation}
M(z_\mathrm{r}) = A \, b_\mathrm{u}^{\rm mock}(z_\mathrm{r}) \, p_\mathrm{u}^{\rm fid}(z_\mathrm{r}),
\end{equation}
where $b_\mathrm{u}^{\rm mock}(z)$ is the source-bias evolution calibrated from the mocks and $A$ is a free overall amplitude fitted independently in each tomographic bin. The black dashed curve shows the corresponding best-fitting bias-weighted fiducial template.}
The coloured points with error bars in Figs.~\ref{fig:post_desy3} to \ref{fig:post_hscy3} show the estimator for $b_\mathrm{u}(z_\mathrm{r})\,p_\mathrm{u}(z_\mathrm{r})$ from Eq.~\ref{eq:bupu}, with $1\sigma$ uncertainties propagated from the posteriors of $A_{rr}$ and $A_{ur}$ (as described in Sec.~\ref{sec:fits}).

Each panel displays two clustering-$z$ solutions: the red points correspond to the full model including magnification terms, while the blue points are obtained from a simplified model that neglects magnification. The former constitutes our baseline analysis, in which magnification-induced correlations are explicitly modelled and marginalised over via the nuisance parameters $p$ and $q$ (Sec.~\ref{sec:magnification}).
\rossi{Some individual $b_\mathrm{u} \, p_\mathrm{u}$ estimates fluctuate below zero, particularly in the lower signal-to-noise ELG regime. We interpret these as statistical fluctuations of the cross-correlation estimator, and retain them to avoid introducing a positivity bias.}



Comparing the cases with and without the magnification correction allows us to assess its amplitude and redshift dependence.  As observed in the mock tests, the model including magnification traces the expectation of the fiducial photo-$z$ calibration more closely and yields slightly higher amplitudes at $z_\mathrm{r}\!\gtrsim\!1$, confirming that magnification effects become increasingly important toward the high-redshift tail of the DESI reference samples.

In redshift ranges where both LRG and ELG tracers are available, our baseline results use LRGs up to $z_\mathrm{r}\simeq1.1$ and ELGs from $z_\mathrm{r}=1.1$ up to the maximum redshift of the reference sample ($z_\mathrm{r}=1.6$). We verified that, in the redshift overlap region, the cross-correlations derived using ELGs yield results consistent with those from LRGs (see Fig.~\ref{fig:lrg_elg_desy3}), as discussed at the end of this section. 

We summarise the performance of the redshift–distribution recovery across the three surveys in Table~\ref{tab:mean_z_stats}, where we report the mean redshift of the inferred distribution, $\bar z_{\rm meas}$, its propagated uncertainty, and the corresponding mean of the fiducial template, $\bar z_{\rm fid}$.  We also list the shifts $\sigma_{\Delta z} \equiv (\bar z_{\rm meas} - \bar z_{\rm fid})/\sigma(\bar z_{\rm meas})$, which provide a compact measure of the agreement with the template.  
To prevent noisy high–redshift tails and small negative fluctuations in $b_\mathrm{u} \, p_\mathrm{u}$ from biasing the first moment, we compute the means within a restricted redshift window $z \in [z_{\min}, z_{\max}]$ defined from the fiducial template: $z_{\min}$ is fixed to the analysis lower bound, while $z_{\max} = \min(z_{\rm max,grid}, z_{\rm peak} + 0.6)$, where $z_{\rm peak}$ is the peak of the fiducial $b_\mathrm{u}(z_\mathrm{r})p_\mathrm{u}(z_\mathrm{r})$ model.  We find that the resulting distribution of $\sigma_{\Delta z}$ values is consistent with statistical errors.  The detailed behaviour for each survey is discussed below.

\paragraph{DES-Y3}

\begin{table*}
    \centering
    \begin{tabular}{lccc}
        \hline\hline
        Survey, tomographic bin &
        $\bar z_{\rm meas} \pm \sigma(\bar z_{\rm meas})$ &
        $\bar z_{\rm fid}$ &
        $\sigma_{\Delta z}$
        \\ 
        \hline
        DES-Y3, tom 0      & $0.281 \pm 0.024$ & 0.314 & $-1.39$ \\
        DES-Y3, tom 1      & $0.465 \pm 0.037$ & 0.496 & $-0.86$ \\
        DES-Y3, tom 2      & $0.774 \pm 0.093$ & 0.760 & $0.15$ \\
        DES-Y3, tom 3      & $1.076 \pm 0.146$ & 0.959 & $0.80$ \\
        KiDS-1000, tom 0   & $0.221 \pm 0.036$ & 0.253 & $-0.89$ \\
        KiDS-1000, tom 1   & $0.424 \pm 0.034$ & 0.397 & $0.79$ \\
        KiDS-1000, tom 2   & $0.481 \pm 0.075$ & 0.566 & $-1.13$ \\
        KiDS-1000, tom 3   & $0.871 \pm 0.085$ & 0.803 & $0.80$ \\
        KiDS-1000, tom 4   & $1.053 \pm 0.108$ & 0.992 & $0.56$ \\
        HSC-Y1, tom 0      & $0.415 \pm 0.038$ & 0.434 & $-0.49$ \\
        HSC-Y1, tom 1      & $0.557 \pm 0.117$ & 0.768 & $-1.81$ \\
        HSC-Y1, tom 2      & $1.099 \pm 0.148$ & 1.085 & $0.10$ \\
        HSC-Y1, tom 3      & $1.401 \pm 0.151$ & 1.319 & $0.54$ \\
        HSC-Y3, tom 0      & $0.420 \pm 0.044$ & 0.453 & $-0.76$ \\
        HSC-Y3, tom 1      & $0.772 \pm 0.089$ & 0.765 & $0.08$ \\
        HSC-Y3, tom 2      & $1.072 \pm 0.121$ & 1.069 & $0.02$ \\
        HSC-Y3, tom 3      & $1.332 \pm 0.121$ & 1.305 & $0.23$ \\
        \hline
    \end{tabular}
       \caption{The mean redshift of the measured $b_\mathrm{u}(z_\mathrm{r})p_\mathrm{u}(z_\mathrm{r})$ distributions and fiducial model, for each survey and tomographic bin. The measured mean redshift $\bar z_{\rm meas}$ and its uncertainty $\sigma(\bar z_{\rm meas})$ are obtained from the discrete estimator $\bar z_{\rm meas} = \sum_i z_i P_i$ with $\sigma^2(\bar z_{\rm meas}) = \sum_i z_i^2 \sigma^2(P_i)$, where $P_i$ are the normalized $b_\mathrm{u} \, p_\mathrm{u}$ measurements in each redshift bin with $\sum_i P_i = 1$.  The fiducial mean redshift $\bar z_{\rm fid}$ is computed from the corresponding normalized model template. We also report the normalized shift $\sigma_{\Delta z} \equiv (\bar z_{\rm meas} - \bar z_{\rm fid}) / \sigma(\bar z_{\rm meas})$.}
    \label{tab:mean_z_stats}
\end{table*}

The recovered redshift probability distributions are shown in Fig.~\ref{fig:post_desy3}.  Across the four tomographic bins, the inferred $b_\mathrm{u} \, p_\mathrm{u}$ closely tracks the model over $z_\mathrm{r}\!\in\!(0.1,1.6)$.  The mean-redshift comparison in Table~\ref{tab:mean_z_stats} shows excellent consistency.  The lowest tomographic bin shows the tightest constraints, reflecting the larger number of pairs and reduced magnification contamination.  Moving to the higher tomographic bins we observe (i) a mild suppression at the high-$z$ tail ($z >1.2$) and (ii) \rossi{increasing uncertainties}, consistent with the reduced tracer density and stronger sensitivity to magnification.

\paragraph{KiDS-1000}

For KiDS-1000 (Fig.~\ref{fig:post_kids}), we report results in five tomographic bins. As shown in Table~\ref{tab:mean_z_stats}, all $\sigma_{\Delta z}$ values are within $\sim 1\sigma$ agreement of the fiducial model.  The first three bins show high S/N detections with good consistency with the template; the highest two tomographic bins are noisier and exhibit a mild low-$z$ deficit / high-$z$ excess, which remains compatible within the $1-2\sigma$ level when accounting for magnification and calibration uncertainties.
 
\paragraph{HSC-Y1}

The HSC-Y1 posteriors (Fig.~\ref{fig:post_hscy1}) show larger uncertainties, driven by the smaller survey overlap area with DESI.  The shape match to the model is good  and consistent with this, Table~\ref{tab:mean_z_stats} reveals that the shifts $\sigma_{\Delta z}$ are within statistical error ranges.


\paragraph{HSC-Y3}

With increased area and depth, HSC-Y3 (Fig.~\ref{fig:post_hscy3}) yields noticeably smaller error bars than HSC-Y1 for a given tomographic bin. The agreement with the template is generally good.
The mean-redshift shifts reported in Table~\ref{tab:mean_z_stats} show that all bins are within $|\sigma_{\Delta z}| < 1$.  
\rossi{For tomographic bin 3 specifically, we find $\bar z_{\rm meas}=1.332\pm0.121$, compared with $\bar z_{\rm fid}=1.305$, corresponding to a shift of only $0.23\sigma$.}
These levels are consistent with the HSC-Y3 photo-$z$ calibration reported in the direct redshift calibration and clustering-$z$ analyses \citep{2023MNRAS.524.5109R}.

\subsubsection{Analysis variations}

\paragraph{Normalization and internal consistency}
By construction, the model curve and the point estimates are both normalized over the set of valid $z_\mathrm{r}$ bins used in the fit. The posterior errors from our MCMC sampler are of the same order of magnitude as the jackknife scatter, indicating that covariance estimates and likelihood approximations are adequate for our binning scheme. We verified that replacing MAP values with posterior means for $(A_{rr}, A_{ur})$ changes $b_\mathrm{u} \, p_\mathrm{u}$ by at most $1\%$ in all bins.


\paragraph{Fitting-range systematics.}
We tested the robustness of our measured $b_\mathrm{u} \, p_\mathrm{u}$ values against the adopted angular scale cuts, varying the fitting range $R_{\min},R_{\max}$. \rossi{As a conservative quality cut, we exclude three BGS reference bins centred at $z\simeq0.23$, $0.28$, and $0.33$, which show unusually large auto-correlation $\chi^2/{\rm dof}$. The fitted $A_{rr}(z)$ amplitudes nevertheless remain smooth across these bins, indicating that this does not correspond to an anomalous feature in the recovered clustering amplitude.} In particular, we explored fitting to larger comoving separation ranges, such as $R>5\,h^{-1}\mathrm{Mpc}$, $1.5< R<10\,h^{-1}\mathrm{Mpc}$, and $3< R<20\,h^{-1}\mathrm{Mpc}$, finding consistent results within the uncertainties. The resulting shifts are negligible over the redshift range retained for cosmological inference.

\paragraph{LRG versus ELG tracer consistency}
\label{sec:lrg_elg_comparison}

\begin{figure*}[t]
\centering
\begin{subfigure}[t]{0.48\textwidth}\centering
\includegraphics[width=\linewidth]{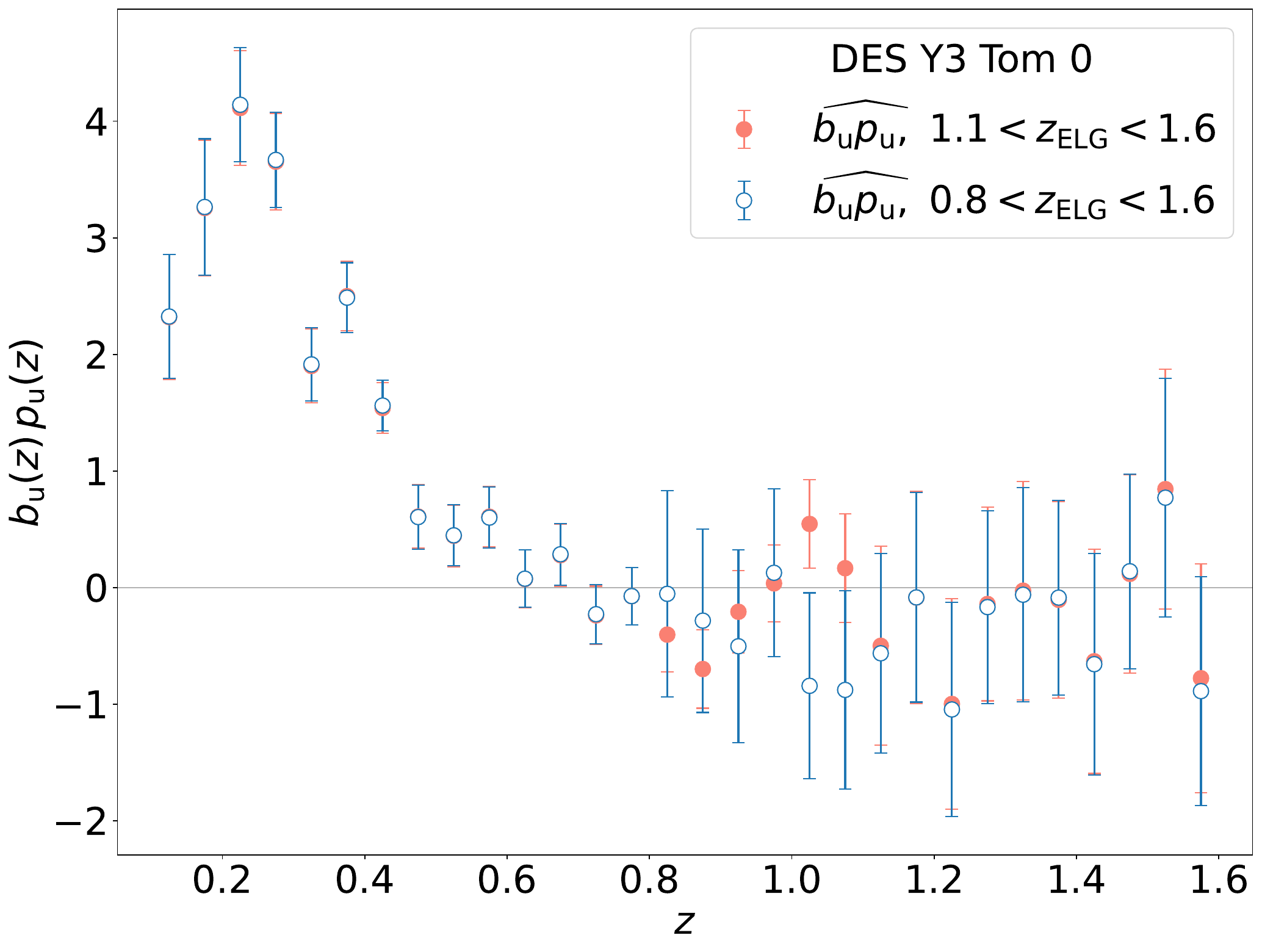}
 \end{subfigure}\hfill
\begin{subfigure}[t]{0.48\textwidth}\centering
\includegraphics[width=\linewidth]{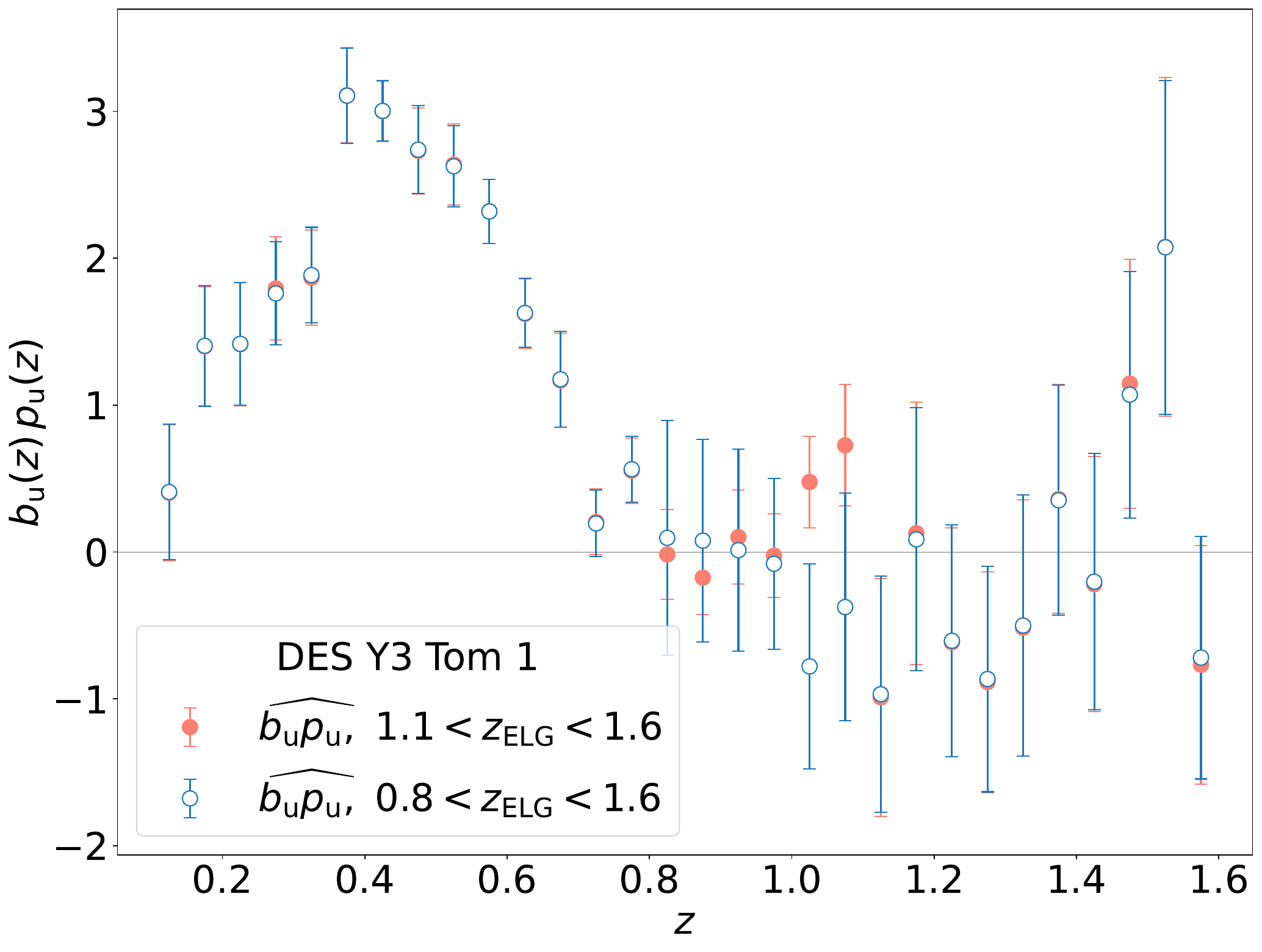}
 \end{subfigure}\\[0.6em]
\begin{subfigure}[t]{0.48\textwidth}\centering
\includegraphics[width=\linewidth]{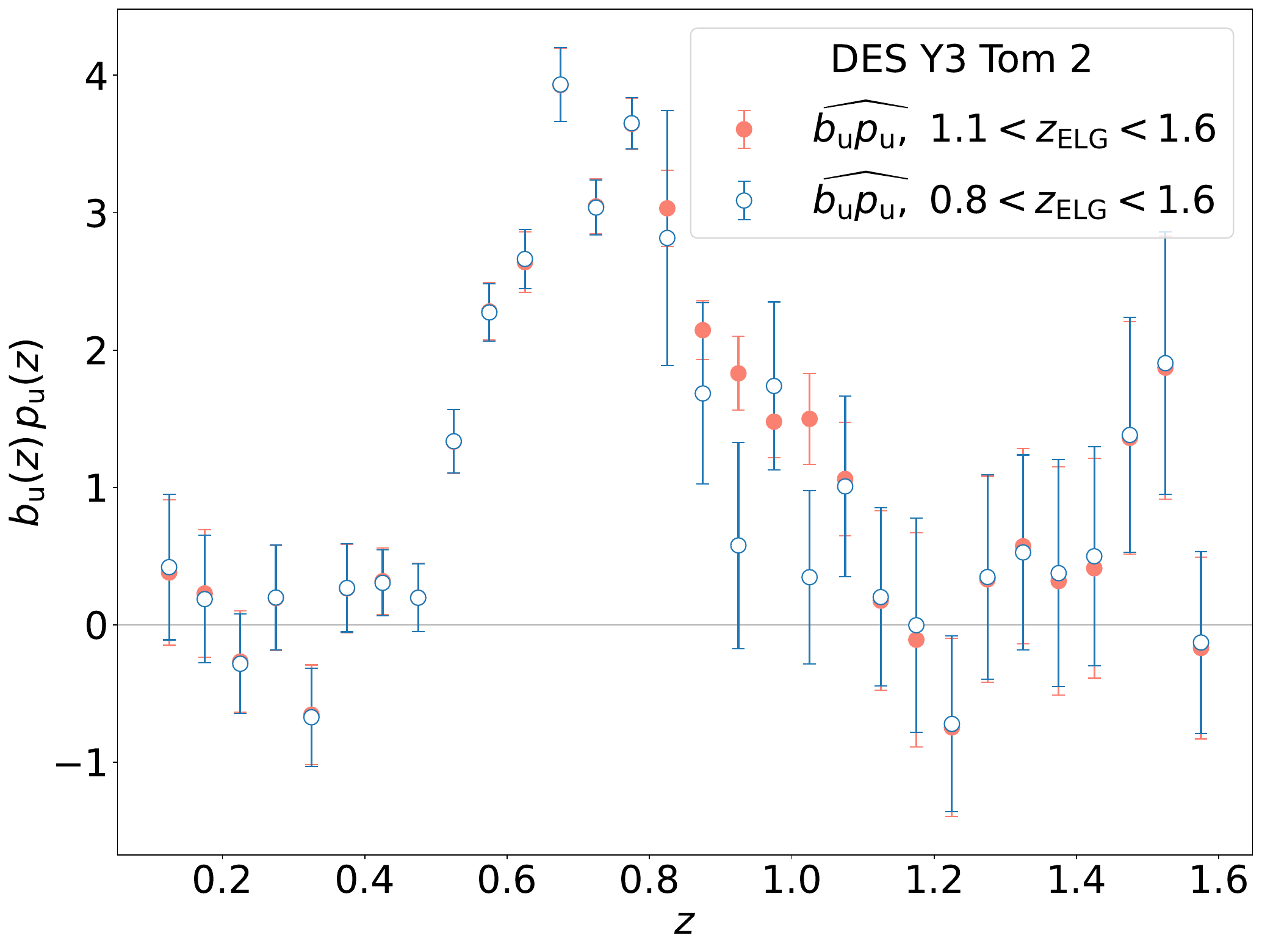}
 \end{subfigure}\hfill
\begin{subfigure}[t]{0.48\textwidth}\centering
\includegraphics[width=\linewidth]{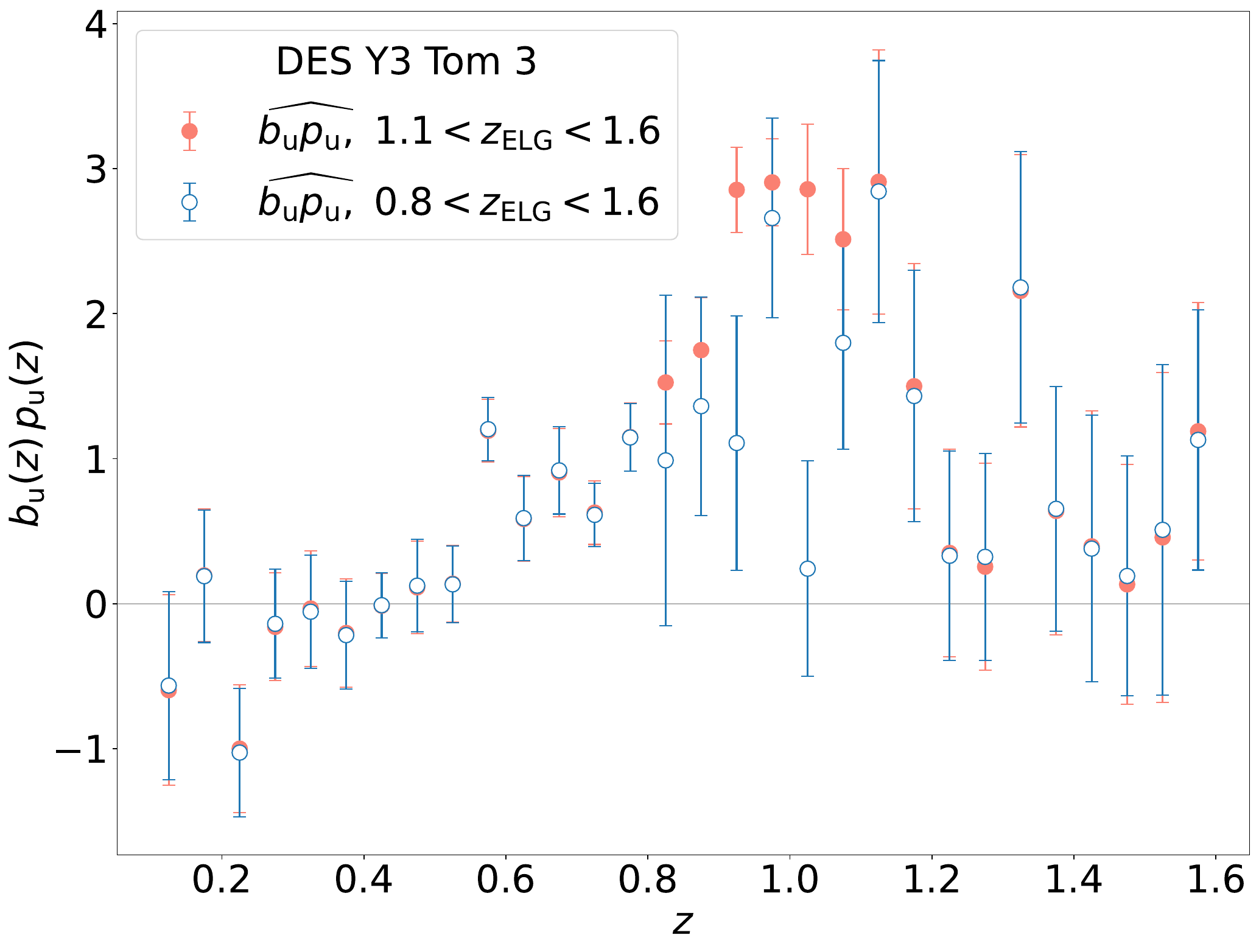}
 \end{subfigure}
\caption{\rossi{Comparison of the recovered $b_\mathrm{u}(z)\,p_\mathrm{u}(z)$ distributions in each DES-Y3 tomographic bin for two choices of the transition redshift between the DESI LRG and ELG reference samples. The filled red points show the fiducial configuration with transition redshift $z_{\rm trans}=1.1$, while the open blue points show the alternative configuration with $z_{\rm trans}=0.8$.}}
\label{fig:lrg_elg_desy3}
\end{figure*}

We tested the internal consistency of the clustering–$z$ results obtained using different spectroscopic tracers in overlapping redshift ranges.  
Specifically, we compared the inferred distributions of $b_\mathrm{u} \, p_\mathrm{u}$ when adopting the default tracer selection:
\[
\text{tracer} = 
\begin{cases}
  \text{BGS}, & z < 0.4,\\[0.3em]
  \text{LRG}, & 0.4 < z < 1.1,\\[0.3em]
  \text{ELG}, & z > 1.1,
\end{cases}
\]
and an alternative configuration in which the LRG/ELG transition occurs at $z = 0.8$ (we note that it is also possible to construct a joint LRG/ELG sample, following \citet{2026JCAP...04..058V}).  This test probes whether the recovered $b_\mathrm{u} \, p_\mathrm{u}$ is robust to the specific tracer choice near the LRG–ELG boundary. 
\rossi{Fig.~\ref{fig:lrg_elg_desy3} compares the clustering-$z$ posteriors obtained for DES-Y3 using two choices of the transition redshift between the LRG and ELG reference samples (noting that the other lensing surveys produced consistent results). The red points show our fiducial configuration, in which LRGs are used up to $z=1.1$ and ELGs at higher redshift, while the blue points show the alternative configuration with the transition moved to $z=0.8$. The inferred $b_\mathrm{u} \, p_\mathrm{u}$ posteriors are stable across all tomographic bins, with the two configurations agreeing within the statistical uncertainties. This indicates that the precise choice of transition redshift between the LRG and ELG reference samples has negligible impact on the recovered clustering-$z$ signal.}


\paragraph{Marginalizing over source bias evolution}
\label{sec:marg-bias}

Since the source galaxy bias $b_\mathrm{u}(z)$ is not independently determined from the cross-correlation measurement, recovering the underlying redshift distribution $p_\mathrm{u}(z)$ requires marginalising over the possible redshift evolution of $b_\mathrm{u}(z)$.  We do this using the mock-inferred source bias evolution shown in Fig.\ref{fig:sourcebias}, noting that the mock has been populated in a manner matching the magnitudes and colours of the photometric samples in each lensing survey.  We propagate the bias evolution and errors by adopting a smooth linear parameterization,
\begin{equation}
b_\mathrm{u}(z)=b_0 + b_1 z ,
\end{equation}
which captures the expected monotonic increase of galaxy bias with redshift, while preventing unphysical bin-to-bin fluctuations in low signal-to-noise regimes.  This form does not impose any specific normalization on $b_\mathrm{u}(z)$; only the relative trend with redshift is constrained.  To regularise the allowed range of $b_0,b_1$ we use Gaussian priors on these parameters set by fitting them to the measurements shown in Fig.~\ref{fig:sourcebias}, where we inflate the width of these priors by a factor of 2 to be conservative. Repeating this process over all reference slices and posterior draws, and normalizing so that $\sum_\mathrm{r} p_\mathrm{u}(z_\mathrm{r}) \, \Delta z = 1$, yields a bias-marginalised estimate of the source redshift distribution.  The resulting uncertainty budget hence includes (i) statistical noise in the clustering measurement, (ii) marginalisation over magnification parameters $(p,q)$, and (iii) uncertainty in the redshift evolution of the source bias.  Overall, the clustering-$z$ posteriors are consistent with the directly-calibrated redshift distributions, within the quoted uncertainties, across all surveys and tomographic bins.

\section{Conclusion}
\label{sec:conclu}

In this study, we have calibrated the redshift distributions of weak-lensing source samples from DES-Y3, KiDS-1000, HSC-Y1 and HSC-Y3 using clustering-based redshift inference with DESI-DR1 spectroscopy as the reference sample. Our methodology combines small-scale angular cross-correlations in narrow spectroscopic redshift slices with a consistent forward model that includes (i) the full DESI BGS, LRG, and ELG samples as reference tracers across $0.1 < z < 1.6$, (ii) explicit marginalisation over magnification terms, and (iii) a treatment of the source galaxy bias evolution based on analyses from realistic mock catalogues.

We validated the robustness of the method using \textsc{Buzzard} mock catalogues including realistic galaxy populations, source weights and photometric redshift errors.  From the mocks, we derived empirical estimates of the mean source bias evolution $b_\mathrm{u}(z)$ and its region-to-region scatter, and we verified that the total error budget on the recovered $b_\mathrm{u}(z)\,p_\mathrm{u}(z)$ -- including jackknife covariance and marginalisation over magnification nuisance parameters -- is consistent with the mock-to-mock variance. The mock tests demonstrate that the clustering-$z$ estimator accurately recovers the bias-weighted redshift distribution, and that our adopted uncertainty model properly captures both statistical and modeling contributions.

Applied to the real data, the clustering-$z$ constraints are consistent with the fiducial redshift distributions of DES-Y3, KiDS-1000, HSC-Y1 and HSC-Y3, largely within quoted uncertainties. We observe that magnification contributes non-negligibly at $z\gtrsim1$, particularly for HSC and the upper tomographic bins of DES and KiDS. Because these effects are explicitly modeled rather than removed or assumed negligible, the resulting posteriors reflect conservative and internally calibrated uncertainty estimates.

The present implementation introduces an uncertainty contribution associated with marginalising over source bias evolution. This is the dominant term in the final error budget at high redshift. Future work will explore strategies to further constrain $b_\mathrm{u}(z)$, for example by combining clustering-$z$ with shear–density cross-correlations, galaxy–galaxy lensing, or halo-occupation based forward-modelling. Such extensions should reduce the width of the redshift posteriors and tighten the cosmological constraining power, while maintaining systematic robustness.

The analysis presented here forms the basis of the redshift calibration pipeline for future DESI-DR2 joint $3\times2$-pt inference, where redshift calibration requirements approach the sub-percent level. The techniques developed in this work are also directly applicable to Stage-IV surveys such as LSST and \textit{Euclid}, where spectroscopic cross-correlations will remain essential for achieving the precision required for dark energy and modified gravity constraints.

\section*{Acknowledgements}

We thank \rossi{two anonymous referees and} Tianqing Zhang for helpful comments on a draft of this paper.

This material is based upon work supported by the U.S. Department of Energy (DOE), Office of Science, Office of High-Energy Physics, under Contract No. DE–AC02–05CH11231, and by the National Energy Research Scientific Computing Center, a DOE Office of Science User Facility under the same contract. Additional support for DESI was provided by the U.S. National Science Foundation (NSF), Division of Astronomical Sciences under Contract No. AST-0950945 to the NSF’s National Optical-Infrared Astronomy Research Laboratory; the Science and Technology Facilities Council of the United Kingdom; the Gordon and Betty Moore Foundation; the Heising-Simons Foundation; the French Alternative Energies and Atomic Energy Commission (CEA); the National Council of Humanities, Science and Technology of Mexico (CONAHCYT); the Ministry of Science, Innovation and Universities of Spain (MICIU/AEI/10.13039/501100011033), and by the DESI Member Institutions: \url{https://www.desi.lbl.gov/collaborating-institutions}. Any opinions, findings, and conclusions or recommendations expressed in this material are those of the author(s) and do not necessarily reflect the views of the U. S. National Science Foundation, the U. S. Department of Energy, or any of the listed funding agencies.

The authors are honored to be permitted to conduct scientific research on I'oligam Du'ag (Kitt Peak), a mountain with particular significance to the Tohono O’odham Nation.

This research made use of the following python packages in addition to those already cited in the manuscript: {\sc astropy} \citep{astropy}, {\sc numpy} \citep{numpy}, {\sc scipy} \citep{scipy} and {\sc matplotlib} \citep{matplotlib}.

\section*{Data availability}

Data points for all the figures are available at \url{https://doi.org/10.5281/zenodo.17945717}.

\bibliographystyle{mnras}
\bibliography{example}

\section*{Affiliations}
\scriptsize
\noindent
$^{1}$ Queensland University of Technology,  School of Chemistry \& Physics, George St, Brisbane 4001, Australia\\
$^{2}$ Centre for Astrophysics \& Supercomputing, Swinburne University of Technology, P.O. Box 218, Hawthorn, VIC 3122, Australia\\
$^{3}$ Lawrence Berkeley National Laboratory, 1 Cyclotron Road, Berkeley, CA 94720, USA\\
$^{4}$ Department of Physics, Boston University, 590 Commonwealth Avenue, Boston, MA 02215 USA\\
$^{5}$ Dipartimento di Fisica ``Aldo Pontremoli'', Universit\`a degli Studi di Milano, Via Celoria 16, I-20133 Milano, Italy\\
$^{6}$ INAF-Osservatorio Astronomico di Brera, Via Brera 28, 20122 Milano, Italy\\
$^{7}$ Department of Physics \& Astronomy, University College London, Gower Street, London, WC1E 6BT, UK\\
$^{8}$ Institut d'Estudis Espacials de Catalunya (IEEC), c/ Esteve Terradas 1, Edifici RDIT, Campus PMT-UPC, 08860 Castelldefels, Spain\\
$^{9}$ Institute of Space Sciences, ICE-CSIC, Campus UAB, Carrer de Can Magrans s/n, 08913 Bellaterra, Barcelona, Spain\\
$^{10}$ Department of Physics and Astronomy, The University of Utah, 115 South 1400 East, Salt Lake City, UT 84112, USA\\
$^{11}$ Instituto de F\'{\i}sica, Universidad Nacional Aut\'{o}noma de M\'{e}xico,  Circuito de la Investigaci\'{o}n Cient\'{\i}fica, Ciudad Universitaria, Cd. de M\'{e}xico  C.~P.~04510,  M\'{e}xico\\
$^{12}$ Department of Astronomy \& Astrophysics, University of Toronto, Toronto, ON M5S 3H4, Canada\\
$^{13}$ Department of Physics \& Astronomy and Pittsburgh Particle Physics, Astrophysics, and Cosmology Center (PITT PACC), University of Pittsburgh, 3941 O'Hara Street, Pittsburgh, PA 15260, USA\\
$^{14}$ Department of Physics, The Ohio State University, 191 West Woodruff Avenue, Columbus, OH 43210, USA\\
$^{15}$ The Ohio State University, Columbus, 43210 OH, USA\\
$^{16}$ University of California, Berkeley, 110 Sproul Hall \#5800 Berkeley, CA 94720, USA\\
$^{17}$ Institut de F\'{i}sica d’Altes Energies (IFAE), The Barcelona Institute of Science and Technology, Edifici Cn, Campus UAB, 08193, Bellaterra (Barcelona), Spain\\
$^{18}$ Departamento de F\'isica, Universidad de los Andes, Cra. 1 No. 18A-10, Edificio Ip, CP 111711, Bogot\'a, Colombia\\
$^{19}$ Observatorio Astron\'omico, Universidad de los Andes, Cra. 1 No. 18A-10, Edificio H, CP 111711 Bogot\'a, Colombia\\
$^{20}$ Center for Astrophysics $|$ Harvard \& Smithsonian, 60 Garden Street, Cambridge, MA 02138, USA\\
$^{21}$ Institute of Cosmology and Gravitation, University of Portsmouth, Dennis Sciama Building, Portsmouth, PO1 3FX, UK\\
$^{22}$ University of Virginia, Department of Astronomy, Charlottesville, VA 22904, USA\\
$^{23}$ Fermi National Accelerator Laboratory, PO Box 500, Batavia, IL 60510, USA\\
$^{24}$ Institute of Astronomy, University of Cambridge, Madingley Road, Cambridge CB3 0HA, UK\\
$^{25}$ Institut d'Astrophysique de Paris. 98 bis boulevard Arago. 75014 Paris, France\\
$^{26}$ IRFU, CEA, Universit\'{e} Paris-Saclay, F-91191 Gif-sur-Yvette, France\\
$^{27}$ Department of Astronomy and Astrophysics, UCO/Lick Observatory, University of California, 1156 High Street, Santa Cruz, CA 95064, USA\\
$^{28}$ Center for Cosmology and AstroParticle Physics, The Ohio State University, 191 West Woodruff Avenue, Columbus, OH 43210, USA\\
$^{29}$ School of Mathematics and Physics, University of Queensland, Brisbane, QLD 4072, Australia\\
$^{30}$ Department of Physics, University of Michigan, 450 Church Street, Ann Arbor, MI 48109, USA\\
$^{31}$ University of Michigan, 500 S. State Street, Ann Arbor, MI 48109, USA\\
$^{32}$ Department of Physics, The University of Texas at Dallas, 800 W. Campbell Rd., Richardson, TX 75080, USA\\
$^{33}$ CIEMAT, Avenida Complutense 40, E-28040 Madrid, Spain\\
$^{34}$ NSF NOIRLab, 950 N. Cherry Ave., Tucson, AZ 85719, USA\\
$^{35}$ Department of Physics and Astronomy, University of California, Irvine, 92697, USA\\
$^{36}$ Department of Physics and Astronomy, University of Waterloo, 200 University Ave W, Waterloo, ON N2L 3G1, Canada\\
$^{37}$ Perimeter Institute for Theoretical Physics, 31 Caroline St. North, Waterloo, ON N2L 2Y5, Canada\\
$^{38}$ Waterloo Centre for Astrophysics, University of Waterloo, 200 University Ave W, Waterloo, ON N2L 3G1, Canada\\
$^{39}$ Department of Physics, American University, 4400 Massachusetts Avenue NW, Washington, DC 20016, USA\\
$^{40}$ Department of Astronomy and Astrophysics, University of California, Santa Cruz, 1156 High Street, Santa Cruz, CA 95065, USA\\
$^{41}$ Departament de F\'{i}sica, Serra H\'{u}nter, Universitat Aut\`{o}noma de Barcelona, 08193 Bellaterra (Barcelona), Spain\\
$^{42}$ Instituci\'{o} Catalana de Recerca i Estudis Avan\c{c}ats, Passeig de Llu\'{\i}s Companys, 23, 08010 Barcelona, Spain\\
$^{43}$ Department of Physics and Astronomy, Siena University, 515 Loudon Road, Loudonville, NY 12211, USA\\
$^{44}$ Space Sciences Laboratory, University of California, Berkeley, 7 Gauss Way, Berkeley, CA  94720, USA\\
$^{45}$ Institute for Astronomy, University of Edinburgh, Royal Observatory, Blackford Hill, Edinburgh EH9 3HJ, UK\\
$^{46}$ Ruhr University Bochum, Faculty of Physics and Astronomy, Astronomical Institute (AIRUB), German Centre for Cosmological Lensing, 44780 Bochum, Germany\\
$^{47}$ Instituto de Astrof\'{i}sica de Andaluc\'{i}a (CSIC), Glorieta de la Astronom\'{i}a, s/n, E-18008 Granada, Spain\\
$^{48}$ Departament de F\'isica, EEBE, Universitat Polit\`ecnica de Catalunya, c/Eduard Maristany 10, 08930 Barcelona, Spain\\
$^{49}$ Department of Physics and Astronomy, Sejong University, 209 Neungdong-ro, Gwangjin-gu, Seoul 05006, Republic of Korea\\
$^{50}$ Max Planck Institute for Extraterrestrial Physics, Gie\ss enbachstra\ss e 1, 85748 Garching, Germany\\
$^{51}$ Department of Physics \& Astronomy, Ohio University, 139 University Terrace, Athens, OH 45701, USA\\
$^{52}$ Sorbonne Universit\'{e}, CNRS/IN2P3, Laboratoire de Physique Nucl\'{e}aire et de Hautes Energies (LPNHE), FR-75005 Paris, France\\
$^{53}$ National Astronomical Observatories, Chinese Academy of Sciences, A20 Datun Road, Chaoyang District, Beijing, 100101, P.~R.~China\\
\normalsize

\end{document}